\documentclass[
 reprint,float,mathtools,
 nofootinbib,
 amsmath,amssymb
]{revtex4-2}

\newcommand*\diff{\mathop{}\!\mathrm{d}}

\newcommand*\gen[1]{\langle #1 \rangle}

\newcommand*\R{\mathbb R}

\newcommand*\te[1]{\text{#1}}

\newcommand*\p[1]{\left(#1\right)}
\newcommand*\ps[1]{\left[#1\right]}

\newcommand*\f[2]{\frac{#1}{#2}}
\newcommand*\mat[2]{\left(\begin{array}{#1}#2\end{array}\right)}

\newcommand*\I{\te{i}}
\newcommand*\pd[3]{\frac{\partial^{#3} #1}{\partial {#2}^{#3}}}
\newcommand*\td[3]{\frac{d^{#3}#1}{d #2^{#3}}}

\newcommand{\del}{\partial}

\usepackage{color} 
\usepackage{hyperref}
\hypersetup{
    colorlinks=true, 
    linktoc=all,  
    linkcolor=black, 
}

\usepackage{physics}
\usepackage{graphicx}
\usepackage{dcolumn}
\usepackage{bm}
\usepackage{fontawesome}
\usepackage{subfloat}
\usepackage{subfigure}
\usepackage{soul}
\usepackage{mathrsfs}

\graphicspath{{./}{./figs/}}

\usepackage{environ}
\NewEnviron{eq}{%
\begin{align}\begin{split}
  \BODY
\end{split}\end{align}
}

\definecolor{mulberry}{rgb}{0.5,0,0.5}
\definecolor{elderberry}{rgb}{0.3,0,0.7}
\definecolor{raspberry}{RGB}{227, 11, 93}
\definecolor{blackberry}{RGB}{100, 10, 65}
\definecolor{CherryRed}{RGB}{210, 4, 45}

\newcommand{\rhos}{\rho_S}
\newcommand{\rhol}{\rho_L}

\newcommand{\ths}{\theta_S}
\newcommand{\thl}{\theta_L}
\newcommand{\Thl}{\Theta_L}
\newcommand{\Ths}{\Theta_S}
\newcommand{\phs}{\phi_S}
\newcommand{\phl}{\phi_L}
\newcommand{\sis}{\sigma_S}
\newcommand{\sil}{\sigma_L}
\newcommand{\sisd}{\dot{\sigma}_S}
\newcommand{\sild}{\dot{\sigma}_L}
\newcommand{\sisdd}{\ddot{\sigma}_S}
\newcommand{\sildd}{\ddot{\sigma}_L}
\newcommand{\vps}{\varphi_S}
\newcommand{\vpl}{\varphi_L}
\newcommand{\vpsd}{\dot{\varphi}_S}
\newcommand{\vpld}{\dot{\varphi}_L}
\newcommand{\vpsdd}{\ddot{\varphi}_S}
\newcommand{\vpldd}{\ddot{\varphi}_L}
\newcommand{\sisb}{\bar{\sigma}_S}
\newcommand{\Phib}{\bar{\Phi}}
\newcommand{\dPhi}{\delta_\Phi}
\newcommand{\dPhid}{\dot{\delta}_\Phi}
\newcommand{\adPhi}{| \delta_\Phi |}
\newcommand{\dPhidd}{\ddot{\delta}_\Phi}
\newcommand{\dsis}{\delta_S}
\newcommand{\adsis}{| \delta_S |}
\newcommand{\dsisd}{\dot{\delta}_S}
\newcommand{\dsisdd}{\ddot{\delta}_S}
\newcommand{\Phid}{\dot{\Phi}}
\newcommand{\Phidd}{\ddot{\Phi}}
\newcommand{\omp}{\omega_{\text{f.o.}}}

\newcommand{\rhoax}{\rho_{\text{ax}}}

\newcommand{\Thzer}{\Theta_0}
\newcommand{\calF}{\mathcal{F}}
\newcommand{\gagg}{g_{a \gamma \gamma}}
\newcommand{\ktw}{\tilde{k}}
\newcommand{\tosc}{t_{\text{osc}}}

\newcommand{\Mpl}{M_{\text{Pl}}}
\newcommand{\eV}{\; \text{eV}}

\newcommand{\GeV}{\; \text{GeV}}

\begin{document}

\title{Friendship in the Axiverse: Late-time direct and astrophysical signatures of early-time nonlinear axion dynamics}
\author{David Cyncynates}
 \email{davidcyn@stanford.edu}
\author{Tudor Giurgica-Tiron}%
 \email{tgt@stanford.edu}
\author{Olivier Simon}%
 \email{osimon@stanford.edu}
\author{Jedidiah O. Thompson}%
 \email{jedidiah@stanford.edu}
\affiliation{%
 Stanford Institute for Theoretical Physics\\ Stanford University\\ 382 Via Pueblo, Stanford, CA 94305, USA
}%
\begin{abstract}
A generic low-energy prediction of string theory is the existence of a large collection of axions, commonly known as a string axiverse. Axions also have a natural cosmological production mechanism, vacuum misalignment, making them well-motivated dark matter (DM) candidates. Much work on axion 
production
has considered the case of a single free axion, but in a realistic axiverse, string axions are expected to be distributed densely over many orders of magnitude in mass, and to interact with one another through their joint potential. 
In this paper, we show that non-linearities in this potential lead to a new type of resonant energy transfer between axions with nearby masses.
This resonance generically transfers energy from axions with larger decay constants to those with smaller decay constants, and leads to a multitude of signatures.  These include enhanced direct detection prospects 
for a resonant pair comprising even a small subcomponent of dark matter,
and boosted small-scale structure
if the pair is the majority of DM.
Near-future iterations of experiments such as ADMX and DM Radio will be sensitive to this scenario, as will astrophysical probes of DM substructure.
\end{abstract}
\maketitle
\tableofcontents
\section{Introduction} \label{sec:intro}
Among the best-motivated 
extensions of the Standard Model (SM)
are \textit{axions}, periodic pseudoscalar fields with an approximate shift symmetry that protects their mass from radiative corrections.\footnote{The term ``axion-like particles'' is also used in the literature.}  The most well-known example is the \textit{QCD axion}, which was originally proposed as a solution to the puzzling smallness of the neutron's electric dipole moment \cite{PhysRevLett.38.1440,peccei1791,PhysRevLett.40.223,PhysRevLett.40.279}.  This is not the only type of axion that can exist though: axions can be quite generic in UV completions of the SM with compact extra dimensions and nontrivial topologies, the principle example of which is string theory \cite{Witten1984,banks1996couplings,Witten2006}. The combined motivation of the QCD axion and string theory lead to predictions of a plenitude of string axions with mass scales spanning many orders of magnitude, a possibility referred to as the \textit{String Axiverse} \cite{arvanitaki2010string}.

A light axion $\phi$ with potential $V(\phi)$ has a natural production mechanism known as the \textit{misalignment mechanism} \cite{preskill1982,abbott1982,dine1982,turner1983coherent}, whereby the axion field is effectively initialized at some finite displacement from the minimum of its potential.  
These potentials are generally expected to be periodic and at leading order are often well-approximated by a cosine:
\begin{equation} \label{eq:cosinePotential}
    V(\phi) = m^2 f^2 \left( 1 - \cos \left( \frac{\phi}{f} \right) \right)\,.
\end{equation}
Here $f$ is the axion’s \textit{decay constant}, which is expected to suppress all couplings of the axion field to the SM \cite{srednicki1985axion,graham2013,pdg2020}.  The periodicity of the axion potential provides a natural 
measure on the space of initial conditions.  In the absence of any dynamic \cite{raymond2019axion,hall2020axion,co2018axion,Huang:2020etx} or anthropic considerations \cite{freivogel2008anthropic,arvanitaki2010}, a reasonable expectation is that the initial condition $\phi_0$ is drawn effectively randomly from the interval $[-\pi f, \pi f )$.  Defining $\Thzer \equiv \phi_0 / f$ we can then compute the present-day energy density in the axion field, yielding:
\begin{equation} \label{eq:misalignmentFraction}
\Omega_{\te{ax}} \approx  0.4 \p{\f{\Thzer}{\pi/2}}^2 \p{\f{m}{10^{-17} \eV}}^{1/2} \p{\f{f}{10^{16} \GeV}}^2\,,
\end{equation}
which receives corrections as $|\Thzer|$ gets very close to $\pi$ \cite{turner1986,lyth1992,strobl1994,kobayashi2013,bae2008update,visinelli2009dark,arvanitaki2020large}.
At these large misalignments, self-interactions from the cosine potential
can play a significant role in the field's evolution
at early times, leading in extreme cases to exponential growth of spatial perturbations and a plethora of associated signatures \cite{turner1986,lyth1992,strobl1994,kobayashi2013,bae2008update,visinelli2009dark,arvanitaki2020large}.

The above discussion of the misalignment mechanism applies to the case of a single axion uncoupled from all other particles in the spectrum.  However the generic prediction of the axiverse is actually many axions, spanning orders of magnitude in both mass $m$ and decay constant $f$.  A more realistic picture of the axiverse is then a sector consisting of $N$ pseudoscalar fields that pick up nonperturbative contributions to their collective potential from $M$ instantons.  We typically expect $M \gg N$ \cite{mehta2021superradiance}, so no axion is expected to be massless.  This results in a generic potential of the form:
\begin{align} \label{eq:axiversePotential}
    V(\phi_1,\dots,\phi_N) = \sum_{i = 1}^M\Lambda_i^4\ps{1 - \cos\p{\sum_{j = 1}^N{\cal Q}_{ij}\f{\phi_j}{f_j} + \delta_i}}\,,
\end{align}
where $\delta_i$ are arbitrary constant phases, $f_i$ are the various decay constants, and $\mathcal{Q}_{i j}$ are rational numbers associated with the axion charges under each instanton \cite{halperin1998axion,easther2006random,bachlechner2015planckian,bachlechner2016axionic,bachlechner2018multiple,demirtas2020kreuzer,mehta2021superradiance}.  The energy scales $\Lambda_i$ are typically exponentially suppressed relative to the UV string scale $\Lambda_{\text{str}}$ by instanton actions $S_i$: $\Lambda_i \sim \Lambda_{\text{str}} e^{- \lambda S_i}$ where $\lambda$ is an $\mathcal{O}(1)$ coupling constant.

In the absence of strong priors on the instanton actions,
the axions are expected to have
an approximately log-flat distribution in mass \cite{arvanitaki2010string}, an expectation that has been confirmed in specific orientifold compactifications of type IIB string theory \cite{mehta2020superradiance,mehta2021superradiance}.  The range of axion masses can easily span several dozen orders of magnitude, from smaller than the current Hubble rate $H_0$ to order $\Mpl$.  The decay constants, meanwhile, are typically more narrowly distributed but can still range over a few orders of magnitude $f \sim 10^{12} - 10^{19} \GeV$ \cite{arvanitaki2010string,halverson2019towards,mehta2021superradiance}.
The number of axions in these compactifications is proportional to the Hodge number of the orientifold and thus can easily be $\mathcal{O}(100s)$,
making
``coincidences'' in axion mass a common occurrence: $\mathcal{O}(100)$ axions distributed log-flat over $\mathcal{O}(60)$ orders of magnitude in mass imply that each axion is on average only a factor of a few away from an axion with a similar mass.  By chance some pairs of axions will be even closer, 
and as we will show, these coincident pairs can be significantly more visible than other axions in the axiverse.

\begin{figure*}
    \centering
    \includegraphics[width = \textwidth]{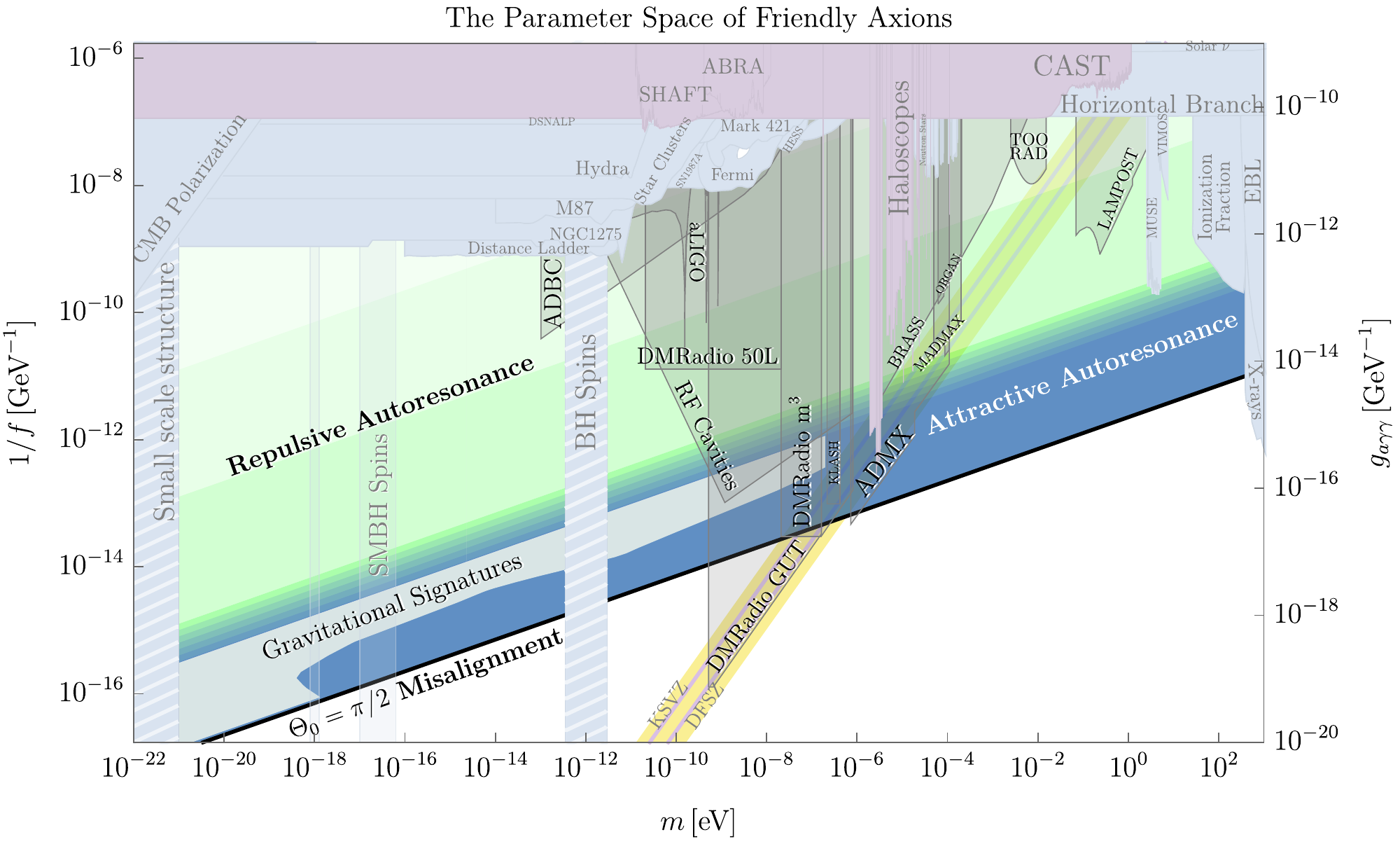}
    \caption{
    Summary of parameter space, constraints, and signatures for a pair of friendly axions undergoing autoresonance. The lower black solid line (``$\Theta_0 = \pi/2$ Misalignment'') corresponds to the decay constant that produces the correct relic abundance for an initial misalignment angle of $\pi/2$ with the simple cosine potential of Eq.~\ref{eq:cosinePotential}.  Autoresonance allows an axion whose parameters lie near this line (i.e.\ an axion that would produce the proper DM abundance in the absence of interactions via misalignment) to efficiently transfer its energy to an axion with a much smaller decay constant.  The blue region (``Attractive Autoresonance'') labels the parameter space accessible to the simple model of Eq.~\ref{eq:twoAxionPotential}. 
    For even smaller values of $f$, nonperturbative structure growth quenches the autoresonant energy transfer in this simple model (see Sec.~\ref{sec:perturbations}), but axion DM with these parameters can still be generated for slightly more complicated axion potentials that include repulsive self-interactions to prevent structure growth (Sec.~\ref{sec:repulsiveSelfInteractions}).  These regions of parameter space are labeled ``Repulsive Autoresonance.''  We also show constraints and projections for various experimental efforts to detect axions and axion DM through the axion-photon coupling $\gagg$~\cite{ciaran_o_hare_2020_3932430,ciaran_o_hare,PhysRevLett.119.031302,armengaud2017constraining,PhysRevD.101.123026,PhysRevD.101.103023,rogers2021strong,buen2020constraints,reynolds2020astrophysical,marsh2017new,dessert2020x,wouters2013constraints,calore2020bounds,ayala2014revisiting,vinyoles2015new,regis2021searching,grin2007telescope,cadamuro2012cosmological,fedderke2019axion,DMRadioGUT,alesini2017klash,stern2016admx,BRASS,lasenby2020microwave,berlin2020axion,berlin2020heterodyne,liu2019searching,michimura2020dance,DMRadiom3,baryakhtar2018axion,nagano2019axion,mcallister2017organ,schutte2021axion}, where we have assumed $\gagg \simeq \frac{\alpha}{4 \pi f}$.  In the friendly scenario, axion DM can be produced with untuned initial misalignment angles and with much stronger couplings to the SM than would be expected based on the decay constant predicted by Eq.~\ref{eq:misalignmentFraction}. We note that these direct detection signatures persist even when the friendly axions make up only a subcomponent of DM (Sec.~\ref{sec:directDetection}). The region labeled ``Gravitational Signatures'' can be probed using DM substructures generated during autoresonance (Sec.~\ref{sec:indirectDetection}).  The horizontal axis of this plot refers to the overall mass scale of the two axions (i.e.\ the parameter $m$ in our potential Eq.~\ref{eq:twoAxionPotential}), while the precise axion masses in the mass basis have additional small dependence on the parameters $\mu$ and $\calF$. As explained in Sec.~\ref{sec:superradiance}, the exclusions from black hole spin measurements extend to arbitrarily small values of $f$ only when viewed as constraints on the specific scenario of the pair of axions being $\mathcal{O}(1)$ of dark matter.  
    }
    \label{fig:new_parameter_space}
\end{figure*}

For concreteness, in this paper we consider a sector consisting of two axions receiving two instanton contributions to their potential:
\begin{align}\label{eq:twoAxionPotentialPhi}
V(\phl, \phs) &= \Lambda_1^4 \left( 1 - \cos \left( \frac{\phs}{f_S} + \frac{\phl}{f_L} \right) \right) \\
&+ \Lambda_2^4 \left( 1 - \cos \frac{ \phl}{f_L} \right)\,, \nonumber
\end{align}
where we will focus on the case where the axion masses are within a factor of $\mathcal{O}(2)$ from each other but the decay constants are not necessarily close.  This potential can be shown in a nicer form by transforming to angular variables $\ths \equiv \frac{\phs}{f_S}$ and $\thl \equiv \frac{\phl}{f_L}$ and then writing the instanton scales as $\Lambda_1^4 \equiv m^2 f^2$, $\Lambda_2^4 \equiv \mu^2 \calF^2 m^2 f^2$, yielding
\begin{align} \label{eq:twoAxionPotential}
    V( \thl , \ths ) &= m^2 f^2 \left[ \left( 1 - \cos \left( \ths + \thl \right) \right) \right. \\
    & + \left. \mu^2\mathcal{F}^2 \left( 1 - \cos \thl \right) \right]\,. \nonumber
\end{align}
Letting $f \equiv f_S$ and $\calF \equiv \frac{f_L}{f_S}$, the mass of $\phi_S$ is $m_S \equiv m$, and that of $\phi_L$ is $m_L \equiv \mu m$.  We will focus on the case where the parameters are in the range $0.75 \lesssim \mu < 1$ and $\calF \gg 1$.  We term such a similar-mass pair ``friendly'' and will refer to $\phl$ and $\phs$ as the ``long'' and ``short'' axion respectively in reference to the size of their decay constants.  We note that $\phl$ and $\phs$ are not exact mass eigenstates, but as discussed in App.~\ref{app:massVsInteractionBasis} they are very nearly mass eigenstates when $\calF \gg 1$.  We will thus neglect this subtlety for the current qualitative discussion but correctly account for it in the main text below.

In the absence of the axion interactions, Eq.~\ref{eq:misalignmentFraction} would suggest that for similar masses and $\mathcal{O}(1)$ misalignments, the long axion will always dominate the late-time energy density of the pair because of its larger decay constant.  This is true for $\mu \lesssim 0.75$, but when the axion masses get within roughly $25\%$ of each other, a new effect occurs and can result in highly efficient energy transfer from the long axion to the short axion.
We identify this new phenomenon as an instance of \textit{autoresonance}, a well-known effect in the mechanics of classical oscillators~\cite{landau_mechanics,bogoliubov,fajansFriedland,rajasekar,glebov}.
Near the bottom of the potential, both axions oscillate with a frequency approximately given by their mass: the long axion at $\mu m$ and the short axion at $m$.  However, because the short axion has a nonlinear potential, its oscillation frequency receives corrections depending on its amplitude.  At $\mathcal{O}(1)$ amplitudes (such as those that may be expected from a random initial misalignment angle), its oscillation frequency can become so detuned from $m$ that it lines up instead with $\mu m$.  At this point the small interaction with the long axion can resonantly drive the short axion and hold it at this fixed amplitude, effectively counteracting the damping effects of Hubble friction.  Locking onto this autoresonance is not a guaranteed process and does depend on the initial misalignment angles, but once it has been established it is extremely stable and persists until nearly all energy has been transferred out of the long axion and into the short axion.  This is by no means a tuned occurrence: As a representative example, for $\mu \sim 0.8$ and $\mathcal{F} \sim 20$, roughly half of the possible combinations of initial misalignment angles result in autoresonance, leading to the late-time energy density of the sector being dominated by the short axion.

The signatures of a period of autoresonance are quite dramatic.  Axion couplings to the SM are generically suppressed by their decay constant, for example they are expected to have couplings to the photon of the form \cite{kim1979weak,shifman1980confinement,dine1981simple,srednicki1985axion,graham2013,tanabashi2018review,pdg2020}:
\begin{equation} \label{eq:axionPhotonCoupling}
    \mathcal{L} \supset - \frac{g_{a \gamma \gamma}}{4} \phi F_{\mu \nu} \tilde{F}^{\mu \nu}\,,
\end{equation}
where $g_{a \gamma \gamma} \sim \frac{\alpha}{4 \pi f}$ with $\alpha \equiv \frac{e^2}{4 \pi}$ the QED fine structure constant.  The short axion (with the smaller decay constant) is thus typically coupled more strongly to the SM than the long axion.  Autoresonance efficiently transfers an axion sector's energy density into a form more easily probed experimentally.  As we summarize in Fig.~\ref{fig:new_parameter_space}, much of the short axion parameter space will be probed with existing and upcoming experiments. 
We emphasize that this enhancement can be observable regardless of whether the friendly pair in question comprises the totality of the DM or only a subcomponent.

In addition, a long period of autoresonance means that the short axion spends a long time under the influence of its nonlinearities.  As shown in Ref.~\cite{arvanitaki2020large} in the context of a single axion model, this can lead to a parametric resonant enhancement in the growth of spatial inhomogeneities of the axion field.  If the axion makes up all of the DM, such inhomogeneities eventually collapse into gravitationally-bound dark matter minihalos that can be probed purely through their gravitational effects.  For simple axion potentials such as Eq.~\ref{eq:cosinePotential}, Ref.~\cite{arvanitaki2020large} found that this required initial misalignments of the order $| \Thzer - \pi | \lesssim 10^{-5}$.  Such a tuning can be motivated by anthropics or dynamical mechanisms \cite{Huang:2020etx}, and in broader classes of axion potentials it can be avoided entirely \cite{arvanitaki2020large}, but similar minihalo phenomenology and signatures can also be reproduced by a friendly autoresonating pair of axions with untuned initial conditions
provided
the friendly pair comprises the entirety of the DM.

The structure of the rest of this paper is as follows: In Sec.~\ref{sec:zeroModes} we outline the dynamics of autoresonance for the spatially homogeneous components of the axion fields in greater detail.
In Sec.~\ref{sec:perturbations} we extend our analysis to inhomogeneities in both fields and show that those in the short axion grow due to a parametric resonance instability.
In extreme cases, inhomogeneities can grow nonperturbatively large during autoresonance, quenching the transfer of energy between the axions.
We then move to discussing signatures of autoresonance in Sec.~\ref{sec:signatures}, going over both the significant effects on direct detection parameter space and the astrophysical and cosmological probes of dense minihalos. In Sec.~\ref{sec:repulsiveSelfInteractions} we broaden our scope somewhat to potentials with repulsive self-interactions,
which do not lead to structure growth but can still support autoresonance.
Finally, in Sec.~\ref{sec:discussion} we summarize the results of this paper and discuss its implications and future directions.

To streamline the presentation we have placed several useful results and derivations in the appendices.  In App.~\ref{app:massVsInteractionBasis} we discuss the difference between the mass and interaction bases for the coupled axion system and show that it has only marginal effects on our analysis.  In App.~\ref{app:autoresInDetail} we give a lengthier analytic treatment of autoresonance for a pair of friendly axions, and we do the same for aspects of perturbative structure growth in App.~\ref{app:perturbations_in_detail}.  App.~\ref{app:nonperturbative_in_detail} concludes with a detailed description and discussion of the numerical simulations used to study the case of nonperturbative structure growth.

Throughout this paper we work in units where $\hbar = c = 1$, and we use the reduced Planck mass $\Mpl \equiv (G/8\pi)^{-1/2} \approx 2.4 \times 10^{18} \GeV$. We use the Planck 2018 results \cite{aghanim2020planck} for our cosmological parameters, taking the dark matter fraction of the universe to be $\Omega_\te{DM} = 0.23$, the scale factor at matter radiation equality $a_\te{eq} = 1/3388$, the present-day Hubble parameter $H_0 = 67.66 \te{(km/s)/Mpc}$, and the Hubble parameter at matter-radiation equality $H_\te{eq} = 2.2\times 10^{-28}\te{eV}$. We work with a mostly negative metric signature $(+,-,-,-)$.

\section{Friendly zero-mode dynamics} \label{sec:zeroModes}

At energies well below its instanton scale, an axion in an expanding universe is well-approximated by a damped harmonic oscillator. Its amplitude decays because of Hubble friction as $a^{-3/2}$, while its energy density falls as $a^{-3}$. The dynamics of our model (Eq.~\ref{eq:twoAxionPotential}) differ from this simple picture in two important ways. First, at early times, the axion field has enough energy that attractive self-interactions of the cosine potential are important, and each axion behaves as a damped \emph{nonlinear} oscillator, with oscillation frequency that is \emph{smaller} than its rest mass. Second, the axions are coupled to one another, allowing energy to flow between them. These two facts lead to the possibility of \emph{autoresonance}, wherein a driven axion may dynamically adjust its frequency to match that of a driver axion. During autoresonance, the driven axion can receive most of the driver's energy, leading to new late time signatures.

We begin by taking appropriate limits of the two-axion model (Eq.~\ref{eq:twoAxionPotential}) to reduce to the equation for a single driven pendulum, which exhibits the same essential behavior.
The equations of motion for the axions $\ths$ and $\thl$ specified by the potential Eq.~\ref{eq:twoAxionPotential} in an FLRW background are
\begin{subequations} \label{eq:eomLSdimensionfull}
\begin{equation}
    \frac{\square}{m^2} \thl + \frac{1}{\calF^2} \sin ( \ths + \thl ) + \mu^2 \sin \thl = 0\,,
\end{equation}
\begin{equation}
    \frac{\square}{m^2} \ths + \sin ( \ths + \thl ) = 0\,,
\end{equation}
\end{subequations}
where $\square \equiv \del_t^2 + 3 H \del_t - \frac{{\bm \nabla}^2}{a^2}$ for a scalar field in FLRW and $H = \frac{1}{2 t}$ during radiation domination.  In this section we are focused on the homogeneous component of both fields, so we will neglect the spatial derivatives and denote the homogeneous components of the fields by $\Ths$ and $\Thl$.  In addition, we will measure time in units of $m^{-1}$, allowing us to write these in a simpler form:
\begin{subequations} \label{eq:eomLS}
\begin{equation} \label{eq:eomThL}
    \del_t^2 \Thl + \frac{3}{2t} \del_t \Thl + \frac{1}{\calF^2} \sin ( \Ths + \Thl ) + \mu^2 \sin \Thl = 0\,,
\end{equation}
\begin{equation} \label{eq:eomThS}
    \del_t^2 \Ths + \frac{3}{2 t} \del_t \Ths + \sin ( \Ths + \Thl ) = 0\,.
\end{equation}
\end{subequations}
In the large-$\calF$ limit, the equation of motion for $\Thl$ decouples from $\Ths$, causing the $\Thl$ field to behave as an independent nonlinear oscillator subject only to Hubble friction.  The solution to such an equation for an $\mathcal{O}(1)$ initial misalignment $\Theta_{L,0}$ and $t \gg 1$ is well-known: $\Thl(t) \propto \Theta_{L,0} t^{-3/4} \cos ( \mu t )$, and at late times this becomes small.  If we expand the $\Ths$ equation of motion in small $\Thl$ we obtain:
\begin{equation} \label{eq:eomThsApprox}
    \del_t^2 \Ths + \frac{3}{2 t} \del_t \Ths + \sin \Ths \approx - \Thl \cos \Ths\,.
\end{equation}
Provided the amplitude of $\Ths$ is not too large, $\cos \Ths$ will be reasonably close to 1, and we can approximate\footnote{This formally corresponds to the limit $\Ths\gg(1/6)\Ths^3\gg(1/2)\Ths^2\Thl$. In practice, this approximation appears to work quite well even when the hierarchy is not very large.}
\begin{equation} \label{eq:eomThsApproxApprox}
    \del_t^2 \Ths + \frac{3}{2 t} \del_t \Ths + \Ths - \f{1}{6}\Ths^3 \approx - \Thl\,,
\end{equation}
which is the equation of motion for a damped, driven pendulum in the small amplitude limit, formally known as a Duffing oscillator.

\begin{figure}
    \centering
    \includegraphics[width = \columnwidth]{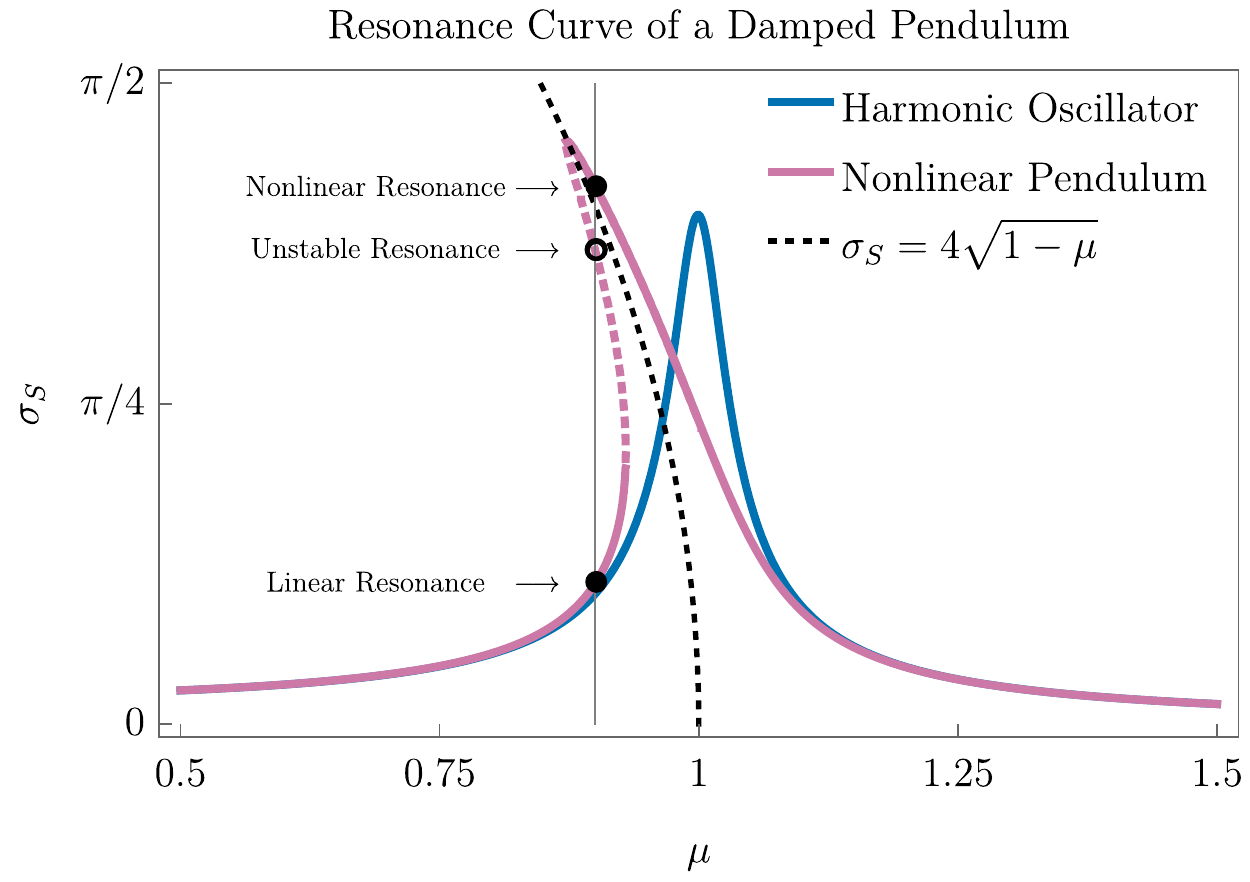}
        \caption{
        Resonance curve (Eq.~\ref{eq:resonanceCurve}) for a pendulum of fundamental frequency $m=1$ driven at an amplitude of $\sigma_d=4\times 10^{-3}$ at a damping of $\gamma=2.5\times 10^{-3}$ (Magenta). The vertical line is drawn for a driving frequency of $\mu=0.9$ and intersects the resonance curve at the three equilibrium solutions. The bottom solution (the linear branch) is stable and well-approximated by the harmonic oscillator resonance curve (Blue). The intermediate solution living on the dashed segment is unstable. The top solution is once again stable, and corresponds to the autoresonant solution for the short axion (with amplitude $\sis$). The Dashed Black curve represents the frequency curve of a free pendulum.}
        \label{fig:landau curves}
\end{figure}

We first consider the left hand side of Eq.~\ref{eq:eomThsApproxApprox} in isolation and in the absence of damping,
\begin{align}
    \del_t^2\Ths + \Ths - \f{1}{6}\Ths^3 = 0\,.
\end{align}
With an oscillatory ansatz $\Theta_S(t) \approx  \sigma_S\cos(\omega t + \delta)$, we find that, due to the attractive self-interactions, the oscillation frequency $\omega$ of the pendulum is a decreasing function of its amplitude $\sigma_S$:
\begin{eq}
\label{eq:duffing frequency}
    \omega(\sigma_S) \approx 1 - \frac{\sigma_S^2}{16} + {\cal O}(\sigma_S^4).
\end{eq}
This fact is key to autoresonance. Because of this effect, the range of frequencies below the fundamental frequency $1$ is now accessible to possible resonances. As we will see below, by driving the pendulum at a frequency $\mu$ below the fundamental, the system can automatically evolve to a new equilibrium amplitude at which $\omega(\sigma_S) \approx \mu$.

We now move to the next stage of complexity by re-introducing constant damping and driving terms,
\begin{eq} \label{eq:LandauOscillator}
    \del_t^2 \Ths + \gamma \del_t \Ths + \Ths - \f16\Ths^3 = \sigma_d\cos(\mu t)\,,
\end{eq}
where $\gamma$ and $\sigma_d$ are the damping and driving coefficients respectively. 
The long-term effect of the driver is best depicted by a {\em resonance curve}, which shows the possible equilibrium amplitudes $\sigma_S$ as a function of the driver's frequency $\mu$. In the absence of the nonlinear term $-\f16\Ths^3$, the oscillator's equilibrium amplitude is unique:
\begin{eq} \label{eq:linearResonanceCurve}
    \sigma_S = \f{\sigma_d}{\sqrt{\p{1 - \mu^2}^2 + \gamma^2 \mu^2}}\,,
\end{eq}
where $1 - \mu^2$ represents the difference between the squares of the oscillator frequency $1$ and the driver $\mu$.
An intuitive trick to extend this resonance curve to the nonlinear oscillator is to replace the fundamental frequency $1$ in Eq.~\ref{eq:linearResonanceCurve} with its amplitude-dependent version in Eq.~\ref{eq:duffing frequency}:
\begin{eq}
\label{eq:resonanceCurve}
    \sigma_S = \f{\sigma_d}{\sqrt{\p{\omega(\sigma_S)^2 - \mu^2}^2 + \gamma^2 \mu^2}}\,.
\end{eq}
By introducing amplitude dependence to the resonance condition, there can now be up to three equilibrium amplitudes for $\Ths$ as a function of the driver frequency $\mu$, which we show in Fig.~\ref{fig:landau curves}. The smallest amplitude corresponds to the regime of linear excitation of the pendulum and is stable to perturbations; we will refer to this solution as the {\em linear branch}. The intermediate amplitude solution is unstable to small perturbations. The third and largest amplitude equilibrium, which we will refer to as the {\em nonlinear branch}, is again stable and, as we will show below, corresponds to autoresonance.

We now return to cosmological scenario of Eq.~\ref{eq:eomThsApproxApprox}, where friction and driving are decaying functions of time. In particular, the damping is given by the Hubble parameter $\gamma\to 3 H(t) \propto t^{-1}$, and the amplitude of the driver follows the cosmological evolution of the long axion, namely $\sigma_d\to\sigma_L(t)\propto t^{-3/4}$. In spite of this time dependence, the notion of a resonance curve is still useful in the cosmological scenario since both damping and driving vary slowly compared to the rapid oscillatory timescale when $t\gg1$, allowing $\sis$ to arrive at a quasi-equilibrium.

\begin{figure}
    \centering
    \includegraphics[width = \columnwidth]{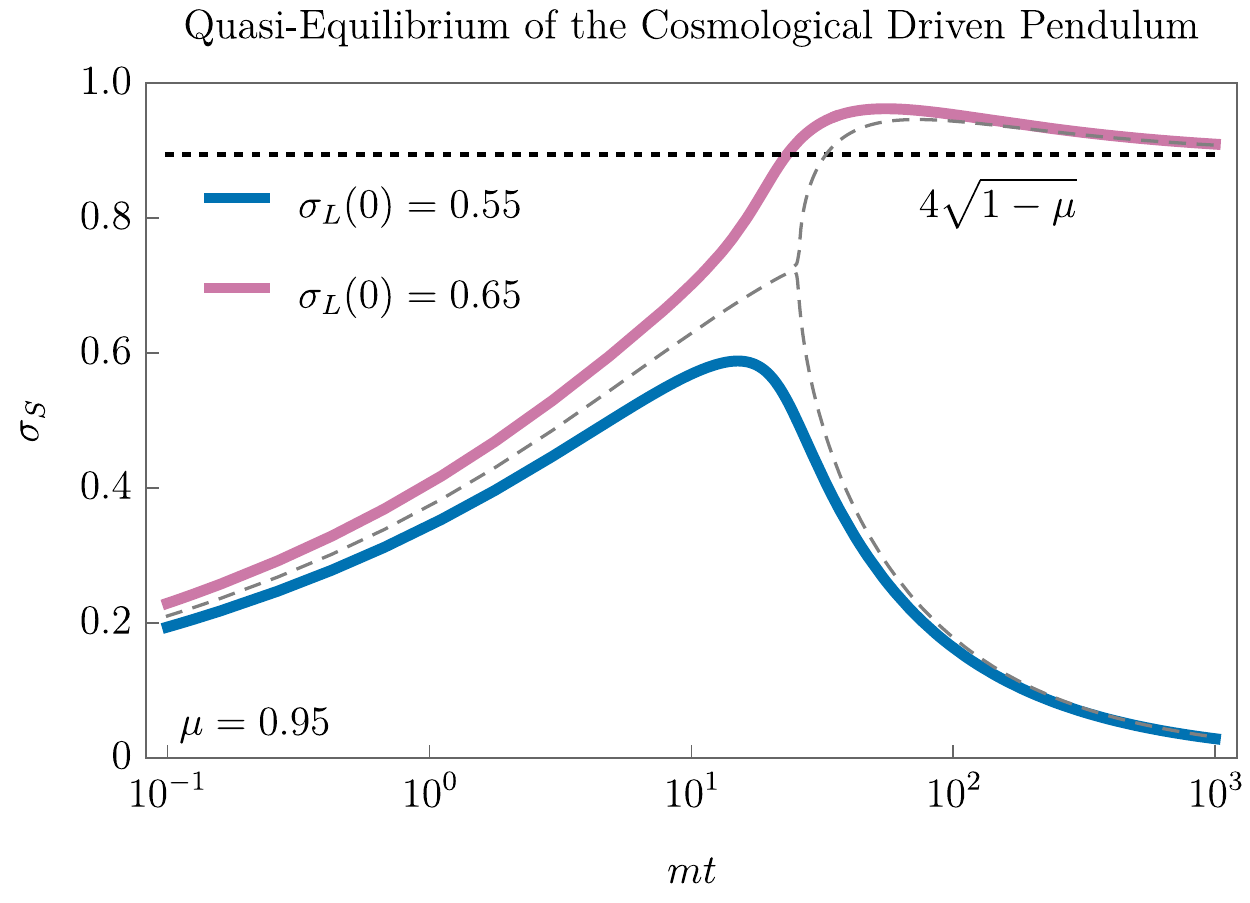}
    \caption{
        Quasi-equilibrium trajectories of the short amplitude $\sigma_S$ as it tracks the time-dependent resonance curve, for two values of the initial driver amplitude $\sigma_L(0)$ and a fixed driver frequency $\mu = 0.95$. For small driver amplitudes (Blue), the short axion never leaves the linear branch of the resonance curve. For large enough driver amplitudes (Magenta), the short axion is smoothly lifted from zero amplitude to the stable nonlinear branch, which converges to the undamped pendulum solution ($\omega (\sis ) = \mu$ with $\omega(\sis)$ given by Eq.~\ref{eq:duffing frequency}). At the critical driving, the two branches are equally accessible as a bifurcation (Gray, Dashed). See App.~\ref{app:NLrescurve}, and in particular Fig.~\ref{fig:resCurveEvolution} for further details.
    }
    \label{fig:equilibriumTrajectory}
\end{figure}

Remarkably, it is the cosmological evolution of $\gamma$ and $\sigma_d$ that is responsible for autoresonance. We show this effect in Fig.~\ref{fig:equilibriumTrajectory}, where we plot the instantaneous equilibrium of $\sis$ at each point in time for two different initial $\Thl$ amplitudes $\sil(0)$ and fixed driving frequency $\mu$. Early on, the system is dominated by friction, and the equilibrium value of $\sis$ is small. At late times, Hubble friction decays faster than the driver, resulting in equilibrium solutions on both the linear branch near zero, and on the nonlinear branch at large amplitude. Whether the short axion is smoothly carried up to the nonlinear branch $\sis \to 4\sqrt{1 - \mu}$, or left on the linear branch where $\sis \to 0$ depends on whether the initial driving amplitude $\sil (0)$ is large enough. The same reasoning can be applied to Eq.~\ref{eq:eomThsApprox} with only slight modifications,  which we discuss in App.~\ref{app:autoresInDetail}.

Thus we have identified a cosmological mechanism for arriving at the nonlinear branch of the resonance curve. This instance of autoresonance is not unique.  For example, Ref.~\cite{fajansFriedland} showed that autoresonance can be induced by sweeping the driver's frequency and applied this effect to a variety of systems, including planetary dynamics and plasma physics. In other words, autoresonance is a generic feature of many driven nonlinear systems where some external parameter varies, and may be a generic feature of the axiverse as well.

\begin{figure}[t]
    \centering
    \includegraphics[width=\columnwidth]{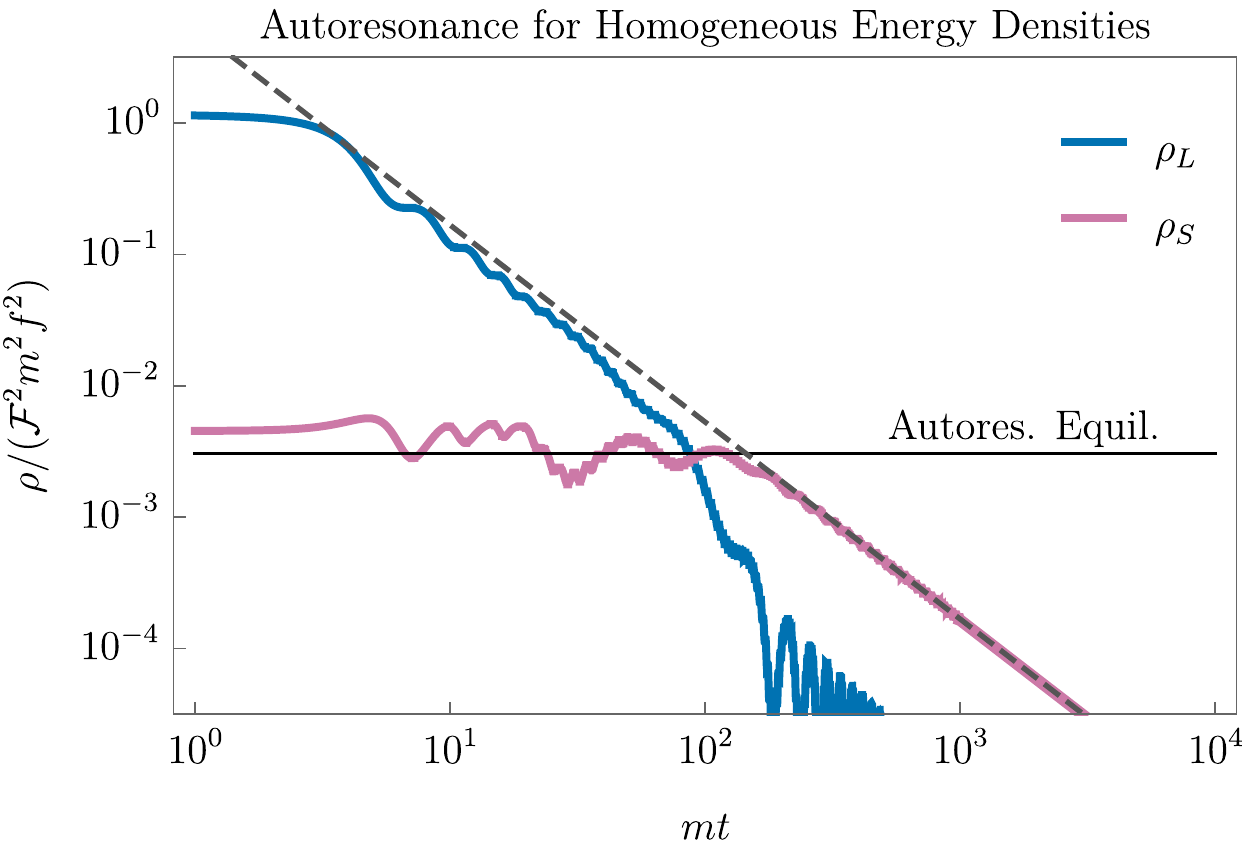}
    \caption{Evolution of energy densities in the short and long axions for generic initial conditions that lead to autoresonance.  The parameters taken here are $\mu = 0.8$, $\calF = 20$, $\Theta_{S,0} = 0.4 \pi$, $\Theta_{L,0} = 0.8 \pi$, although the qualitative features are similar for broad ranges of initial conditions within the ``friendly'' band $0.75 \lesssim \mu < 1$.  $\calF$ sets the rough initial ratio of energy densities in the short and long modes but does not play any significant role in determining whether the system lands on autoresonance provided it is somewhat large ($\calF \gtrsim 5$).  The short axion energy density is held approximately constant at a value determined by the equilibrium amplitude of Eq.~\ref{eq:duffing frequency} (labeled ``Autores.\ Equil.'') until the long axion no longer has enough energy density to drive the autoresonance.  Note that the final energy densities are not equal, but rather the short axion ends up with virtually all of the system's energy density.  At late times, the mass mixing of the two axions leads to rapid flavor oscillations in the long axion's energy density.  Rotating to the mass basis (see App.~\ref{app:massVsInteractionBasis}) removes these.}
    \label{fig:sampleAutoresZeroModes}
\end{figure}

We now return to the full system of Eq.~\ref{eq:eomLS}, which describes the homogeneous part of the coupled axion system of Eq.~\ref{eq:twoAxionPotential} in an FLRW background.
For some range of values of $\mu$, $\calF$, and initial misalignment angles $\Ths$ and $\Thl$, the system autoresonates, with $\Ths$ dynamically adjusting its amplitude so that its frequency matches the driver frequency $\mu$, and then remaining at this amplitude until backreaction onto $\Thl$ eventually cuts off the autoresonance.  For a representative choice of parameters this can be seen concretely in Fig.~\ref{fig:sampleAutoresZeroModes}.  The physics of this autoresonance is quite rich, and in App.~\ref{app:autoresInDetail} we develop a formalism that lets us quantitatively understand many details about it, but for the remainder of this section we focus on three questions.  First, at what amplitude is the short field held during autoresonance?  Second, assuming the system begins to autoresonate, what eventually cuts it off (i.e.\ how long does it last) and what is the final energy density in the short axion field?  And third, what range of parameters ($\mu$, $\calF$, and the initial misalignment angles) lead to autoresonance?

The first question is also the simplest to answer. If a nonlinear oscillator is being autoresonantly driven in its steady state, its amplitude will be chosen such that its frequency approximately matches the driver frequency.  In the case of two friendly axions discussed here, the short axion is driven by the long axion, which oscillates with frequency $\mu$ in its linear regime (i.e.\ once $\Thl \ll 1$).  As discussed above, the frequency of a cosine oscillator as a function of its amplitude $\sis$ is given by Eq.~\ref{eq:duffing frequency}.  During autoresonance, the amplitude of $\Ths$ will remain fixed at $\omega ( \sis) \approx \mu$.  For $\mu = 0.8$ for example, this evaluates to $\sis \approx 1.82$.

This ``locking'' of the $\Ths$ amplitude has important cosmological effects.  Hubble friction operates to steadily dilute the total axion energy density, but because $\Ths$ is autoresonantly held at fixed amplitude, its energy density does not decrease.  As a result, there is a steady transfer of energy from the long axion to the short axion, and the relative partition of energy between the two fields shifts as the universe evolves.  If both axions have $\mathcal{O}(1)$ initial misalignment angles,
then at $H \sim m$ we have that $\rhos \sim m^2 f^2$ and $\rhol \sim \mu^2 \calF^2 m^2 f^2$.  As time goes on, $\rhos$ remains roughly constant but $\rhol$ decreases $\propto a^{-3} = t^{-3/2}$.  Thus after approximately a time
\begin{equation} \label{eq:tequalDens}
    t_{\text{eq}} \equiv \frac{C_\te{eq}}{m} (\mu \calF )^{4/3} \,,
\end{equation}
the short and long axion energy densities will have equalized, where $C_\te{eq}$ is an order $1$ constant.  Autoresonance is still maintained for some time after this, although from this point on the energy loss in the long field is dominated by the transfer to the short field rather than Hubble friction.  This continues until autoresonance is cut off.

That autoresonance must eventually be cut off is clear from energetics; the short axion amplitude cannot remain constant forever.  Our second principal question is what causes this cutoff, and the answer lies in the equation of motion for $\Thl$ (Eq.~\ref{eq:eomThL}).  In our above first pass, we neglected the $\calF^{-2} \sin ( \Ths + \Thl )$ term in the large-$\calF$ limit, but in truth this approximation is only valid when the amplitude of $\Thl$ remains somewhat large.  If we expand in small $\Thl$ and retain the first-order contribution from the $\calF^2$ term we obtain:
\begin{equation}
    \del_t^2 \Thl + \frac{3}{2t} \del_t \Thl + \Thl ( \mu^2 + \frac{1}{\calF^2} \cos \Ths ) + \frac{1}{\calF^2} \sin \Ths = 0\,,
\end{equation}
and so we can see that when $\calF^{-2} \sin \Ths \sim \mu^2 \Thl$, backreaction will significantly affect the frequency of $\Thl$.  This is a somewhat decent proxy for when autoresonance ends, which predicts a maximum ratio of the amplitudes $\sis$ and $\sil$ of the short and long axions:
\begin{equation}
    \left. \frac{\sis}{\sil} \right|_{\text{late-time}} \sim \mu^2 \calF^2 \gg 1\,.
\end{equation}
Defining the homogeneous energy density in each axion by
\begin{align}\label{eq:arbitraryPartition}
\notag
    \rho_S&=f^2\p{\f12(\del_t\Ths)^2 + m^2\p{1 - \cos(\Ths + \Thl)}}\\&\approx \f12 m^2 f^2\sis^2 \,,\\
\notag
    \rho_L&=\calF^2f^2\p{\f12(\del_t\Thl)^2 + \mu^2m^2\p{1 - \cos\Thl}}\\&\approx \f12 \mu^2 m^2 \calF^2f^2\sil^2\,,
\end{align}
where the approximations are only valid when $\Thl \ll \Ths$ (the expectation after a period of autoresonance), we then have,
\begin{equation} \label{eq:finalRhoRatioAutoresNaive}
    \left. \frac{\rho_S}{\rho_L} \right|_{\text{late-time}} \sim \mu^2 \calF^2 \gg 1\,.
\end{equation}
Once autoresonance ends, the two axions behave as uncoupled fields with the exception of a small mass mixing, which can be rotated away by shifting to the mass basis.  The details of this transformation are discussed in App.~\ref{app:massVsInteractionBasis}, but the important result is that for $\calF \gg 1/\p{1 - \mu^2}$ the rotation angle is quite small.  The resulting flavor oscillations, however, do have a small effect, which we take into account in App.~\ref{app:autoresInDetail}.  This yields a more precise estimate for the final energy density ratio which is given in App.~\ref{app:autoresInDetail}.
For $\calF \gg 1/(1 - \mu^2)$ this ratio is well-approximated by:
\begin{equation} \label{eq:finalRhoRatioAutores}
    \left. \frac{\rho_S}{\rho_L} \right|_{\text{late-time}} \sim 4\calF^2 (1 - \mu)^2 \gg 1\,.
\end{equation}
This ratio then remains approximately constant as the universe evolves, since both $\rho_S$ and $\rho_L$ redshift $\propto a^{-3}$.

Although it is a simple heuristic, Eq.~\ref{eq:finalRhoRatioAutores} is extremely important, and highlights one of the main results of this paper: if autoresonance occurs, $\Thl$ transfers nearly all of its energy density into $\Ths$, which then dominates the late-time axion energy density.  The short axion can thus have far more energy density than would seem possible using the misalignment mechanism with $\mathcal{O}(1)$ misalignments for all fields.  Because $\Ths$ has a smaller decay constant, it will also generically have larger couplings to the SM.
As we will discuss in Sec.~\ref{sec:signatures}, these larger couplings can be probed by direct detection experiments even when the friendly pair makes up only a subcomponent of the dark matter.

In actuality Eq.~\ref{eq:finalRhoRatioAutores} is a decent heuristic but there are a few additional effects which can modify the final result significantly.  The first is the fact that when the initial conditions of the axions cause an autoresonance to occur, they typically also excite oscillations about the steady-state autoresonance.  These lead to a variance of the final ratio in Eq.~\ref{eq:finalRhoRatioAutores} of up to a few orders of magnitude.  We devote App.~\ref{app:autoresInDetail} to a more detailed study of autoresonance that touches on such effects, although analytic results are limited in precision by the nonlinearity of the dynamics.  In all such cases, however, the vast majority of the axion energy density ends up in the short field, so this effect only significantly affects the final abundance of the long field (a small subcomponent of the total axion energy density).  
The second and by far most significant effect is that of spatial inhomogeneities in the short field.  These can be resonantly amplified during autoresonance and, if they grow large enough, can cut off the autoresonance before the full $\mathcal{O}(\calF^2)$ ratio of Eq.~\ref{eq:finalRhoRatioAutores} is achieved.  We discuss these effects in Sec.~\ref{sec:perturbations}.

\begin{figure}[t]
    \centering
    \includegraphics[width = \columnwidth]{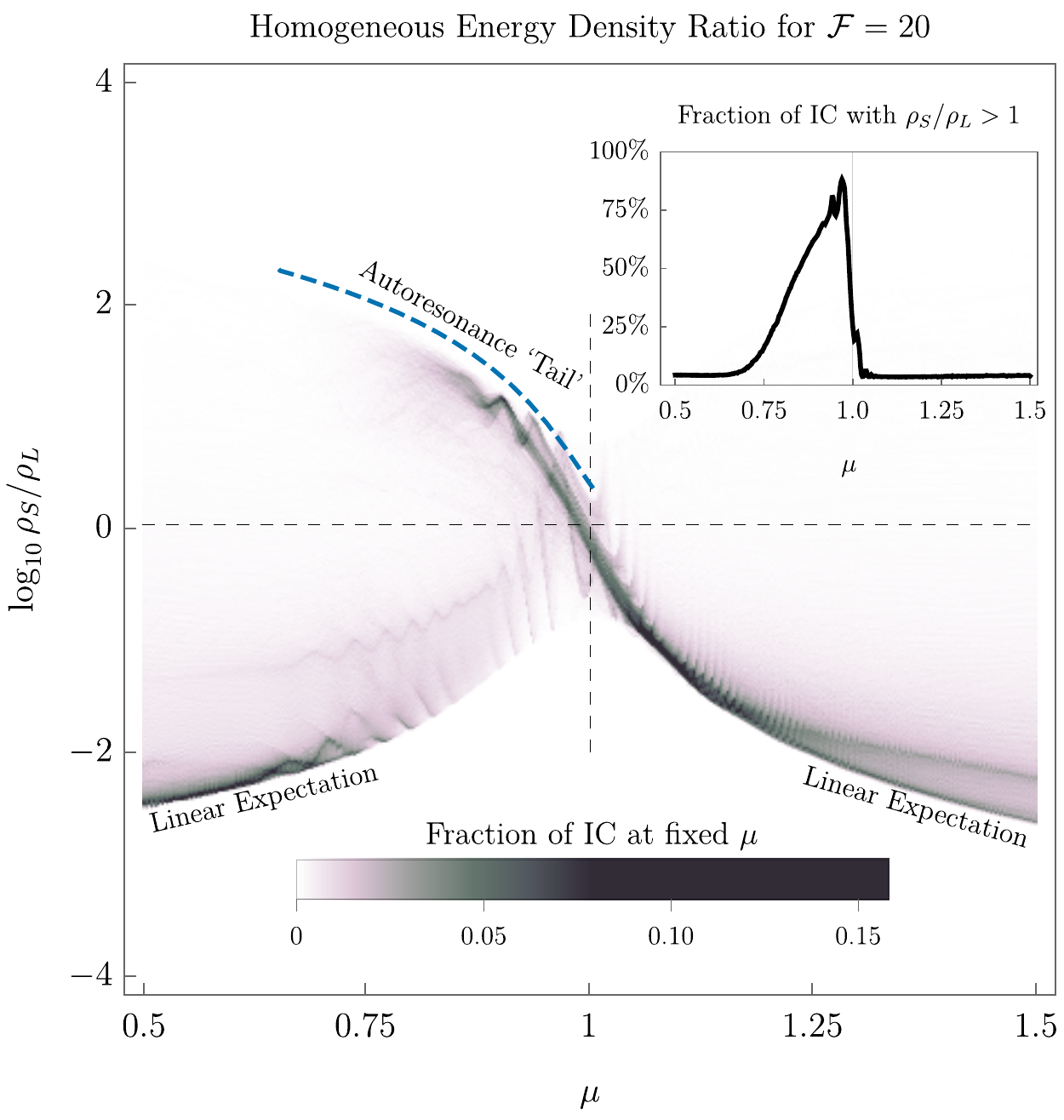}
    \caption{The relic density ratio of the short axion $\rho_S$ to the long axion $\rho_L$ in the model two-axion system of Eq.~\ref{eq:twoAxionPotential}. A vertical slice of this plot at fixed $\mu$ should be read as a histogram, with darker colors representing a higher likelihood of choosing initial conditions (IC) uniformly sampled from $(\Theta_S(0), \Theta_L(0))\in[-\pi,\pi]\times[-\pi,\pi]$ corresponding to that density ratio. For $\mu\geq 1$ and $\mu\leq 0.75$, most initial conditions lead to $\rho_S\ll \rho_L$ (lower dark bands), as na\"{i}vely expected for two uncoupled axions. For $0.75 \lesssim \mu < 1$, a period of autoresonance causes $\rho_S$ to dominate the relic abundance (wispy dark features pointing towards the upper left). We plot the analytical estimate for the shape of the autoresonance tail in dashed blue (see App.~\ref{app:autoresInDetail}).
    \textbf{\underline{Inset:}} An integrated version of this plot that shows, for each value of $\mu$, the total fraction of initial misalignment angles that result in the short axion dominating the late-time energy density in the axion sector.
    }
    \label{fig:autoresParamSpace}
\end{figure}

\begin{figure}
    \centering
    \includegraphics[width = \columnwidth]{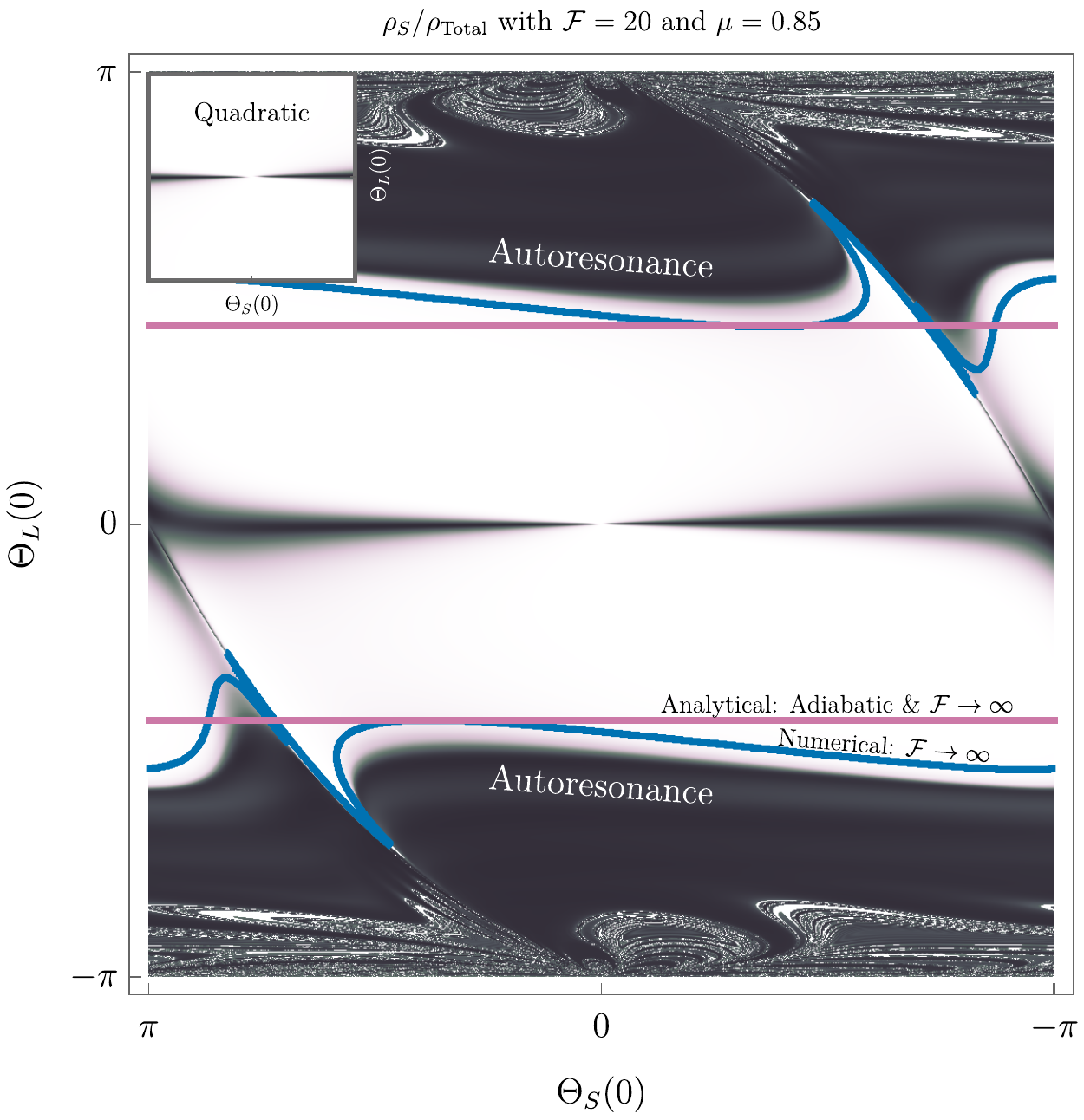}
    \caption{A representative plot of the late-time relative abundance of the short axion $\Ths$ compared to the total axion energy density, as a function of initial misalignment angles for both $\Ths$ and $\Thl$. Black regions correspond to initial angles for which $\Ths$ dominates the final relic abundance.  It is clear that this happens in two qualitatively distinct regions: when $\Thl(0)$ is tuned close to zero and when $\abs{\Thl(0)}$ is above some threshold, which for these parameters is roughly $\pi / 2$. The latter corresponds to those initial misalignment angles which land on autoresonance and thus lead to a nearly complete transfer of energy density from $\Thl$ to $\Ths$. The former is simply explained by the linearized dynamics, as shown in the inset. The autoresonance cutoff predicted in the adiabatic $\calF \to \infty$ limit (Eq.~\ref{eq:analyticalAutoresonanceCutoff}) is displayed in Magenta. The numerical $\calF\to\infty$ cutoff is displayed in Blue, which differs from the adiabatic prediction in that it accounts for transient $\Ths$ motion (see main text for details).  At very large initial long axion misalignments, a fractal-like structure emerges due to chaotic dynamics in the coupled system, which we discuss in App.~\ref{app:chaos}.
    \textbf{\underline{Inset:}} The same plot obtained by discarding all terms in the potential $V(\Thl, \Ths)$ of Eq.~\ref{eq:twoAxionPotential} higher than quadratic order in the fields.  In this case, the upper and lower regions completely disappear because autoresonance relies on the self-interactions of the short axion to achieve frequency-matching between the long and short fields.}
    \label{fig:energyPaisley}
\end{figure}

With this central result we can pass onto our third principal question: what range of parameters ($\mu$, $\calF$, and the initial misalignment angles) lead to autoresonance?  Let us first consider the effect of the decay constant ratio $\calF$. Because the dynamics of autoresonance are mainly determined by the $\calF \to \infty$ limit of the axion equations of motion (Eq.~\ref{eq:eomLS}), the precise value of $\calF$ does not play a big role in determining whether autoresonance will occur, although it must be somewhat large ($\calF \gtrsim 5$) to trust the above analytic results.  Numerically, we find that there are potentially-observable effects on gravitationally-bound structures for $\calF \gtrsim 3$, which we discuss further in Sec.~\ref{sec:perturbations}.

The mass ratio $\mu$ of the axions plays a much larger role.  For the attractive self-interactions of $\Ths$ discussed in the bulk of this paper, autoresonance requires $\mu < 1$, since the driving frequency must be less than the fundamental frequency of the driven field (i.e.\ the long axion's mass must be slightly smaller than the short axion's).  However if the hierarchy of masses is too large, autoresonance ceases to be possible.  Intuitively, this is because as the masses get further apart, the amplitude of the short axion predicted by Eq.~\ref{eq:duffing frequency} gets larger and larger.  Eventually, the approximation $\cos \Ths \sim 1$ in Eq.~\ref{eq:eomThsApproxApprox} fails, and the effects of this destroy the possibility of autoresonance.  As we discuss in App.~\ref{app:autoresInDetail}, this predicts a minimum value of $\mu \gtrsim 0.64$ to achieve autoresonance.  In practice, very few initial conditions lead to autoresonance for $\mu \lesssim 0.75$ (see inset of Fig.~\ref{fig:autoresParamSpace}), so the range $0.75 \lesssim \mu < 1$ is a useful notion of how ``friendly'' two axions must be to see significant effects of the kind we have described. We have studied this question numerically in the finite $\calF$ limit, and summarize our results in Fig.~\ref{fig:autoresParamSpace} and in particular its inset.  We find that for $\mu$ in the ``friendly'' band $0.75 \lesssim \mu < 1$, $\mathcal{O}(50\%)$ of the space of initial misalignment angles result in autoresonance, which leads to the short axion dominating the late-time energy density whenever it happens. 

For fixed $\mu$ and $\calF$, we can gain a better understanding of which initial misalignment angles lead to autoresonance by using the resonance curve techniques discussed above.
In App.~\ref{app:autoresInDetail} we show that all $\Ths(0)$ will be brought to autoresonance by sufficiently large $\Thl(0)$ in the large $\calF$ and small $1-\mu$ limits (see Eq.~\ref{eq:analyticalAutoresonanceCutoff} and surrounding discussion).
In Fig.~\ref{fig:energyPaisley}, we show a representative scan over initial misalignment angles for the parameters $\mu = 0.85$ and $\calF = 20$.  For initial $|\Theta_{L,0}| \gtrsim \pi / 2$, nearly all values of $\Theta_{S,0}$ end up autoresonating, directing nearly all the axion energy density into the short field. Fig.~\ref{fig:energyPaisley} also displays the large-$\calF$ autoresonance thresholds: the Magenta contour represents the adiabatic prediction (Eq.~\ref{eq:analyticalAutoresonanceCutoff}), which one should compare to the numerical Blue contour. These thresholds differ because the numerical contour accounts for initial transient $\Ths$ oscillations that depend mildly on the misalignment angles, while the analytical approximation assumes that all transients have died out. These differences vanish as we take $\mu$ closer to 1, where the adiabatic approximation becomes exact.

\section{Spatial fluctuations} \label{sec:perturbations}

In the previous section, we described the phenomenon of autoresonance in the two-axion potential of Eq.~\ref{eq:twoAxionPotential}. Autoresonance causes the short axion to undergo sustained, large-amplitude oscillations by drawing energy from the long axion. At these large amplitudes, $\ths$ experiences strong attractive self-interactions which can lead to the growth of large density perturbations in the axion field during radiation domination.  If the friendly pair comprises a sizable fraction of DM, these perturbations collapse early during matter domination, leading to a multitude of present-day astrophysical signatures.  The mechanism at play is a form of parametric resonance, quite similar to that studied in Ref.~\cite{arvanitaki2020large}.  In this section we generalize that study to our case of coupled axions.  We begin in Sec.~\ref{sec:perturbations_one} by considering a one-axion analogue of the friendly axion system that contains most of the relevant physics of perturbation growth. We then show in Sec.~\ref{sec:perturbations_two} that the results of this analogue model apply almost without modification to the case of friendly axions, and we arrive at analytic expressions for the growth rate of the short axion perturbations.  In Sec.~\ref{sec:NonpertAutoresonance} we proceed to a preliminary numerical study of autoresonance in the presence of non-perturbative $\ths$ fluctuations. Our $3+1d$ numerical simulations provide evidence that the autoresonant energy transfer of Sec.~\ref{sec:zeroModes} can be cut off early if $\ths$ fluctuations grow sufficiently large, significantly changing the predictions of the homogeneous theory. Finally, in Sec.~\ref{sec:newtonianPert} we conclude by describing the Newtonian formalism to evolve the density perturbations to the present day and discuss the late-time axion halo spectrum.  In this final section we treat only the case where the friendly axions constitute all of the DM.  We expect qualitatively similar effects if the pair constitute a significant ($\gtrsim\mathcal{O}(1\%)$) fraction of the DM, but we leave this case to future work.

\subsection{Invitation: A single axion model of perturbation growth} \label{sec:perturbations_one}

In the standard misalignment picture, the axion $\phi$ starts out displaced by order $f$ from its vacuum expectation value. The axion begins oscillating at $H \sim m$ and quickly loses energy to Hubble friction, diluting to approximately one fifth of its initial amplitude over a single oscillation. At such small amplitudes, self-interactions are weak, and the axion's potential is well-approximated by a free quadratic. If, however, the axion starts very close to the top of the cosine, then oscillations are delayed, and Hubble friction is tiny by the time the axion starts oscillating. It thus takes a long time for the axion to damp down from its large initial amplitude. The consequence of this \emph{large misalignment} is that the axion probes the nonlinear part of the potential for an extended period of time. The now-accessible many-to-one interactions convert the non-relativistic spectrum of axion fluctuations into semi-relativistic modes through parametric resonance. The resulting density fluctuations can then collapse into small scale structure, leading to an abundance of late-time signatures \cite{arvanitaki2020large,PhysRevD.96.023507,PhysRevD.96.063522}.

It turns out that fine-tuned initial conditions are not necessary for such effects if the axion has a more complicated potential. For example, Ref.~\cite{arvanitaki2020large} also studied monodromy-inspired potentials that flatten at large field values, effectively extending the cosine plateau. We can obtain a similar effect if a single axion's potential receives contributions from two instantons:
\begin{align} \label{eq:oneAxionPotential}
    V(\phi) &= m^2 f^2\left[ \p{1 - \cos\p{\f{\phi}{f} + \delta}} \right. \\
    & \left. + \mu^2\calF^2\p{1 - \cos\p{\f{\phi}{\calF f}}} \right] \,, \nonumber
\end{align}
where in this setup $\calF$ is an integer\footnote{A potential of this form can naturally arise from a general axiverse potential such as that of Eq.~\ref{eq:axiversePotential}, and in that context $\calF$ is just the ratio of the axion's integer charges $\mathcal{Q}$ under two different instantons.  $\calF$ can thus in general be any rational number rather than only an integer, but this does not change any of the qualitative features of the analysis and so we neglect it here.} and $\delta\in[0,2\pi)$ is a generic phase offset. Like the two-axion potential of Eq.~\ref{eq:twoAxionPotential}, this potential is comprised of a ``short'' and a ``long'' instanton (first and second lines respectively), whose ratio of periods is $\calF$. For parameters $\calF \gtrsim 3$ and $\mu\sim 1$, the resemblance goes further. Since the fundamental period of the field is $(- \pi \calF f , \pi \calF f)$, an untuned initial misalignment angle is $\phi / f \sim \mathcal{O}(\calF)$.  After a time $\tosc \sim \frac{1}{m} ( \mu \calF )^{4/3}$, the axion amplitude will have diluted to the scale of the small instanton ($\phi / f \sim \mathcal{O}(1)$) and it will feel strong self-interactions.  This delay is completely analogous to the time it takes for $\ths$ to fall off the autoresonance (Eq.~\ref{eq:tequalDens}).  In addition, Hubble friction has already decreased significantly by this time, and $\tosc$ is thus functionally equivalent to the delay time of oscillations during large misalignment \cite{arvanitaki2020large}.  At this point the self-interactions can lead to rapid perturbation growth.

We study the axion perturbations in the background of the perturbed FLRW metric
\begin{align}\label{eq:perturbedFLRW}
    \diff s^2&= (1 + 2\Phi)\diff t^2 - a^2(1 - 2 \Phi)\delta_{ij}\diff x^i\diff x^j\,,
\end{align}
where $\Phi(t,{\bf x}) = \sum_{\bf k}\Phi_{k}(t)e^{\I {\bf k}\cdot{\bf x}}$ is the adiabatic scalar perturbation generated by inflation.
Planck measurements of the CMB are consistent with a nearly scale-invariant dimensionless power spectrum  ${\cal P}_\Phi(t \to 0) = \gen{\Phi_{k,0}\Phi_{k,0}}\approx 2.1\times 10^{-9} (k/k_\star)^{n_s - 1}$, where $n_s\approx 1-0.03$ is the spectral tilt and $k_\star\approx 0.05 \te{Mpc}^{-1}$ is the pivot scale \cite{aghanim2020planck}. Because we lack measurements below $k = 1\te{ Mpc}^{-1}$, and for simplicity, we assume a scale-invariant power spectrum for the remainder of the text $\gen{\Phi_{k,0}\Phi_{k,0}}= 2.1\times 10^{-9}$.

We separate the axion field $\theta(t,{\bf x})\equiv \phi/(\calF f)$ into a homogeneous component and a spatially varying perturbation
\begin{align}
    \theta(t,{\bf x}) = \Theta(t) + \sum_{{\bf k}}e^{\I {\bf k}\cdot {\bf x}}\delta\theta(t,{\bf k})\,,
\end{align}
where ${\bf k}$ is the comoving wavenumber. To make our notation simpler, we re-scale the comoving wavenumber by defining
\begin{align}\label{eq:ktwDef}
    \tilde k^2&\equiv\f{1}{2 m H_\te{rad}(t)}\f{k^2}{a(t)^2}\,,
\end{align}
where $a(t)\propto t^{1/2}$ is the scale-factor during radiation domination, $H_\te{rad}^2=8\pi G^2 \rho_\te{rad}$, and $\rho_\te{rad}\propto a(t)^4$ is the energy density in radiation.  Note that with this definition $\ktw$ is dimensionless and constant in time, and $\ktw \sim 1$ corresponds to those modes that enter the horizon at $H \sim m$.
The zero-mode obeys the equation
\begin{align}
    \del_t^2\Theta + \f{3}{2t}\del_t\Theta + \f{1}{(\calF f)^2}V'(\calF f \Theta) = 0\,,
\end{align}
and the perturbation obeys the linearized equation
\begin{widetext}
\begin{align}
\label{eq:oneAxionPerturbations}
    \del_t^2\delta\theta(t,\tilde k) + 3H\del_t\delta\theta(t,\tilde k) + \p{\f{m}{t}\ktw^2 + \f{1}{ (\calF f)^2}V''(\calF f \Theta(t))}\delta\theta(t,\tilde k)={\cal S}(t,\tilde k)\,,
\end{align}
\end{widetext}
where primes indicate differentiation with respect to $\Theta$, and the perturbation initial conditions are set by inflation, which after many $e$-folds has flattened the axion field so that $\delta\theta(0,{\bf x}) = 0$ to high precision. ${\cal S}(t,\tilde k)$ is a small source representing the effect of the adiabatic scalar perturbations to the metric on the axion field:
\begin{align}
    {\cal S}(t,\tilde k)&=2\ps{\f{t_k}{t}\td{\Phi_k}{t_k}{}\del_t\Theta + \Phi_k V'(\Theta)}\,,
\end{align}
where 
\begin{align}\label{eq:Phik0}
    \Phi_k&=3\Phi_{k,0}\p{\f{\sin t_k}{t_k^3} - \f{\cos t_k}{t_k^2}}\,,\\
    t_k^2&\equiv\f{2}{3}\f{m}{H_\te{rad}}\tilde k^2\,.
\end{align}

Unlike misalignment in the cosine potential (Eq.~\ref{eq:cosinePotential}), the two scales of Eq.~\ref{eq:oneAxionPotential} mean that misalignment takes place in two parts. In the first epoch, the axion has a large amount of energy coming from the larger of the two instantons (the long instanton). These initial oscillations have kinetic energy density many times larger than the small instanton, and the axion rolls over the short instanton's wiggles without noticing them. The second epoch begins once the axion's energy matches the small instanton scale at a time $t = t_\te{osc}$. At this point, strong self interactions from the short instanton lead to the parametric resonant growth of perturbations.

More quantitatively, the story of misalignment in the two-instanton potential (Eq.~\ref{eq:oneAxionPotential}) is as follows. At early times when $H \gg m$, the axion remains fixed at its untuned initial condition $\Theta = \Theta_0 = {\cal O}(1)$, where it acts as a cosmological constant.
After Hubble friction dilutes below the mass scale, the zero-momentum mode starts oscillating and the axion energy density dilutes like matter. After just one oscillation, $\Theta$ is small enough that the self-interactions caused by the large instanton are negligible, and we can approximate the equation for $\delta\theta$ as
\begin{align}
    \del_t^2\delta\theta + 3 H\del_t\delta\theta + m^2\p{\f{\tilde k^2}{mt} +  \mu^2 + \cos(\calF \Theta + \delta))}\delta\theta&\approx{\cal S}\,.
\end{align}
Although the self-interactions of the long instanton are no longer relevant, it still dominates the energy density of $\Theta$, $\rho\sim\f12\mu^2 m^2 \calF^2 f^2$. Thus, when the axion is rolling past the bottom of the potential, we can approximate $\dot\Theta\sim \mu m$, and the short instanton acts as a parametric driver at integer multiples of the fundamental frequency $\calF\dot\Theta \approx \calF \mu m$. Because the mass of $\delta\theta$ is order $\mu m \ll \calF\dot\Theta$, these rapid parametric oscillations do \emph{not} induce parametric resonance, and $\delta\theta$ remains small during this early phase.

The axion does not begin to feel strong self-interactions until its energy density has diluted to the scale of the small instanton, 
\begin{align}
    \rho(t)\approx  (\mu m)^2 (\calF f)^2 \Theta_0^2(m t)^{-3/2} = m^2 f^2\Theta_0^2\p{\f{ t}{t_\te{osc}}}^{-3/2}\,,
\end{align}
at a time $t=t_\te{osc}\approx (\mu\mathcal{F})^{4/3}/m$. At this point, the amplitude of the zero mode oscillations has damped to $\Theta\sim 1/\mathcal{F}$, and $\Theta$ acts as a parametric driver with frequency at integer multiples of $\mathcal{F}\dot\Theta\sim m$. Now that the parametric driver and the perturbation frequency are both order $m$, $\delta\theta$ will experience a period of exponential growth due to a parametric resonance instability. 

As we will derive in App.~\ref{app:perturbations_in_detail}, the growth rate of the axion perturbations is controlled by a single parameter, the frequency shift $\delta\omega$ of the zero-mode oscillations, defined by the relationship
\begin{align}
    \delta\omega(\sigma)\equiv\omega(\sigma) - \omega(0)\,,
\end{align}
where $\sigma$ and $\omega(\sigma)$ are the amplitude and frequency of the homogeneous mode $\Theta$. The sign of $\delta\omega$ characterizes the net-repulsive or attractive interactions of the potential over the range of a complete $\Theta$ oscillation. Consider, for example, the case of a repulsive (positive) quartic interaction. The interaction increases the potential at larger amplitudes, causing the axion to turn around faster than it would have in a quadratic potential, reducing the period of oscillation. Similar reasoning applies to attractive quartic and to cubic interactions, which both work to increase the oscillation period.\footnote{Cubic interactions are always net-attractive, since the axion always spends more time on the attractive side of the potential.} Thus, net-repulsive interactions have $\delta\omega > 0$ and net-attractive interactions have $\delta\omega < 0$.

The instantaneous exponential growth rate $\Gamma(t,\ktw)$ of the axion perturbation $\delta\theta(t, \ktw)$ amplitude at comoving wavenumber $\ktw$ is (see App.~\ref{app:perturbations_in_detail}):
\begin{align}
\label{eq:growthRate}
    \Gamma(t,\ktw) = \Re\ps{-\f{3}{4 t} + \abs{\delta\omega}\sqrt{1 - \p{1 + \f{\ktw^2}{2 t \delta\omega }}^2}}\,,
\end{align}
where the $-3/4t$ is due to Hubble friction.  We can see that for repulsive self-interactions ($\delta\omega > 0$), the growth rate is always negative, and thus density perturbations do not grow through parametric resonance. Consequently, the late-time signatures of repulsive interactions are completely characterized by the analysis of Sec.~\ref{sec:zeroModes}, offering a clean benchmark model of autoresonant dark matter which we describe further in Sec.~\ref{sec:repulsiveSelfInteractions}.
On the other hand, attractive self-interactions, for which $\delta\omega < 0$, \emph{do} grow density perturbations, which we describe below and calculate in detail in App.~\ref{app:perturbations_in_detail}.

We can estimate the size of the $\delta\theta$ by integrating the growth rate
\begin{align} \label{eq:oneParticleGrowthRate}
    \gen{\delta\theta(t,\tilde k)^2}\approx\gen{\delta\theta(t_\te{init},\tilde k)^2}\exp\ps{2\int_{t_\te{init}}^t\diff t'\Gamma(t',\tilde k)}
\end{align}
where $t_\te{init} \approx t_\text{osc}$ is the earliest time where $\Gamma\geq 0$, and
\begin{align}\label{eq:dthetaInit}
    \gen{\delta\theta(t,\tilde k)^2}&\approx \f{\Phi_{k,0}^2}{\p{1 + \f{mt\tilde k^2}{\pi^2}}^2}\,,
\end{align}
is an empirical formula for the amplitude of $\delta\theta$ before perturbations start growing \cite{arvanitaki2020large}.
Because the leading-order frequency shift is always quadratic in the zero-mode amplitude $\delta\omega\propto\sigma^2$, we can parametrize the frequency shift's time evolution as $\delta\omega(t) = \delta\omega_\te{osc}(t/t_\te{osc})^{-3/2}$. As we show in App.~\ref{app:perturbations_in_detail}, the resulting scalar perturbations are maximized at $\tilde k = \tilde k_\te{max}$, with corresponding integrated growth rate
\begin{align}
    \tilde k_\te{max}^2&\approx-0.622\delta\omega_\te{osc} t_\te{osc}\,,\\
    \label{eq:oneParticleIntegratedGrowth}
    \lim_{t\to\infty}\int_{t_\te{osc}}^t\diff t'\Gamma(t',\tilde k_\te{max})&\approx -1.45\delta\omega_\te{osc} t_\te{osc} - 2.8\,,
\end{align}
where we have taken $t_\te{init} = t_\te{osc}$, and $-2.8$ corresponds to the suppression from Hubble damping.

To summarize, the axion only starts to experience parametric resonance once it has damped to the short instanton scale. The early period of large-amplitude oscillations only serves to delay parametric resonance to a late-enough time that it is not immediately quenched by Hubble friction. In the following section, we will study perturbations in the two-axion model Eq.~\ref{eq:twoAxionPotential}, and we will find that the results of this section carry over to the period after autoresonance ends, and in addition that autoresonance provides a mechanism for mode growth even during the early phase of large amplitude oscillations, leading to enhanced total perturbation growth.

\subsection{Perturbation growth during autoresonance}\label{sec:perturbations_two}

In this section, we quantify mode growth during the early phase of autoresonance, where the zero-mode physics is quite different from that of Sec.~\ref{sec:perturbations_one}. Nonetheless, the single-axion model (Eq.~\ref{eq:oneAxionPotential}) introduced in the previous section shares important features with the friendly axion model (Eq.~\ref{eq:twoAxionPotential}), and the same framework for parametric resonance is easily extended to this case. Importantly, we will find that autoresonance is a period of significant parametric resonance, which accounts for exactly one third of the total mode growth, lasting only $2\%$ of the total growth time. This is the consequence of the large, constant amplitude oscillations that are the hallmark of autoresonance.

The equations of motion for the density perturbations of the short and long axion are
\begin{widetext}
\begin{subequations}
\label{eq:perturbation_gr}
\begin{equation}
    \del_t^2\delta\ths + 3 H\del_t\delta\ths  + m^2\p{\f{1}{m t}\ktw^2 + \cos(\Ths + \Thl)}\delta\ths + m^2 \cos(\Ths + \Thl)\delta\thl = {\cal S}_S\,,\\
\end{equation}
\begin{equation}
    \del_t^2\delta\thl + 3 H\del_t\delta{\theta}_{L}  + m^2\p{\f{1}{m t}\ktw^2 + \mathcal{F}^{-2}\cos(\Ths + \Thl) + \mu^2\cos\Thl}\delta\thl + m^2 \mathcal{F}^{-2}\cos(\Ths + \Thl)\delta\ths = {\cal S}_L\,.
\end{equation}
\end{subequations}
\end{widetext}
where ${\cal S}_{S,L}$ represent how the metric fluctuations source the scalar perturbations of $\ths$ and $\thl$ respectively (see App.~\ref{app:perturbations_in_detail}). In the large-$\calF$ limit, we can see that $\delta\thl$ will behave just as in ordinary misalignment in a single cosine potential. Therefore, we approximate $\delta\thl\to 0$ and consider $\delta\ths$ in isolation. We further approximate $\Ths + \Thl\approx \Ths$, since $\Thl$ damps quickly to small amplitudes while $\Ths$ is locked by autoresonance. Thus, the equation for the short axion perturbation becomes
\begin{align} \label{eq:perturbation_singleApprox}
    \del_t^2\delta\ths + 3H\del_t\delta\ths + m^2\p{\f{1}{m t}\ktw^2 + \cos\p{\Ths}}\delta\ths \approx {\cal S}_S \,.
\end{align}
This is of the same form as Eq.~\ref{eq:oneAxionPerturbations}, and therefore our expression for the growth rate is exactly Eq.~\ref{eq:growthRate}, where the frequency shift is now given by the condition for autoresonance $\delta\omega(\sigma_S(t)) =\delta\omega_\te{osc}= \mu - 1$ for $t < t_\te{osc}$. In this case, $t_\text{osc} = C_\te{osc}(\mu{\cal F})^{4/3}/m$ is the time at which autoresonance ends and nearly-harmonic decaying $\Ths$ oscillations begin.  $C_\text{osc}$ is an $\mathcal{O}(1)$ constant that depends on initial conditions.
We now integrate the growth rate to arrive at the magnitude of $\delta\ths$ at the end of autoresonance
\begin{align}
    \gen{\delta\theta_S(t_\te{osc},k)^2}\approx\gen{\delta\theta(t_\te{init},k)^2}\exp\ps{2\int_{0}^{t_\te{osc}}\diff t'\Gamma(t',k)}\,.
\end{align}

The fastest growing mode starts growing at $t_\te{init}\approx 0.155 t_\te{osc}$, with comoving wave number $\tilde k_\te{max}$ and integrated growth rate
\begin{align}
    \tilde k_\te{max}^2&\approx-0.622\delta\omega_\te{osc} t_\te{osc}\,,\\
    \label{eq:autoresonantGrowth}
    \int_{t_\te{init}}^{t_\te{osc}}\diff t'\Gamma(t',\tilde k_\te{max})&\approx -0.725\delta\omega_\te{osc} t_\te{osc} - 1.4\,,
\end{align}
where $-1.4$ originates from Hubble damping.

After the end of autoresonance, $\sigma_S$ decays as $t^{-3/4}$ and $\delta\omega(\sigma_S(t)) = \delta\omega_\te{osc} (t/t_\te{osc})^{-3/2}$, just as in Sec.~\ref{sec:perturbations_one}. At this point, we have reduced the two-axion perturbation equations Eq.~\ref{eq:perturbation_gr} to a single-axion equation Eq.~\ref{eq:perturbation_singleApprox}, and we may directly apply the results of Sec.~\ref{sec:perturbations_one}, leading to the post-autoresonance integrated growth rate
\begin{align}
    \tilde k_\te{max}^2&\approx-0.622\delta\omega_\te{osc} t_\te{osc}\,,\\
    \label{eq:twoParticlePostAutoGrowth}
    \lim_{t\to\infty}\int_{t_\te{osc}}^t\diff t'\Gamma(t',\tilde k_\te{max})&\approx -1.45\delta\omega_\te{osc} t_\te{osc} - 2.8\,.
\end{align}
Notice that the spectrum of axion perturbations produced during autoresonance is peaked in the same location as the post-autoresonance perturbations. As a result, the total growth from both the fixed-amplitude autoresonance and the subsequent decaying-$\Ths$ oscillations is just the sum of Eq.~\ref{eq:autoresonantGrowth} and Eq.~\ref{eq:twoParticlePostAutoGrowth}
\begin{align}
    \lim_{t\to\infty}\int_{0}^{t}\diff t'\Gamma(t',\tilde k_\te{max})&\approx -2.175\delta\omega_\te{osc} t_\te{osc} - 4.2\,.
\end{align}

\begin{figure}[t]
    \centering
    \includegraphics[width = \columnwidth]{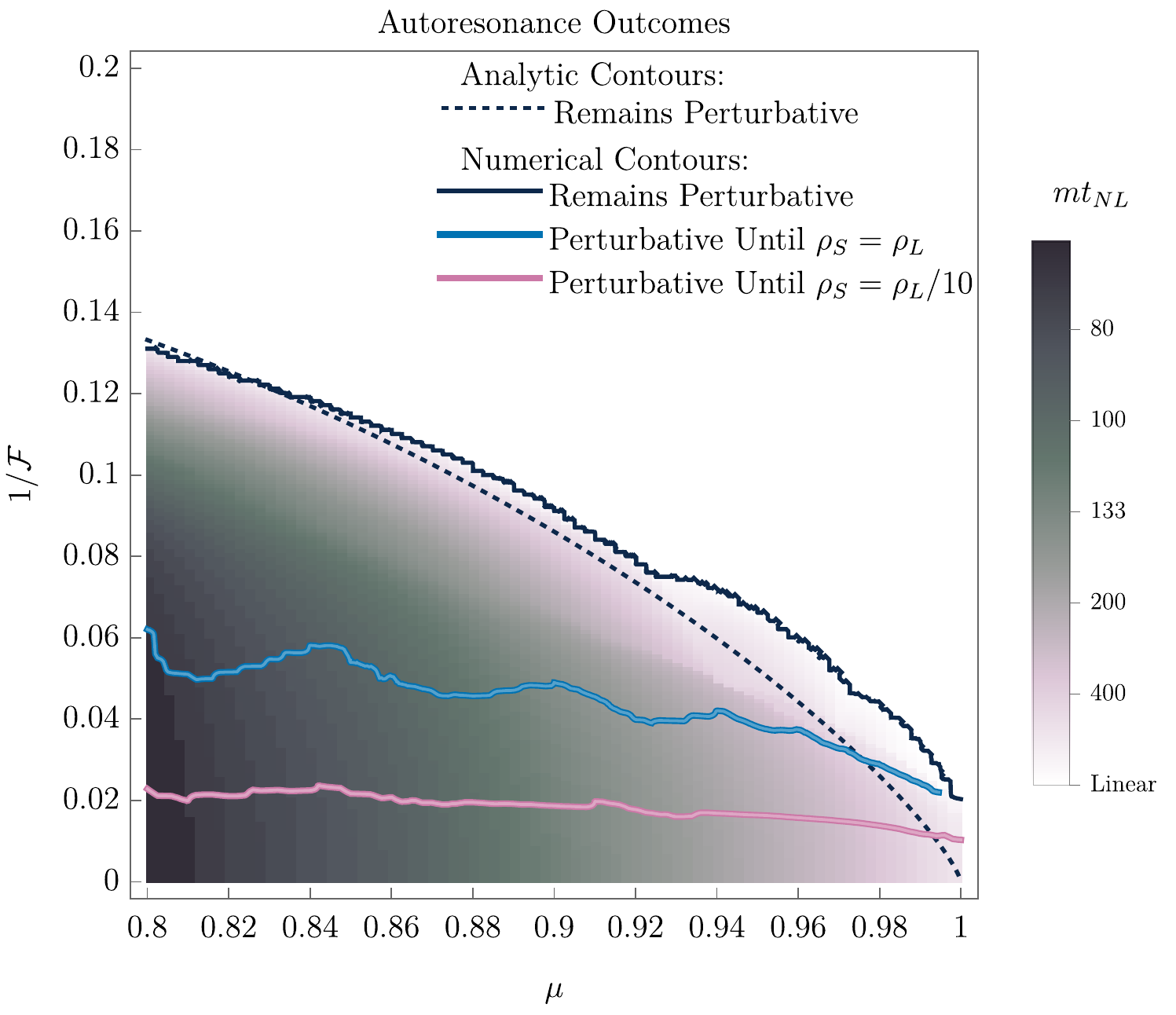}
        \caption{The time until the onset of nonlinearity, obtained for the specific initial conditions $\theta_S(0) = 0,\theta_L(0) = 0.8\pi$, chosen because they lead to autoresonance for the entire set of scanned $(\mu,\calF)$. The criterion for nonlinearity is that a single mode crosses $\delta\ths\geq 1$. Above the Solid Black contour, the axion remains perturbative indefinitely. The Dotted Black contour is the corresponding analytical estimate using the techniques of Sec.~\ref{sec:perturbations_two}. Above the Blue contour, the axion only becomes non-perturbative after the energy densities of $\ths$ and $\thl$ have equalized. Below this, modes become nonlinear even earlier, but above the Magenta contour modes remain linear until $\ths$ has at least $1/10$ the energy density of $\thl$.}
        \label{fig:autoresonance_outcomes}
\end{figure}

The linear analysis of this section allows us to predict a late-time spectrum of DM halos provided all perturbations remain small (Sec.~\ref{sec:newtonianPert}).  However, it is possible that a density perturbation grows non-perturbatively large, at which point this analysis breaks down.  We treat this numerically in the next section, where we find that non-perturbative structures can also quench the autoresonant transfer of homogeneous energy density described in Sec.~\ref{sec:zeroModes}.
We summarize the distinction between the perturbative and non-perturbative regions in Fig.~\ref{fig:autoresonance_outcomes}, where the colors indicate the time at which modes become nonlinear. In the white regions, all modes remain linear and the conclusions of Sec.~\ref{sec:zeroModes} go through unchanged. In the colored regions, the various contours indicate the different stages of parametric resonance at which modes become nonlinear. For modes becoming nonlinear after the end of autoresonance, we can safely apply the results of Sec.~\ref{sec:zeroModes}. For parameters where modes become nonlinear before the end of autoresonance, we must instead turn to the techniques of Sec.~\ref{sec:NonpertAutoresonance}.

\subsection{Nonperturbative structures during autoresonance} \label{sec:NonpertAutoresonance}
\begin{figure}
    \centering
    \includegraphics[width = \columnwidth]{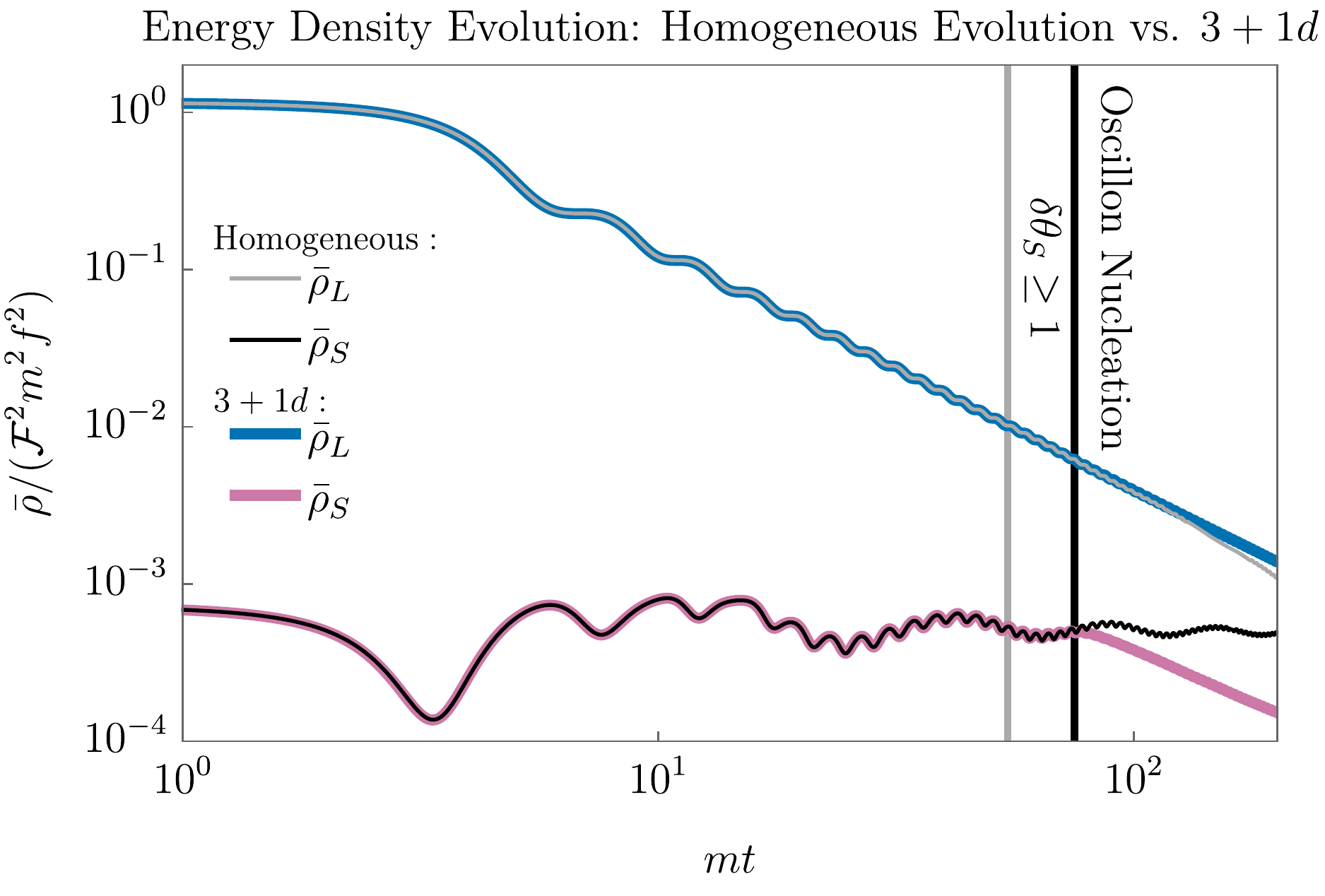}
    \caption{Comparison of the energy densities of the long and short axions from a homogeneous calculation (Sec.~\ref{sec:zeroModes}) versus the corresponding $3+1$ dimensional lattice simulation (see App.~\ref{app:nonperturbative_in_detail} for details). Here, ${\mathcal F} = 50$ and $\mu = 0.8$, with initial conditions $\theta_L(0) = 0.8\pi$ and $\theta_S(0) = 0$ chosen to lie in the autoresonance band. The vertical gray line represents the point beyond which $\theta_S$ fluctuations become non-perturbative, although $\rho_S$ does not  yet deviate significantly from the homogeneous expectation. Once these large $\ths$ fluctuations collapse under their own attractive self-interactions at the vertical black line, the autoresonant energy transfer stops, and both species dilute approximately like cold matter.}
    \label{fig:Numerical3+1}
\end{figure}

Autoresonance holds the homogeneous field $\Ths$ at large amplitudes for a long time, causing the spatial perturbations $\delta\ths$ to undergo a long period of exponential growth through parametric resonance. When these perturbations become ${\cal O}(1)$, the notion of the homogeneous mode $\Ths$ breaks down, and the conclusions of Sec.~\ref{sec:zeroModes} no longer apply. In order to get a sense of what happens in this nonlinear regime, we have performed a preliminary numerical investigation for a small set of Lagrangian parameters and initial conditions, which we describe in detail in App.~\ref{app:nonperturbative_in_detail}. Here we summarize our early results, which suggest that non-perturbative structure shuts down autoresonance, generically leading to a smaller final energy density in $\ths$ than predicted by Sec.~\ref{sec:zeroModes}.

We simulate two axions in the potential Eq.~\ref{eq:twoAxionPotential} in the background of the perturbed FLRW metric Eq.~\ref{eq:perturbedFLRW} where all fields are required to satisfy periodic boundary conditions. The results of one such simulation are given in Fig.~\ref{fig:Numerical3+1}. Because of the non-perturbative fluctuations in $\ths$, there is no unique way to partition the energy densities between $\ths$ and $\thl$, so we make the following choice:
\begin{align}\notag
    \bar\rho_S&=\f{f^2}{V}\int\diff V\biggl[\f12(\del_t\ths)^2 + \f12(\nabla\ths)^2 \biggr.\\&\hspace{1cm}\biggl.+m^2  (1 - \cos(\ths + \thl))\biggr]\,,\\
    \notag
    \bar\rho_L&=\f{f^2 \calF^2}{V}\int\diff V\biggl[\f12(\del_t\thl)^2 + \f12(\nabla\thl)^2 \biggr.\\&\biggl.\hspace{1cm}+ m^2 \mu^2(1 - \cos\thl)\biggr]\,.
\end{align}
where $V$ is the simulation volume.  Even after the onset of non-perturbative $\ths$ fluctuations (marked by the vertical gray line), the $\ths$ energy density only deviates slightly from the homogeneous prediction. This deviation remains small until the perturbations begin collapsing under their own attractive self-interaction, which we mark with a vertical black line. The objects nucleating from this nonlinear collapse are \emph{oscillons}: long-lived spherically symmetric scalar configurations held together by attractive self-interaction \cite{kudryavtsev1975solitonlike,makhankov1978dynamics,gleiser1994pseudostable,kolb1994nonlinear,salmi2012radiation,amin2012oscillons,kawasaki2020oscillon,olle2020recipes,zhang2020classical,cyncynates2021structure}. At this point, both $\rho_S$ and $\rho_L$ diverge from the prediction of Sec.~\ref{sec:zeroModes}, and simultaneously begin diluting (almost) like cold matter. Unexpectedly, we observe the final energy density ratio $\rho_S/\rho_L$ to scale like $t^{0.17}$, although it is unclear whether this scaling persists until the energy densities equalize, or whether it is a numerical artifact. In our later estimates of direct detection prospects, we assume that the energy density ratio is fixed after oscillon nucleation, which is conservative since we are mainly interested in the detection of $\rho_S$.

In spite of this numerical uncertainty, there is a possible physical explanation for why oscillon nucleation may end autoresonance. Consider that for $\ths$ to sustain autoresonance in any given region of space, $\ths$'s amplitude must remain locally large enough that its frequency can remain locked to $\mu$. At early times, $\ths$ fluctuations are dominated by a single momentum mode $\ktw = \ktw_\te{max}$, whose wavelength is typically much longer than the Compton wavelength of the axion field. As this mode grows, a fixed fraction of the comoving volume is at a large enough amplitude for autoresonance, even after $\delta\ths(\ktw_\te{max})$ becomes much larger than unity. After a short time, these comoving regions of space collapse into oscillons with a fixed physical size much smaller than the scale of $\ktw_\te{max}$. At this point the long-wavelength perturbations at $\ktw_\te{max}$ have lost much of their amplitude to gradient energy and to radiation production, and most of space is below the autoresonance threshold. While the large-amplitude oscillons may in principle still remain autoresonant with $\thl$, the $\ths$ energy density now dilutes like matter, since the comoving number density of oscillons is approximately conserved, and the non-autoresonant parts of space cannot become autoresonant.

We do, however, emphasize the need for higher resolution simulations to confirm our results and intuition. Even though it is physically reasonable that non-perturbative structure cuts off autoresonance, the opposite possibility also offers exciting observational prospects. If autoresonance is not cut off, then the short axion may become even more visible at smaller $f_S$ (larger $\calF$), offering enhanced direct detection prospects. On the other hand, if our numerics are confirmed, then the resulting oscillons may have parametrically enhanced lifetimes, leading to interesting present-day signatures of their own.  We do not perform a full analysis of this possibility here, but we do discuss it further in Sec.~\ref{sec:discussion}.

\subsection{Newtonian evolution and gravitational collapse} \label{sec:newtonianPert}

A long time after parametric resonance has concluded, the axion field is firmly non-relativistic and can be well-approximated by its Newtonian evolution. If the friendly pair comprises a majority of the dark matter, the over-dense regions begin to collapse under their own gravity and virialize at the onset of matter domination, leading to the formation of axion minihalos, which eventually comprise galactic substructure. In this section, we extend the formalism of Ref.~\cite{arvanitaki2020large} to describe this process in the case of two friendly axions. For concreteness, in this section we assume the friendly pair makes up all of the dark matter.

After parametric resonance, the axion fields are best described in the mass basis
\begin{align}
    \nu_h &\equiv \phi_S\cos\eta  + \phi_L\sin\eta \,,\\ 
    \nu_l &\equiv \phi_L\cos\eta  - \phi_S\sin\eta \,,
\end{align}
where the $\nu$ basis is related to the old basis by the rotation angle $\eta$, and the states $\nu_h$ and $\nu_l$ have corresponding heavy and light masses $m_h$ and $m_l$, all defined in App.~\ref{app:massVsInteractionBasis}.
When $\mathcal{F}\gg (1 - \mu^2)^{-1}$, the mass-eigenstates $\nu_h$ and $\nu_l$ are \emph{mostly} comprised of $\phi_S$ and $\phi_L$ respectively. The fields $\nu_h$ and $\nu_l$ may be broken down into a homogeneous background and perturbations
\begin{align}
    \nu_{h,l} &= N_{h,l}(t) + \sum_{{\bf k}}e^{\I {\bf k}\cdot{\bf x}}\delta\nu_{h,l}(t,{\bf k})
\end{align}
yielding the corresponding relative density perturbations $\rho_{h,l} = \bar\rho_{h,l}(1 + \delta_{h,l})$,
\begin{align}
\label{eq:mass_basis_density}
    \delta_{h,l}&=\f{\del_t N_{h,l}\del_t\delta\nu_{h,l} + m_{h,l}^2 N_{h,l}\delta\nu_{h,l}}{\f12(\del_t N_{h,l})^2 + \f12 m_{h,l}^2 N_{h,l}^2}\,.
\end{align}
where $\bar\rho_{h,l}$ is the average density of $\nu_{h,l}$ respectively.

\begin{figure}
     \centering
     \includegraphics[width =\columnwidth]{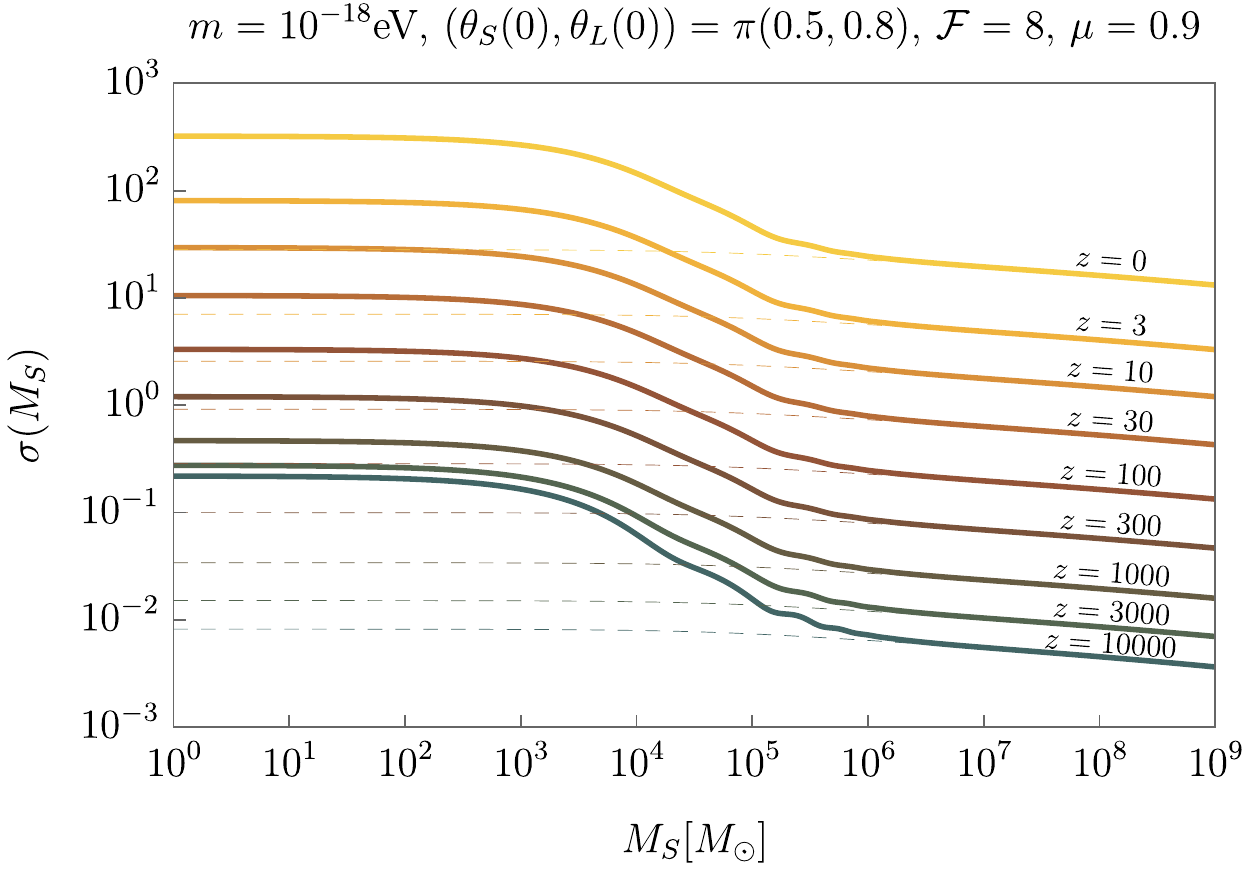}
     \includegraphics[width =\columnwidth]{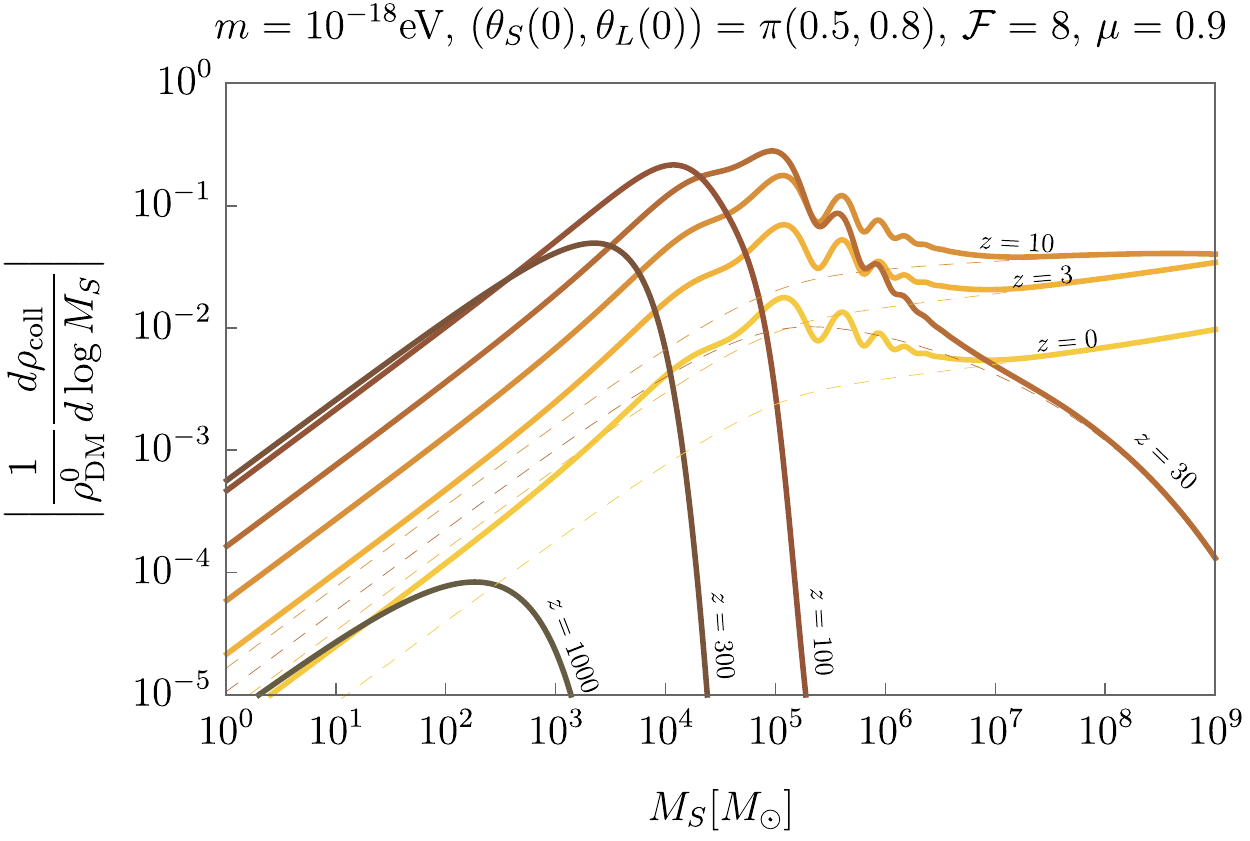}
    \caption{The standard deviation of the density perturbations (top) and the differential fraction of collapsed structures (bottom) at a given smoothing mass $M_S$. The mass scale $m = 10^{-18}\te{ eV}$ is chosen to enable direct comparison with Fig.~7 of Ref.~\cite{arvanitaki2020large}, where a $10^{-10}$ tuning of the initial misalignment angle is necessary to achieve comparable density fluctuations. The thin dashed lines correspond to the same density fluctuations and collapsed fraction for a non-self-interacting scalar of the same mass $m = 10^{-18}\te{ eV}$.}
    \label{fig:density_fluctuation}
\end{figure}
\begin{figure}
    \centering
    \includegraphics[width = \columnwidth]{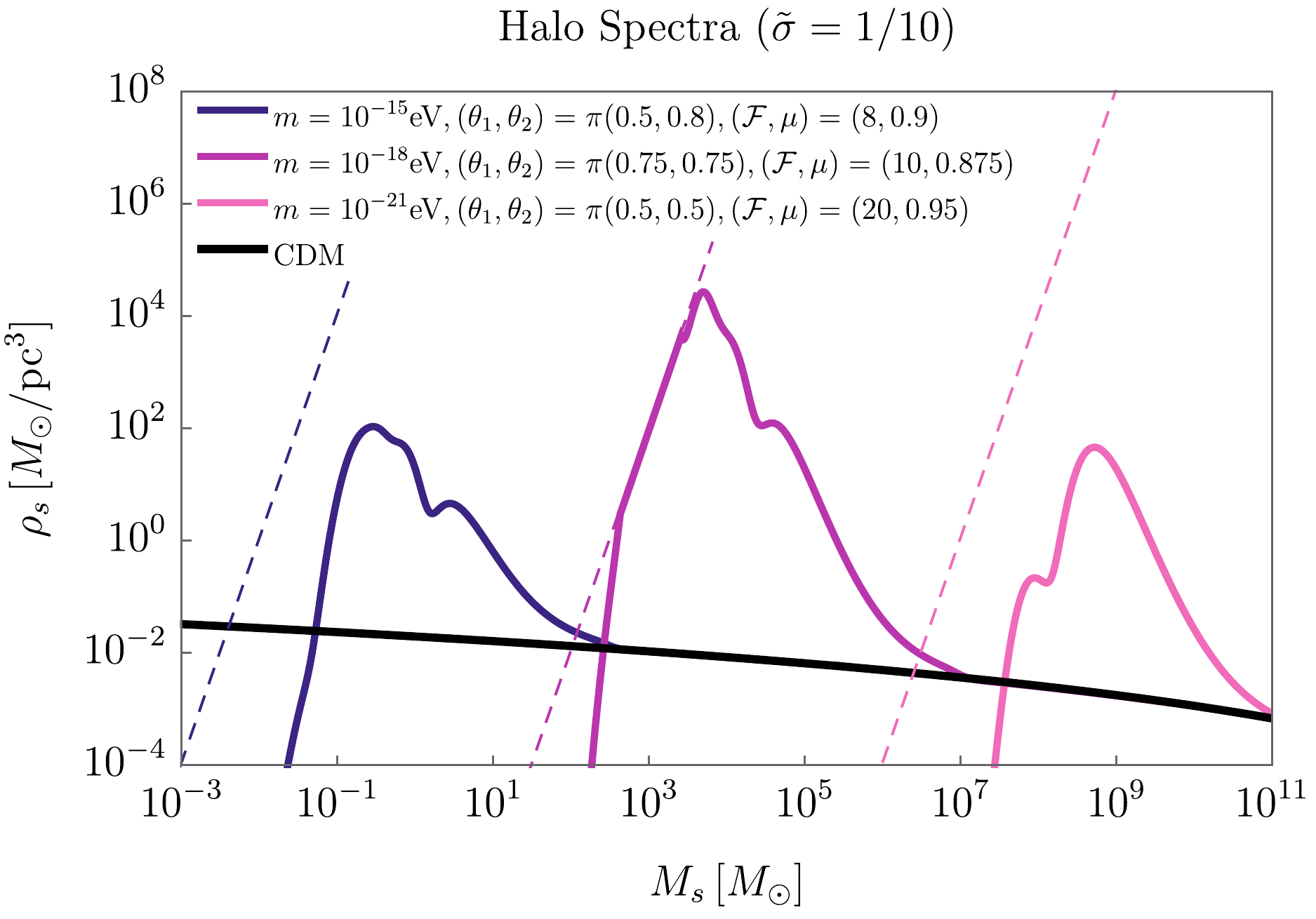}
    \caption{The halo spectrum $\rho_s$ versus scale mass $M_s$ in the friendly axion model with initial misalignments and Lagrangian parameters chosen to be representative of what one might expect to find in the axiverse. The three masses $m$ chosen for this plot match those of Fig.~8 in Ref.~\cite{arvanitaki2020large} in order to allow for direct comparison.  Note the large enhancement of subhalo density relative to the CDM expectation. The dashed lines correspond to the density of a soliton, a gravitationally-bound scalar field configuration supported by kinetic pressure, which represents the densest stable collapsed axion structure of a given mass. The soliton mass-density relationship is given by $\rho_s\approx 0.067 G^3 m^6 M_s^4$ \cite{ChavanisNumeric}.}
    \label{fig:halo_spectra}
\end{figure}

Following Ref.~\cite{arvanitaki2020large}, we now change variables from $t$ to $y \equiv a(t)/a_\te{eq}$, where $a_\te{eq}$ is the scale factor at matter-radiation equality. The
density fluctuations deep inside the horizon $\tilde k^2\gg H/m$ then obey the Newtonian equations of motion
\begin{align} \label{eq:newtonian_density}
\notag
    0&=(1 + y)\delta''+\p{\f{3}{2} + \f{1}{y}} \delta' \\&- \p{\f{3}{2y} F_G - \f{1}{y^3}\tilde k^2 C_s^2 - \f{1}{y^2}\tilde k^4 C_Q^2}\delta\,,
\end{align}
where we have defined the vector of relative density perturbations $\delta \equiv(\delta_h,\delta_l)^T$, and primes denote differentiation with respect to $y$. The matrices of Eq.~\ref{eq:newtonian_density} are defined
\begin{align}
    C_s^2&= \f{3\sqrt{2}m H_\te{eq} \Mpl^2}{\bar\rho_h + \bar\rho_l}\mat{cc}{\f{\lambda_{hh} \bar\rho_h}{16m_h^4}&\f{\lambda_{hl}\bar\rho_l}{8m_h^2 m_l^2}\\\f{\lambda_{hl}\bar\rho_h}{8m_h^2 m_l^2}&\f{\lambda_{ll}\bar\rho_l}{16m_l^4}}\,,\\
    C_Q^2&= \mat{cc}{m^2/m_h^2&0\\0&m^2/ m_l^2}\,,\\
    F_G&=\f{1}{\bar\rho_h + \bar\rho_l}\mat{cc}{\bar\rho_h&\bar\rho_l\\\bar\rho_h&\bar\rho_l}\,,
\end{align}
where $\lambda_{hh,hl,ll}$ are the quartic interactions in the mass basis of even parity (corresponding to interactions with even numbers of both species), whose full expressions are given in App.~\ref{app:massVsInteractionBasis}. The matrices $C_s$ and $C_Q$ are coefficients representing the strength of self-interactions and kinetic pressure respectively, which together comprise the effective speed of sound. The matrix $F_G$ represents the attractive force of gravity.  These equations may then be numerically integrated to late times.

Having solved for the full history of the linear density perturbations, we can now describe the nonlinear collapse of these density perturbations into small-scale structures. The formalism to describe nonlinear gravitational collapse is well-known \cite{press1974formation} and worked out in detail in Ref.~\cite{arvanitaki2020large}, which we summarize here for completeness.

In the extended Press-Schechter formalism, a local overdensity is considered to have collapsed if it exceeds the critical overdensity $\delta_c = 1.686$ \cite{bardeen1986statistics}. In the two-axion model, the total DM overdensity in momentum space is 
\begin{align}
    \delta(t,{\bf k}) \equiv \f{\bar\rho_h(t)\delta_h(t,{\bf k}) + \bar\rho_l(t)\delta_l(t,{\bf k})}{\bar\rho_h(t) + \bar\rho_l(t)}\,.
\end{align}
To obtain a distribution for the density perturbations in position space, we smooth the density field $\delta(t,{\bf x}) \equiv (2\pi)^{-3}\int\diff^3 k e^{\I {\bf k}\cdot{\bf x}}\delta(t,{\bf k})$ over a radius $R_S$ using the spherical top-hat window function $W(R_S,{\bf x}) = \Theta(R_S - |{\bf x}|)(3/(4\pi R_S^3))$:
\begin{align} \label{eq:smoothed_density}
    \delta(t,{\bf x},R_S)\equiv\int\diff^3 x' W(R_S,{\bf x} - {\bf x}')\delta(t,{\bf x}')\,.
\end{align}
The mass contained within the smoothing radius is $M_S = (4\pi/3)\rho_\te{DM}^0 R_S^3$, where $\rho_\te{DM}^0 = 3.3 \times 10^{-8} M_\odot/\te{pc}^3$ is the average dark matter density in the present-day universe. 

Assuming that the density perturbations obey a Gaussian distribution, the differential collapsed fraction of energy density per unit mass is
\begin{align}
    \f{1}{\rho_\te{DM}^0}\td{\rho_\te{coll}}{\log M_S}{} = \sqrt{\f{2}{\pi}}\f{\delta_c}{\sigma(M_S)}\abs{\td{\log\sigma(M_S)}{\log M_S}{}}e^{-\f{\delta_c^2}{2\sigma^2(M_S)}}\,,
\end{align}
where the density fluctuation variance is $\sigma^2(M_S) = \gen{\delta(t,{\bf x},R_S)^2}$. We plot the variance and differential collapsed fraction in Fig.~\ref{fig:density_fluctuation} for a representative set of initial conditions and Lagrangian parameters for a mass scale $m\approx 10^{-18}\te{eV}$ to allow for direct comparison to figure 7 of Ref.~\cite{arvanitaki2020large}. We see that an early period of autoresonance has enhanced structure at the mass scale $M_S\approx 10^5 M_\odot$, which collapses significantly earlier than the larger-scale structure comprising entire galactic halos.

In Ref.~\cite{arvanitaki2020large}, the authors point out two downsides of Press-Schechter theory. First, $\delta(t,{\bf x},R_S)$ can be large even if there is no structure at the scale $R_S$, so long as there is structure at larger scales. Second, the differential collapsed fraction does not count substructure. To remedy this, they propose the use of a smoothing function in momentum space which isolates structures of scale $R$,
\begin{align}
    W({\bf k},R) = \f{1}{\p{2\pi \tilde\sigma^2}^{1/4}}\exp\ps{-\f{\te{log}(|{\bf k}| R/\pi)^2}{4\tilde\sigma^2}}\,.
\end{align}
Using Eq.~\ref{eq:smoothed_density} with this new window function, we compute the variance $\sigma^2(M_S)$,
\begin{align}
    \sigma^2(M_s)&=\int\diff\log k \,|\delta(t,k)|^2 |W({\bf k},R_s)|^2\,.
\end{align}
Structures at a given mass scale $M_s$ are considered to have collapsed at a time corresponding to the scale factor $a_\te{coll}(M_s)$ when a 1-$\sigma$ overdensity exceeds $\delta_c$, where $M_s = \rho_\te{DM}^0(4\pi/3)R_s^3$.  The resulting collapsed structure has a well-known density roughly 200 times the ambient density at the time of collapse $\rho_s \approx 200\times \rho_\te{DM}^0 a_\te{coll}^{-3}$. We plot the resulting halo spectra in Fig.~\ref{fig:halo_spectra} for three representative sets of initial conditions and Lagrangian parameters, where we have chosen mass scales that match those in Fig.~8 of Ref.~\cite{arvanitaki2020large} to enable direct comparison. This halo spectrum peaks at a scale mass determined by the $\tilde k_\te{max}$ in Sec.~\ref{sec:perturbations_two}, which is well-approximated by:
\begin{align}
\nonumber
    M_s &\sim \f{4}{3}\pi \rho_\te{DM}(H = m/2) \p{\f{2\pi}{m \tilde k_\te{max}}}^{3} \\
    &\sim 1.2 \times 10^{4} M_\odot \p{ \f{10^{-19} \eV}{m} }^{3/2} \left( \frac{5}{\tilde k_\te{max}} \right)^3\,.
\end{align}

\section{Signatures} \label{sec:signatures}

So far, we have primarily focused on the early-time dynamics of a pair of friendly axions, but in this section we turn to the late-time observable effects of these dynamics.  Broadly they fall into two categories.

First, autoresonance can facilitate a significant transfer of energy density from an axion with a large decay constant to an axion with a much smaller decay constant.  Since the axion's couplings to the SM are generically suppressed by its decay constant, axions produced via autoresonance can be coupled significantly more strongly to the SM than axions produced via the usual misalignment mechanism, and can be observable even if they make up only a small subcomponent of DM.  We discuss this point and outline future detection prospects in Sec.~\ref{sec:directDetection}.

The second broad class of observable effects are indirect gravitational signatures.  As discussed in Sec.~\ref{sec:perturbations}, an era of autoresonance can lead to significant growth of density fluctuations that can collapse into gravitationally-bound structures earlier than would be predicted by $\Lambda$CDM, as shown in Fig.~\ref{fig:density_fluctuation}.  This collapse requires that the pair of friendly axions make up the entirety of dark matter, but if this happens such structures can be detectable through their gravitational effects.  The halo substructure turns out to be quite similar to that produced by the mechanism of Ref.~\cite{arvanitaki2020large}, so the techniques discussed therein for detecting such structures apply here as well.  We briefly review these in Sec.~\ref{sec:indirectDetection}.  Finally, both the long and short axions can potentially be constrained by black hole superradiance; we comment on this in Sec.~\ref{sec:superradiance}. The reach of all signatures discussed in this section are summarized in Fig.~\ref{fig:attractiveSensitivity} for the case where the friendly axions are the DM, and in Fig.~\ref{fig:rescaledSensitivity} for the case where they are only a subcomponent.

\subsection{Enhanced direct detection prospects} \label{sec:directDetection}

The most striking effect of axion friendship to significantly improve the prospects of probing an axiverse in direct detection experiments. In the absence of interactions, all axions with similar masses would be equally detectable provided they all started at similarly untuned initial misalignment angles. An axion with a smaller decay constant $f$ will have a smaller present-day abundance, but its stronger coupling to the SM precisely cancels this out when it comes to observability. Quantitatively, haloscope experiments couple to the combination $\gagg^2\rhoax$, where the axion-photon coupling is expected to be of order $\gagg \simeq \alpha/4\pi f$ with $\alpha$ the QED fine-structure constant. An axion of a given mass $m$ will thus be detectable to an experiment with sensitivity:\footnote{This expression and the analysis of this section refer to experiments that probe the axion through its coupling to photons.  There are other potential axion couplings that can be probed which are subject to similar analyses, but we do not discuss them here.}
\begin{equation} \label{eq:targetSensitivityNaive}
    \left( \gagg^2 \frac{\rhoax}{\rho_{\text{DM}}^0} \right)^{1/2}_{\text{na\"ive}} \sim 2.3 \times 10^{-17} \GeV^{-1} \left(\frac{\Theta_0}{\pi/2} \right) \left(\frac{m}{10^{-7} \eV}\right)^{1/4}\,,
\end{equation}
where we have normalized to the current universe-average DM density, $\Theta_0$ is its initial misalignment, and this formula receives logarithmic corrections near $\Theta_0 = \pi$.  Note importantly that Eq.~\ref{eq:targetSensitivityNaive} is independent of the decay constant. For this reason, in this naive scenario, an axion haloscope experiment sensitive to a wide range of masses is unlikely to see any axiverse axion until it reaches the sensitivity threshold of Eq.~\ref{eq:targetSensitivityNaive}. However, once it does reach this point it may see several axion signals at the same time, even from axions which make up only a small subcomponent of the DM.

In contrast, we have seen that for a pair of friendly axions in the axiverse, autoresonance can transfer nearly all of the energy density from the long axion (with the larger decay constant $f_L$) to the short axion (with the smaller decay constant $f_S$).  This results in a ``best of both worlds'' scenario: if autoresonance completes, the short axion's energy density is set by $f_L$ while its coupling to the SM is set by $f_S$.  This makes the short axion much more observable, enhancing its signal strength relative to Eq.~\ref{eq:targetSensitivityNaive}:
\begin{multline} 
    \left( \gagg^2 \frac{\rhoax}{\rho_{\text{DM}}^0} \right)^{1/2}_{\text{friendly}} = \calF \left( \gagg^2 \frac{\rhoax}{\rho_{\text{DM}}^0} \right)^{1/2}_{\text{na\"ive}} \\
    \sim 4.6 \times 10^{-16} \GeV^{-1} \left( \frac{\calF}{20} \right) \left( \frac{\Theta_0}{\pi/2} \right) \left( \frac{m}{10^{-7} \eV} \right)^{1/4}
    \label{eq:targetSensitivityAutores}
\end{multline}
where $\Theta_0$ refers to the long axion's initial misalignment angle.  Although there may only be a few pairs of friendly axions in the axiverse which end up autoresonating, these few pairs (or, more precisely, the short axion in each of these pairs) may become the one most visible to direct detection experiments.

\begin{figure}[t]
    \centering
    \includegraphics[width = \columnwidth]{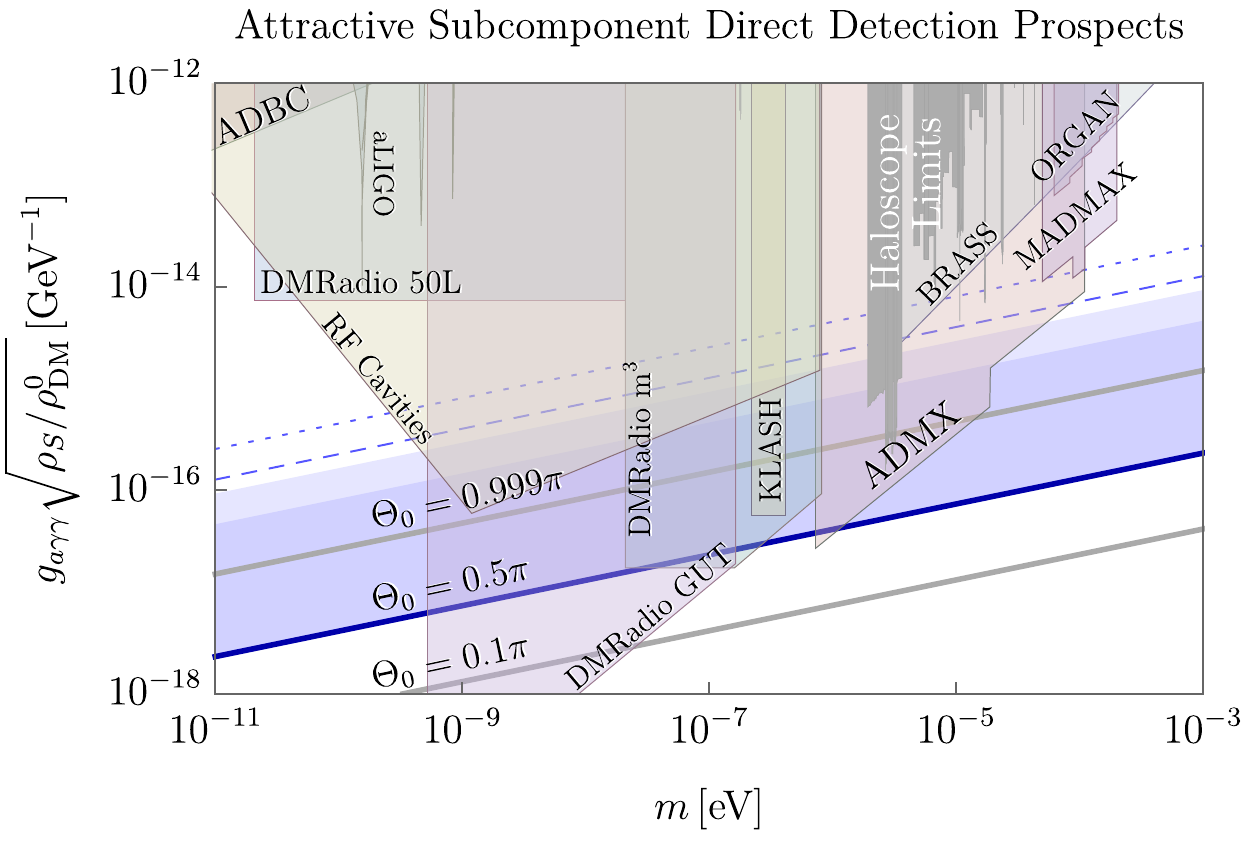}
    \caption{Enhanced direct detection prospects for a short axion, assuming that the friendly pair comprises a small fraction of the total dark matter energy density. The darker blue band shows the prospects for $\mu = 0.8$ and $\Thl(0) = 0.5\pi$ in the large $\calF$ limit, where the possible enhancement saturates for $\calF \gtrsim 20$ due to the formation of nonperturbative structure (Sec.~\ref{sec:NonpertAutoresonance}). For $\mu = 0.99$, the possible enhancement saturates for $\calF \gtrsim 40$ (light blue band). As $\calF$ decreases below the saturation value, the visibility decreases linearly with $\calF$. This enhanced visibility should be compared to that of a single free axion with initial misalignment $\Theta_0 = 0.5\pi$ (middle solid blue line). The dashed and dotted blue lines are the sensitivity prospects for $\mu = 0.8$ and $\mu = 0.99$ respectively in the large $\calF$ limit with $\Thl(0) = 0.9\pi$. Because the friendly pair makes up only a subcomponent of DM, its overdensities do not collapse under self-gravity, and minihalos never form. Thus, an $\mathcal O(1)$ fraction of $\rhos$ in the galaxy is ambient (as opposed to clumped) and will pass through direct detection experiments. As a result, the direct detection prospects are improved relative to those in  Fig.~\ref{fig:attractiveSensitivity}. This plot was made using limits compiled in \cite{ciaran_o_hare_2020_3932430,ciaran_o_hare,PhysRevLett.119.031302,armengaud2017constraining,PhysRevD.101.123026,PhysRevD.101.103023,rogers2021strong,buen2020constraints,reynolds2020astrophysical,marsh2017new,dessert2020x,wouters2013constraints,calore2020bounds,ayala2014revisiting,vinyoles2015new,regis2021searching,cadamuro2012cosmological,fedderke2019axion,DMRadioGUT,alesini2017klash,stern2016admx,BRASS,lasenby2020microwave,berlin2020axion,berlin2020heterodyne,DMRadiom3}.}
    \label{fig:rescaledSensitivity}
\end{figure}

For fixed $\calF$, the enhancement to the signal strength (Eq.~\ref{eq:targetSensitivityAutores}) does not depend on whether the friendly pair makes up all of DM or only a subcomponent, but this distinction can still matter for direct detection due to the formation of spatial structure.  The subcomponent case is simpler, and we summarize the enhancement to direct detection prospects in Fig.~\ref{fig:rescaledSensitivity}.  Any experiment whose projected sensitivity intersects the blue regions (set by different values of $\calF$) will be able to probe any friendly axion pair in their mass range with large-enough $\calF$.
Attractive autoresonance may thus be visible to many proposed experiments such as ADMX \cite{stern2016admx}, DM Radio \cite{DMRadiom3,DMRadioGUT}, HAYSTAC \cite{zhong2018results}, KLASH \cite{alesini2017klash}, superconducting RF cavities \cite{lasenby2020microwave,berlin2020axion,berlin2020heterodyne}, and, optimistically, BRASS \cite{BRASS} and MADMAX \cite{beurthey2020madmax}.

\begin{figure*}
    \centering
    \includegraphics[width = \textwidth]{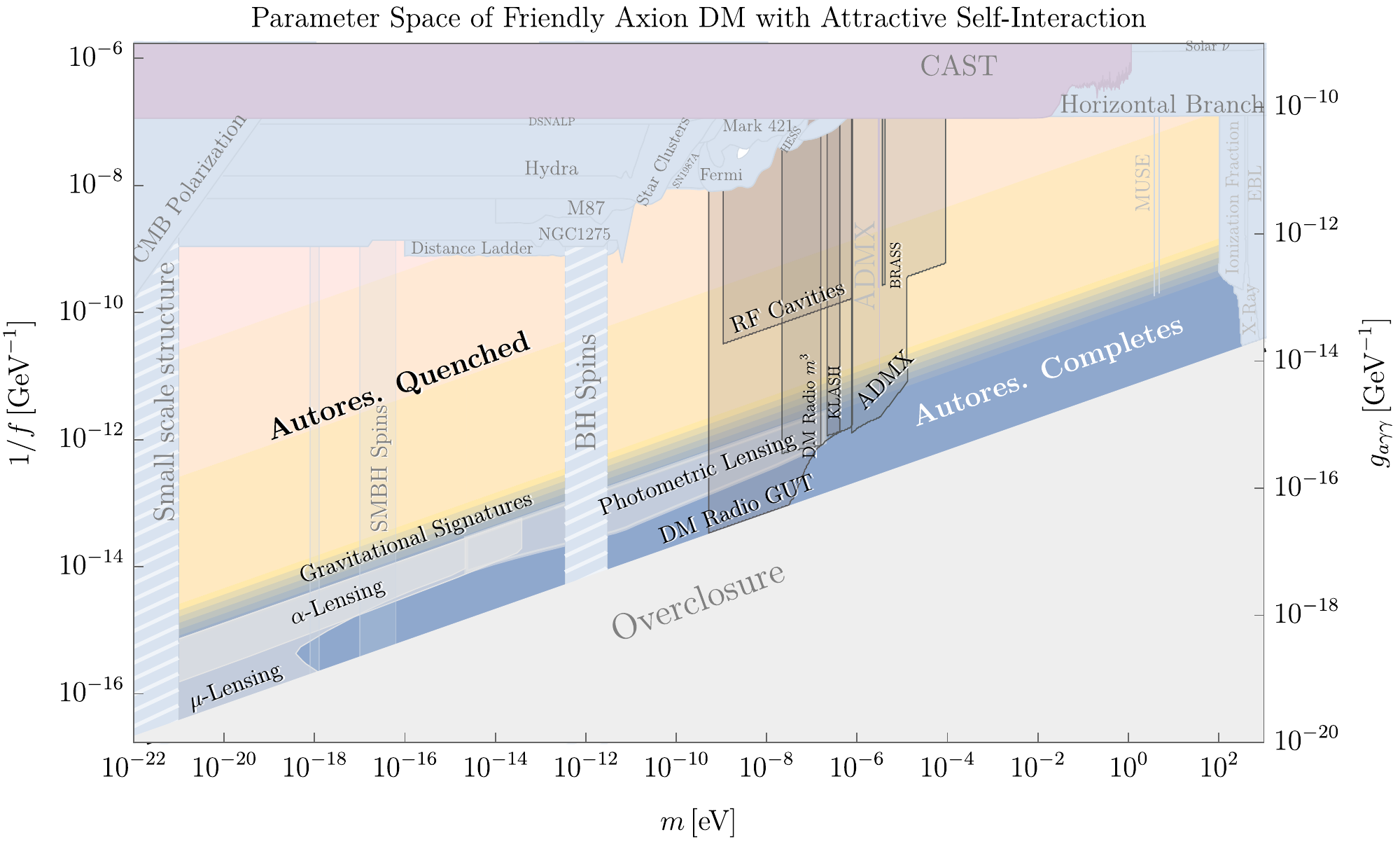}
    \caption{Summary of parameter space, constraints, and signatures for friendly axions in the concrete model of Eq.~\ref{eq:twoAxionPotential} for $\mu = 0.8$ and representative initial conditions that result in autoresonance. This plot is for the case where the friendly pair makes up the entirety of DM, and the axes $m$ and $f$ refer to the mass and decay constant of the short axion specifically. In the region labeled ``Autores.\ Completes,'' autoresonance lasts long enough that nearly all of the axion energy density is in the form of $\ths$, while in the region labeled ``Autores.\ Quenched,'' nonperturbative structure halts autoresonance early and the short axion makes up only a subcomponent.  Throughout, we assume that the short axion has a coupling to photons of size $\gagg \simeq \frac{\alpha}{4 \pi f}$ and we plot direct detection constraints and projections based on this coupling.  Even when $\ths$ is only a subcomponent, it can be a very visible subcomponent due to its enhanced coupling to the SM.
    The regions labeled ``Gravitational Signatures'' are discussed in Sec.~\ref{sec:indirectDetection} and elaborated on in Fig.~\ref{fig:gravitational_signatures}.  The regions labeled ``BH Spins'' and ``SMBH Spins'' refer to BH superradiance constraints discussed in Sec.~\ref{sec:superradiance}. This plot was made using limits compiled in \cite{ciaran_o_hare_2020_3932430,ciaran_o_hare,PhysRevLett.119.031302,armengaud2017constraining,PhysRevD.101.123026,PhysRevD.101.103023,rogers2021strong,buen2020constraints,reynolds2020astrophysical,marsh2017new,dessert2020x,wouters2013constraints,calore2020bounds,ayala2014revisiting,vinyoles2015new,regis2021searching,cadamuro2012cosmological,fedderke2019axion,DMRadioGUT,alesini2017klash,stern2016admx,BRASS,lasenby2020microwave,berlin2020axion,berlin2020heterodyne,DMRadiom3,zhong2018results,beurthey2020madmax}.
    }
    \label{fig:attractiveSensitivity}
\end{figure*}

If the friendly pair comprises the totality of DM, the situation is slightly more complicated.
In this case, as discussed in Sec.~\ref{sec:perturbations}, the self-interactions of $\ths$ can result in the growth of density perturbations that gravitationally collapse earlier than they would have in $\Lambda$CDM and thus form dense axion minihalos.  The region where these structures remain perturbative until most of the axion energy density is in the short axion is labeled ``Autores.\ Completes'' in Fig.~\ref{fig:attractiveSensitivity}, but even in this case anywhere from 95--99\% of the dark matter can reside in these minihalo structures.\footnote{\label{foot:frac}We estimate the ambient dark matter fraction by computing the collapsed fraction in structures whose mass is smaller than that of the Milky Way, and subtracting that from the total collapsed fraction at the present day. This calculation neglects several important effects, including tidal stripping, which may boost the ambient dark matter component. The resulting ambient fractions we found were all between 1\% and 30\%, and we quote 1\% to be conservative.} If the minihalos are numerous enough that one may expect at least one encounter with a detector during its experimental runtime, then the experimental sensitivity is not significantly changed by such substructure, although for a resonant experiment the scanning strategy may need to be modified to maximize the likelihood of scanning the correct frequency during a minihalo encounter \cite{arvanitaki2020large}.  This is generally the case for axions with mass $m \gtrsim 10^{-3} \eV$, where the minihalos are light and therefore extremely numerous.  For smaller axion masses, where the minihalos are heavy and fewer in number, direct detection experiments are sensitive only to the ambient background fraction of DM.  To be conservative, we assume an ambient fraction of only $1\%$ when computing the projected sensitivity of experiments to short axions lighter than $10^{-3} \eV$.

For larger decay constant ratios $\calF \gtrsim 20$, $\ths$ can grow nonperturbative fluctuations during autoresonance.  In this case, detailed simulations are required to understand the full dynamics of autoresonance, but our initial numerical explorations provide tentative evidence that the autoresonant energy transfer is quenched shortly after the $\ths$ field becomes nonperturbative.  Most of the friendly pair's energy density remains in the long axion, but the short axion's energy density is still boosted compared to the ``single axion misalignment'' expectation of Eq.~\ref{eq:misalignmentFraction}.
In addition, if autoresonance is quenched, the overall density fluctuations in the dark sector cease their parametric resonant growth before becoming $\mathcal{O}(1)$.  The large fluctuations in the $\ths$ energy density only lead to $\mathcal{O}(\calF^{-2} (m t_{\te{NL}})^{3/2})$ fluctuations in the total axion energy density, where $t_\te{NL}$ is the time it takes for $\delta \ths$ to become $\mathcal{O}(1)$.
These fluctuations can in principle still seed early collapse during matter domination, but computing their precise effects is difficult due to the uncertainties inherent in the nonlinear collapse of the $\ths$ field.

We adopt a conservative strategy to estimating the sensitivity of future direct detection experiments in the event that autoresonance is quenched.
We take the short axion energy density $\rhos$ to be given by its value at the point that autoresonance ends (i.e.\ the point at which the $\ths$ perturbations become nonlinear), redshifted as matter to late times. 
Nonperturbative $\ths$ fluctuations at the end of autoresonance correspond to ${\mathcal O}(\rho_S/(\rho_S + \rho_L))$ perturbations in the total matter energy density, which remain approximately frozen during radiation domination and grow linearly with the scale factor during matter domination. They then undergo Newtonian collapse at a scale factor $a$ given by:
\begin{align}
    \f{\rho_S}{\rho_S + \rho_L}\f{a}{a_\te{eq}} = \delta_\te{c}\,.
\end{align}
If these structures collapse before the present-day ($a < 1$), some of the $\ths$ and $\thl$ energy densities will reside in dense minihalo structures that may transit an experiment only rarely.  To be conservative, we quote an ambient fraction of only $1\%$ (see Footnote~\ref{foot:frac}).  If these structures have not yet collapsed by the present-day ($a > 1$), we consider an $\mathcal{O}(1)$ fraction of our local halo's $\ths$ density to reside in an ambient component.  This occurs for a density ratio at least as small as
\begin{equation}
    \frac{\rhos}{\rhos+\rhol} \lesssim \delta_\te{c} a_\te{eq} \approx \frac{1}{2000}\,.
\end{equation}

The effects of substructure can thus be viewed as occurring for three distinct ranges of $\calF$. For $\calF \lesssim 20$, autoresonance completes and $\rho_S$ dominates the dark matter density, although its fluctuations suppress the ambient component, reducing overall direct detection sensitivity relative to the case where the friendly pair collectively makes up only a DM subcomponent. For $\calF \gtrsim 20$, $\rho_S$ begins to drop by $\calF^{-2}$, but this is exactly counteracted by its enhanced coupling $\propto\calF^2$. For even larger $\calF \gtrsim 20\sqrt{2000}\approx 900$,  $\rho_S$ comprises an ${\mathcal O}(1/2000)$ subcomponent or less, and its fluctuations no longer lead to early collapse, boosting overall detectability relative to when $\calF \lesssim 900$. Altogether, these effects result in the direct detection prospects of Fig.~\ref{fig:attractiveSensitivity} for the case where the friendly pair makes up all of DM.

\subsection{Gravitational signatures of substructure} \label{sec:indirectDetection}

\begin{figure}[t]
    \centering
    \includegraphics[width = \columnwidth]{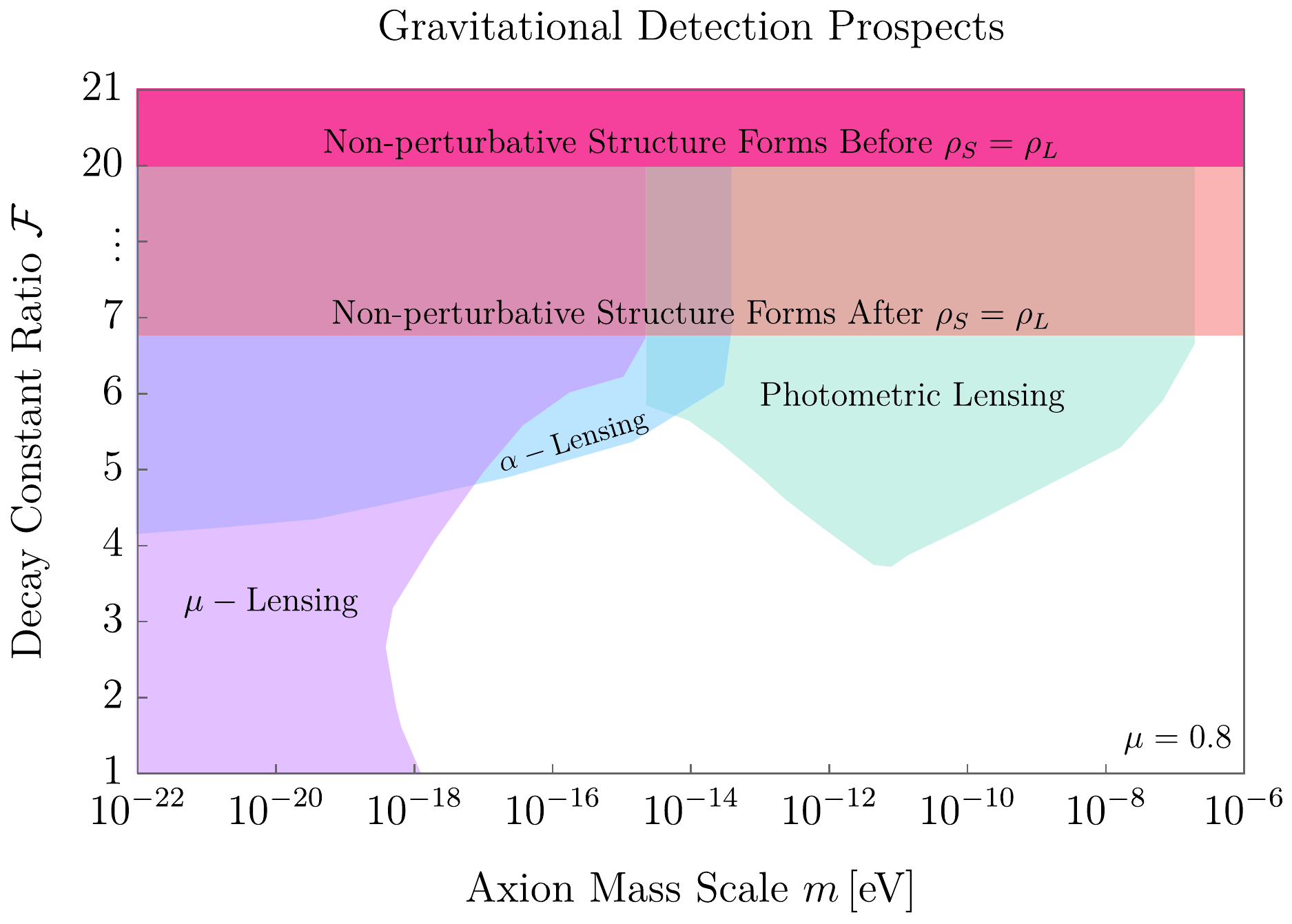}
    \caption{
    Gravitational detection prospects for short-axion DM substructure.  This plot was generated for $\mu = 0.8$, but does not have significant dependence on $\mu$ or the initial misalignment angles (provided they result in autoresonance).  The Purple ``$\mu$-lensing'' and Blue ``$\alpha$-lensing'' regions show projected sensitivities of future telescopes to weak astrometric lensing of local stars (correlated distortions in their velocities with SKA and their accelerations with $\textit{Theia}$ respectively)~\cite{Tilburg_2018}.  The Teal ``photometric lensing'' region may be probed through brightness fluctuations of a critically-lensed distant star~\cite{Dai_2020,arvanitaki2020large}.  Inside the Peach region, nonperturbative structures form during radiation domination, making this region subject to theoretical uncertainties about how this substructure will resolve today.  Nonetheless, we expect that $\mathcal{O}(1)$ density fluctuations will collapse immediately after matter-radiation equality and lead to similar direct detection prospects as for the perturbative region below.  In the Hot Pink region at the top, nonperturbative structure quenches autoresonance before the two axion energy densities equalize; in this region the short axion is a subcomponent and gravitational detection prospects die off quickly as $\calF$ increases.
    }
    \label{fig:gravitational_signatures}
\end{figure}

As discussed in Sec.~\ref{sec:perturbations}, 
if the friendly axion pair makes up a majority of the dark matter then
autoresonance can lead to DM substructures that are denser than predicted by $\Lambda$CDM.  In this respect it is quite similar to the mechanism of Ref.~\cite{arvanitaki2020large}, and indeed the halo mass spectrum predicted by that mechanism is quite similar to the one that emerges from a period of autoresonance.  We are thus able to adapt their subhalo detection projections to the case studied here, and we summarize the results in Fig.~\ref{fig:gravitational_signatures}.  We dedicate the rest of this section to a brief review of the two most relevant signatures, suppressing others which are interesting but slightly less sensitive.  For a more complete treatment we refer the reader to Ref.~\cite{arvanitaki2020large} and the references cited therein.

The first class of indirect signatures we focus on are \textit{astrometric lensing signatures}.  A dense, heavy halo passing through our line-of-sight weakly lenses all background stars, and the lensing pattern is correlated across all stars behind the halo.  A telescope with good angular resolution and a wide field-of-view can in principle look for such correlated deflections and infer the presence of an intervening weak lens.  In practice, since the true positions of individual stars are unknown, it is impossible to observe the correlations of the stars' angular positions on the sky, but as the lens moves it will induce correlated proper motion and proper acceleration of the background star field. A high-angular-resolution experiment that periodically measures the positions of a large number of stars can search for such correlated motions, either with templates or by looking for global correlations.  Several such astrometric experimental efforts either exist (\textit{Gaia}~\cite{Gaia2018}, HST~\cite{HSTastrometric}) or are planned (\textit{Theia}~\cite{Theia2017}, WFirst~\cite{WFirstAstrometry}, SKA~\cite{SKAastrometry}, TMT~\cite{Skidmore_2015}).  Ref.~\cite{Tilburg_2018} worked out dense subhalo sensitivity projections for \textit{Gaia} and \textit{Theia}, and we report these in Fig.~\ref{fig:gravitational_signatures} for the halo mass spectrum predicted in Sec.~\ref{sec:perturbations}.

Another potential class of observable signatures are those associated with \textit{photometric microlensing}.  The basic idea is to monitor a distant star and look for changes in its brightness that would indicate a gravitational lens passing through the line of sight.  This technique has been used to place constraints on extremely compact objects (such as primordial black holes), but in general it is harder to use it for dilute, gravitationally-bound subhalos because they only lens weakly and thus have minute effects on a star's observed brightness.  To deal with this, Ref.~\cite{Dai_2020} has proposed using highly-magnified stars that are only observable because they lie close to a critical gravitational lensing caustic of a galaxy cluster.  If the DM in the galaxy cluster is composed of subhalos, the virial motion of these subhalos will add Poissonian noise to the position of the star, which has an amplified impact on the star's brightness.  This noise has a characteristic frequency and amplitude that depends on the DM halo mass spectrum, and Ref.~\cite{Dai_2020} suggests the observation of this noise can probe DM substructure.  Ref.~\cite{arvanitaki2020large} has made projections of the sensitivity of such a technique for gravitationally-bound subhalos and we report these in Fig.~\ref{fig:gravitational_signatures} for the halo spectrum calculated in this paper.  It should be noted that these projections are subject to potentially significant uncertainties associated with the galactic evolution (and tidal stripping) of such gravitationally-bound subhalos, and we caution that proper simulations must be done to confirm them.

For $\calF \gtrsim 20$, perturbations in the short axion field can grow nonperturbative and quench the autoresonance before the majority of the axion energy density is transferred to $\ths$.  In this case, even though there are large fluctuations in the short axion field, the overall density fluctuations are small because the majority of the axion energy density is still in $\thl$.  Structures thus collapse gravitationally at roughly the same time they would have in $\Lambda$CDM, and all gravitational signatures of autoresonance disappear.  We show this in Fig.~\ref{fig:attractiveSensitivity}, where the gravitational signatures appear only in the region where $\ths$ can compose the totality of dark matter.

\subsection{Superradiance signatures and constraints} \label{sec:superradiance}

The phenomenon of black hole superradiance (SR), by which the angular momentum of an initially rapidly rotating black hole (BH) is transferred to a cloud of bound axions generated around the BH, can be used to constrain axions at ultralight masses by measuring the age and spin of astrophysical BHs \cite{zeldovich1971,arvanitaki2010string, arvanitaki2010,brito2014,arvanitaki2014,brito2015,cardoso2018constraining,simon2020,mehta2021superradiance}. SR bounds are quite unique in that they are more constraining for an axion which has \emph{small} interactions, as interactions tend to slow down the extraction of angular momentum from the BH into the cloud. Even a single axion with a potential typified by Eq.~\ref{eq:cosinePotential} inevitably has \emph{self}-interactions, which at leading order are quartic with dimensionless coupling $\lambda \sim m^2/f^2.$ As one moves towards values of $f$ smaller than $\sim 10^{15}\GeV$ in axion parameter space, the growth of the SR cloud is cut off at perturbative values of $\theta$ and angular momentum can no longer efficiently be extracted from the BH \cite{simon2020}. \par
For the case of the coupled short and long axions studied here, as long as the evolution remains perturbative in $\theta_S,\theta_L$, SR is better studied in the mass basis in which flavor oscillations are removed (App.~\ref{app:massVsInteractionBasis}). For $\mathcal F \gg (1-\mu^2)^{-1}$, the heavy state $\nu_h \approx \phi_S$ has quartic self-interactions $\lambda_{hh} \approx m^2/f^2$, while the light state $\nu_l \approx \phi_L$ has quartic self-interactions $\lambda_{ll} \approx (\mu/\mathcal F)^2 \lambda_{hh}$.  As emphasized previously, in the scenario in which the friendly axion pair is DM, the light state (i.e.\ the long axion) must fall within the region of parameter space that would yield the correct present-day DM density in the absence of friendly interactions (i.e.\ within a band centered on the ``$\Theta_0=\pi/2$ Misalignment'' line of Fig.~\ref{fig:new_parameter_space}). The coupling $\lambda_{ll}$ is therefore fixed. Depending on the value of $\mathcal F$, the self-coupling $\lambda_{hh}$ of the heavy state (i.e.\ the short axion) may or may not be small enough that the SR bounds apply to the short axion directly. If $\mathcal F$ is large enough that the short axion cannot be constrained by SR, then the scenario of two friendly axions being DM is still constrained by SR bounds on the long axion (one can check that cross-couplings do not change those bounds in that limit). For this reason, we have shown the SR bounds from astrophysical BHs on Figs.~\ref{fig:new_parameter_space} and \ref{fig:attractiveSensitivity} as extending to arbitrarily large $\mathcal F$, since they exclude a long axion living near the ``$\Theta_0=\pi/2$ Misalignment'' line within that mass range. \par
Because of the complicated merger history of supermassive BHs and the larger uncertainties on their measured parameters, it is difficult to make a definite claim that a lack of spindown implies the absence of an axion in the spectrum. A more detailed understanding of merger histories and better measurements could make supermassive BHs robust probes of  axions in the $10^{-18}-10^{-16}\te{eV}$ mass range in the future. We show this region on Fig.~\ref{fig:new_parameter_space} and Fig.~\ref{fig:attractiveSensitivity} in a lighter shade to reflect this uncertainty. \par
We note that there is a somewhat tuned---but not entirely excluded---scenario in which \emph{neither} DM axion can be constrained by SR bounds on BH spins. If $\mu$ is close enough to unity that $\mathcal F \ll (1-\mu^2)^{-1}$, one can have that $\lambda_{ll} \simeq \lambda_{hh} \simeq m^2/f^2$  and all mass states have comparable self-interactions. In the interaction basis, this can be explained by observing that strong mixing between the two axions causes the long axion to inherit the strong self-interactions of the short axion via flavor oscillations. One might view this as the spindown signatures of an axion with a nominally large decay constant being masked by the presence of a closely resonant axion with a small decay constant.\par 
If the friendly axion pair is a subcomponent of DM, the long axion is not required to live near the ``$\Theta_0=\pi/2$ Misalignment'' line of Fig.~\ref{fig:new_parameter_space}. In this case, both axions can have small enough decay constants to evade SR spin bounds. Rather than rapidly extracting the angular momentum from a BH and storing it in a SR cloud, axions with small decay constants form smaller clouds that slowly transfer angular momentum directly from the BH to spatial infinity in the form of coherent axion waves that could be detected on Earth by planned nuclear magnetic resonance 
experiments \cite{simon2020}. The signal strength on Earth of these small clouds scales as the axion mass to the fourth power, but does not scale with the decay constant. It is therefore possible that small clouds of both short and long axions exist simultaneously around a BH and emit axion waves at nearby frequencies $\sim m$ and $\sim \mu m$ that are similarly detectable. A more detailed study of cross-cloud interactions would be necessary to fully understand this scenario.

\section{Repulsive self-interactions} \label{sec:repulsiveSelfInteractions}
\begin{figure}
    \centering
    \includegraphics[width = \columnwidth]{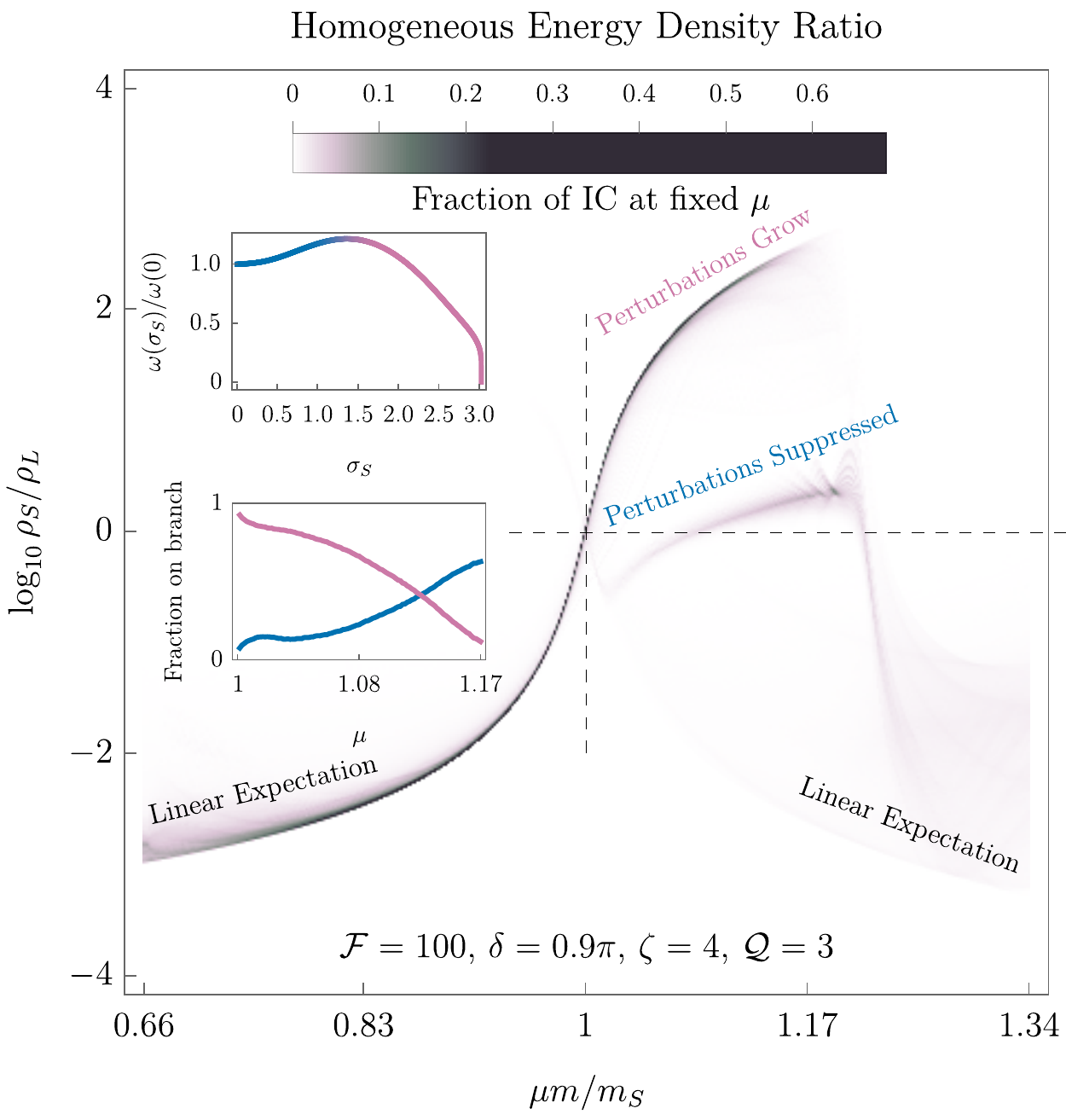}
    \caption{The distribution of late-time energy density ratios $\rho_S/\rho_L$, as defined by Eq.~\ref{eq:kineticDensity} in the potential Eq.~\ref{eq:repulsivePotential}. For each choice of $\mu$, the initial conditions $(\Ths(0),\Thl(0))\in[-\pi,\pi]\times[-\pi,\pi]$ are sampled uniformly, and the results are binned by the final density ratio $\log\rho_S/\rho_L$. This figure should be compared to Fig.~\ref{fig:autoresParamSpace}. For $m_S\leq\mu m\lesssim 1.17 m_S$, (note, $m_S \approx 3 m$ is the short axion mass) there are two $\Ths$ amplitudes that can autoresonate with $\Thl$, corresponding to the upper and lower tails visible in the upper right.
    \textbf{\underline{Top Inset:}} The frequency versus amplitude curve for $\Ths$, showing that small amplitudes experience net-repulsive self-interactions, which suppress perturbation growth (Blue), and larger amplitudes experience net-attractive self-interactions, which enhance growth (Magenta). The two autoresonant tails correspond to the two solutions $\sigma_S$ of the equation $\omega(\sigma_S) = \mu$ for $\mu\geq\omega(0)$. \textbf{\underline{Bottom Inset:}} The fraction of initial misalignment angles landing on each branch. Note, the total probability of landing on either nonlinear branch does not equal 1 because one may also land on the linear branch, where the short axion does not autoresonate.}
    \label{fig:repulsiveDensityRatio}
\end{figure}

So far our analysis has been focused on the axion potential of Eq.~\ref{eq:twoAxionPotential}, which has attractive self-interactions for $\ths$.  This is often the case in the most minimal axion potentials, because instanton contributions typically enter the potential as cosines, which have negative (i.e.\ attractive) quartic interactions.  However this is not a universal rule, and repulsive self-interactions can exist in axion models \cite{fan2016ultralight,mehta2021superradiance}.  In this section we summarize the phenomenology when the short axion has repulsive self-interactions.  As we will see, autoresonance can occur with few differences from the attractive case.  Importantly, however, repulsive self-interactions can prevent all structure growth during autoresonance, implying that autoresonance cannot be cut off early by non-perturbative structures.  Therefore, if the system lands on autoresonance, it is guaranteed to complete the energy transfer, further enhancing signatures at large decay constant hierarchies $\calF\gg 20$, for which attractive self-interaction signatures would be saturated (see Fig.~\ref{fig:rescaledSensitivity}). Future direct detection experiments such as ADBC \cite{liu2019searching}, DANCE \cite{michimura2020dance}, DM Radio 50L \cite{DMRadiom3}, LAMPOST \cite{baryakhtar2018axion}, aLIGO \cite{nagano2019axion}, ORGAN \cite{mcallister2017organ}, and TOORAD \cite{schutte2021axion} may therefore see a self-repulsive short axion, even though they cannot access the parameter space relevant to an attractive theory.

To make our discussion concrete, consider the following axiverse-inspired potential with repulsive $\theta_S$ self-interactions
\begin{align}
\label{eq:repulsivePotential}
    V(\ths,\thl)&=m^2 f^2\p{\zeta^2(1 - \cos(\ths + \thl)) \right.\\\notag&\left. + (1 - \cos({\cal Q}\ths + \delta)) + \mu^2 \calF^2(1 - \cos\thl)}\,.
\end{align}
For small $\ths$ amplitudes, interactions are repulsive if $1<{\cal Q}\lesssim\zeta\lesssim {\cal Q}^2$ and $3\pi/4\lesssim\delta\lesssim5\pi/4$, and repulsive autoresonance may occur if $\mu^2\gtrsim \zeta^2 - {\cal Q}^2$, and ${\cal F}\gg \zeta$. 

A good diagnostic of autoresonance is to measure the late-time energy density ratio of $\ths$ and $\thl$ as in Fig.~\ref{fig:autoresParamSpace}. As before, it is often helpful to think about the energy density ratio in the interaction basis, since it is this quantity that late-time signatures depend on. However, the partition of energy between the two fields becomes ambiguous beyond the scale of flavor oscillations. A useful choice is the time-average of the corresponding kinetic term
\begin{align}
\label{eq:kineticDensity}
    \rho_S\approx \gen{(\del_t\Ths)^2}\,,\hspace{0.5cm}\rho_L\approx \calF^2 \gen{(\del_t\Thl)^2}\,.
\end{align}
This estimate generalizes easily to theories with a large number of fields and instantons, provided the mass matrix is close to diagonal. We plot the late time energy density ratios in Fig.~\ref{fig:repulsiveDensityRatio} for a representative set of parameters, which is meant to be compared to Fig.~\ref{fig:autoresParamSpace}. This plot shows two important distinguishing features. First, autoresonance occurs for driver frequencies above the short rest mass $\mu > \omega_S(0)$, and not below as in the case of attractive self-interactions. This is a consequence of repulsive self-interactions, which cause the short axion's frequency to increase with an increase in its amplitude (see inset of Fig.~\ref{fig:repulsiveDensityRatio}). Second, there are two apparent autoresonance bands in Fig.~\ref{fig:repulsiveDensityRatio} as opposed to the single band in Fig.~\ref{fig:autoresParamSpace}. This is again a consequence of the nontrivial dependence of frequency on amplitude. Because $\Ths$ is a periodic variable, the repulsive self-interactions that take place at small amplitudes cannot continue to arbitrary field displacements. Thus, the positive frequency shift that occurs at small amplitudes must eventually turn around and decrease, ultimately passing through zero as shown in the inset of Fig.~\ref{fig:repulsiveDensityRatio}. Therefore, every possible positive frequency shift in the potential Eq.~\ref{eq:repulsivePotential} is achieved at two separate amplitudes $\sigma_S$. Depending on the initial conditions, the driver $\Thl$ of a particular frequency $\mu$ may drive $\Ths$ at one of two possible amplitudes, giving rise to the two autoresonant tails.

These two tails, while both the consequence of repulsive self-interactions, lead to very different phenomenology. Let us first consider the small amplitude tail (Blue). Here, the result of the small-amplitude formalism for computing the perturbation growth rate Eq.~\ref{eq:growthRate} goes through unchanged: perturbations do not grow because the frequency shift $\delta\omega$ is positive (see App.~\ref{app:perturbations_in_detail}). 

At larger amplitudes (Magenta), the motion of the zero-mode $\Ths$ is no longer well approximated by its motion near the bottom of the potential, and the formalism of App.~\ref{app:perturbations_in_detail} no longer applies. Even though we cannot analytically quantify the growth rate of modes beyond the small amplitude approximation, we may gain some qualitative intuition through the following considerations. Recall from our discussion in Sec.~\ref{sec:perturbations_one} that perturbations are agnostic to features in the potential below the kinetic energy of $\Ths$. Therefore, the relevant features of the potential for perturbation growth occur near the turnaround points where kinetic energy vanishes. At these points, the potential behaves as locally attractive if increasing $\sis$ decreases $\omega(\sis)$, and locally repulsive if it increases $\omega(\sis)$. In other words, the relevant quantity for mode growth is $\omega'(\sis)$. This argument predicts that autoresonance on the large amplitude tail (Magenta) of Fig.~\ref{fig:repulsiveDensityRatio} for which $\omega'(\sis) < 0$, representing net-attractive self-interactions, drives the growth of large perturbations. We have confirmed this intuition with numerical simulations.

The observational prospects for repulsive autoresonance are striking. Spatial perturbations to the axion field do not grow, and so autoresonance is not quenched even for $\calF \gg 20$. This implies that the boost to direct detection signal strength (Eq.~\ref{eq:targetSensitivityAutores}) can be quite large if such large hierarchies of decay constants exist in the axiverse.\footnote{This itself is a question worthy of future study.  At least some concrete realizations of the axiverse result in decay constant distributions that are spread only $1-2$ orders of magnitude about a central value \cite{mehta2021superradiance}.} Such strongly coupled relics provide important targets for direct detection experiments probing mass ranges where both the expectation Eq.~\ref{eq:targetSensitivityNaive} and that of attractive autoresonance are out of reach. These observational implications motivate us to take the possibility of repulsive autoresonance seriously, even though the potential Eq.~\ref{eq:repulsivePotential} is repulsive over a relatively small range of parameters. Whether repulsive interactions remain relatively rare in realistic axiverse potentials is an open question, and our model serves as motivation to study this question further.

\section{Discussion and future directions} \label{sec:discussion}

In this paper, we have studied the dynamics of coupled axion dark matter, and in particular the case of a pair of axions with nearby masses. We have shown that one axion can dynamically adjust its amplitude so that its frequency matches that of another and then remain fixed at this amplitude for cosmologically-relevant times, avoiding the damping effects of Hubble friction 
long enough
to dominate the energy density in the axion sector.  This frequency-matching is a form of autoresonance, and within the concrete model of this paper, it is a common phenomenon provided the long axion mass $m_L$ is within around $25\%$ of the short axion mass: $0.75 m_S \lesssim m_L < m_S$.  This gives a good notion of how ``friendly'' two axions must be to see the effects we have described, and such a coincidence of masses is unsurprising in an axiverse with $\mathcal{O}(100s)$ of axions distributed log-flat in mass.

If autoresonance does occur, the energy transfer typically runs from the axion with a larger (``long'') decay constant to the axion with a smaller (``short'') decay constant, meaning that the effect on the axion sector is generally to make it more detectable by direct detection experiments.
This alone is a very exciting prospect, and various experiments such as ADMX, DM Radio, and HAYSTAC will probe significant regions of parameter space of friendly axions, independent of the friendly pair's total energy density.  In addition, if the pair makes up all of DM, we have shown that autoresonance for a potential with attractive self-interactions can lead to a parametric-resonance-driven growth of spatial perturbations in the axion field, which can then collapse at early times and form dense axion minihalos.  For axion masses $m \lesssim 10^{-7} \eV$, these have gravitational signatures that can be probed with near-future experiments. 
If autoresonance lasts for a long time (which occurs for large hierarchies of axion decay constants), spatial perturbations can grow nonperturbative and the analytic formalism developed here breaks down, but our preliminary numerical results suggest that the autoresonance is quenched.
Still in this case, the short axion receives a significant boost to its energy density.
The various signatures discussed are complementary, and in some parts of parameter space multiple signatures may be observed, allowing a concrete identification of the friendly axion scenario from other mechanisms which may predict similar minihalo spectra.

There remain several natural questions about this mechanism.  The first is
whether the QCD axion, which has a temperature-dependent mass, can receive an energy density enhancement due to resonances with an axiverse. It turns out that autoresonant (i.e.\ nonlinear) energy transfer to the QCD axion is impossible: by the time the QCD axion nears its zero-temperature mass $m_a$, any would-be friend has already lost too much energy for the nonlinear interactions in the QCD potential to be accessible, putting autoresonance out of reach. \par
On the other hand, linear resonances are accessible to the QCD axion: as the QCD axion mass increases through the masses of other axions in an axiverse, an level-crossing may lead to energy transfer to or from the QCD axion. The possibility of the QCD axion generating a cosmological abundance of axiverse axions has been explored in Refs.~\cite{daido2015domain,kitajima2015resonant,daido2016level,ho2018enhanced}. 
We note that it is similarly possible for an axiverse axion to transfer its initial energy density to the QCD axion, leading to QCD axion DM signatures at large masses $m_a\gtrsim \te{meV}$, well above the range expected from an $\mathcal{O}(1)$ initial misalignment angle.

A second 
natural question is what happens if the decay constant hierarchy $\calF$ is large enough that spatial fluctuations grow nonperturbative during autoresonance and collapse into oscillons---compact axion structures bound by self-interactions.  We have performed numerical simulations in this regime that indicate autoresonance is quenched by oscillon formation, but they are limited in their resolution.  Further simulations are necessary to verify our results, but it is worth pointing out that oscillons can potentially have intriguing signatures of their own which we have not worked out here.
Oscillons in general do not have very long lifetimes, but may live significantly longer in the background of a long axion condensate that can resonantly drive them (see Refs.~\cite{friedland2003emergence,friedland2005excitation} for examples of driven nonlinear field equations).  Energy conservation suggests that in this case the oscillon's lifetime may be extended parametrically to:
\begin{equation}
    m T_{\text{driven}} \propto (m T_{\text{vacuum}})^{4/3}\,,
\end{equation}
where $m$ is the axion mass and $T_{\text{vacuum}}$ is the lifetime of an oscillon with fundamental frequency $\mu m$.
For potentials with somewhat long-lived oscillons already (see e.g.\ Refs.~\cite{salmi2012radiation,amin2012oscillons,kawasaki2020oscillon,olle2020recipes,zhang2020classical,cyncynates2021structure}), this enhancement would allow them to survive to matter-radiation equality even at larger axion masses $m \gtrsim 10^{-15} \eV$ for the longest lived oscillons \cite{cyncynates2021structure}.  At late times, if such an oscillon is in a galactic halo of $\thl$ DM and it can remain locked to the virialized $\rho_L$ background, then the only upper bound on its lifetime comes from exhausting the entire halo energy density.  Since even a small subcomponent of oscillons can be detected \cite{prabhu2020resonant,buckley2021fast,prabhu2020optical}, this is an important case to study further.

\begin{figure}
    \centering
    \includegraphics[width = \columnwidth]{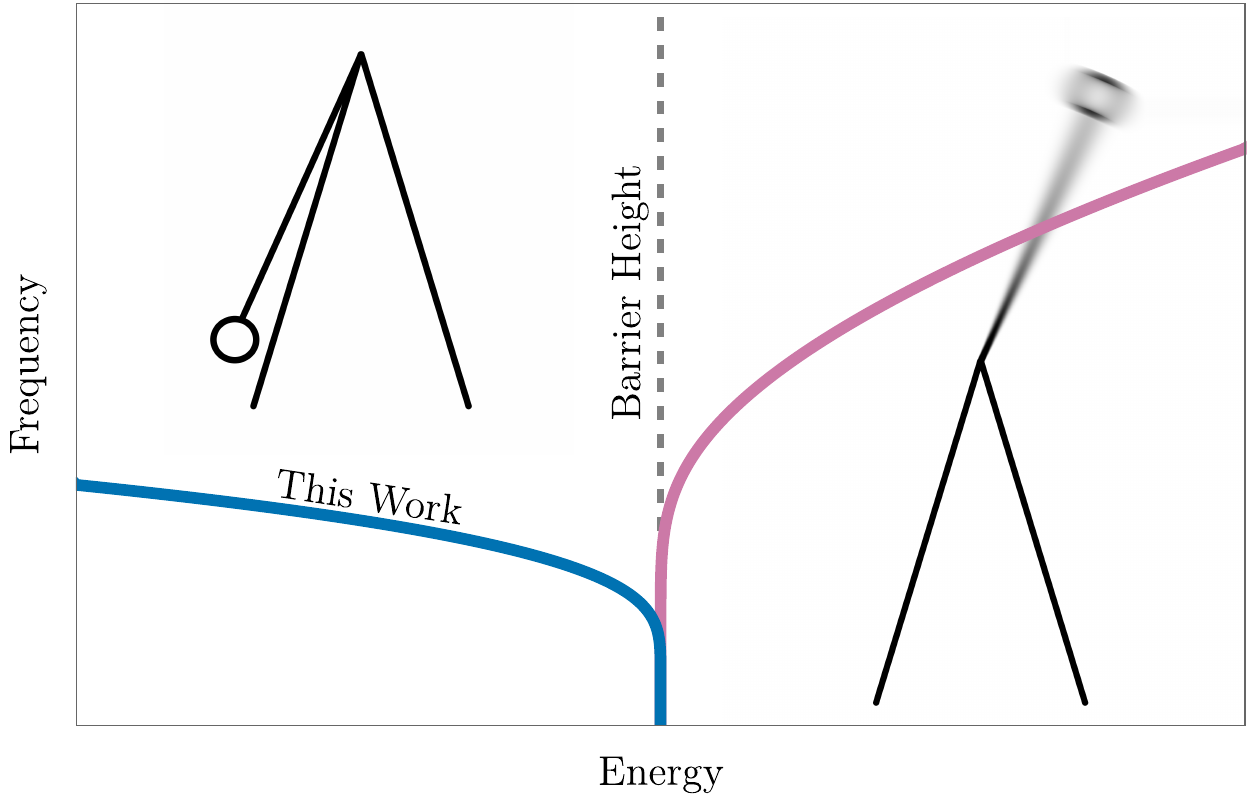}
    \caption{
       The frequency of a classical pendulum versus its energy. There are two distinct regimes. First, at energies below the barrier height, the pendulum oscillates around its equilibrium angle, at a frequency which decreases with energy (Blue). At energies above the potential barrier height, the pendulum completes full rotations. In this regime, it is the pendulum's velocity which oscillates around an `equilibrium value,'  and the oscillation frequency increases with energy (Magenta). In this paper, we have described how a driver can lock onto the low-energy branch of this curve through autoresonance. The high-energy branch opens up the possibility of autoresonance and associated signatures over a larger frequency range.
    }
    \label{fig:swing_resonance}
\end{figure}

\begin{figure}[t]
    \centering
    \includegraphics[width = \columnwidth]{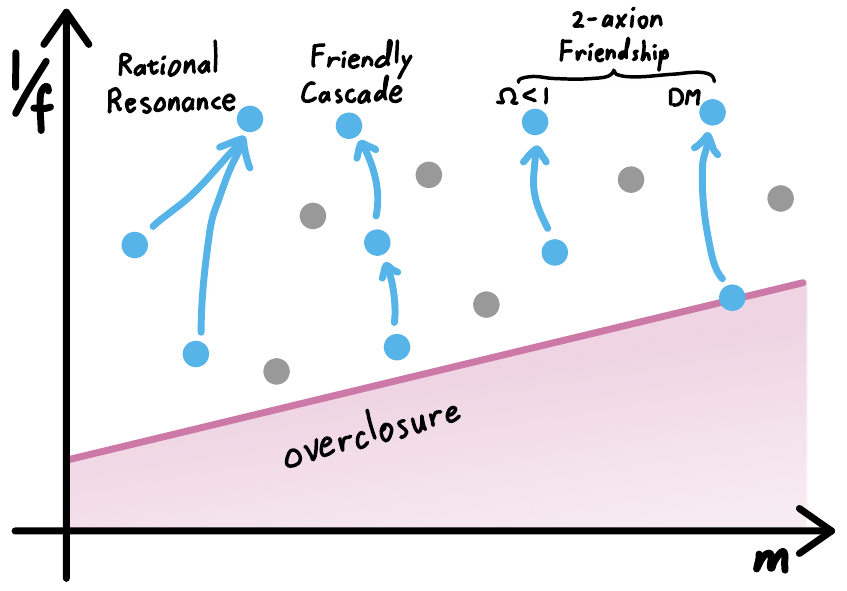}
    \caption{
       Some possibilities for energy redistribution in the axiverse.  Each axion in the axiverse is represented as a point in the mass-decay constant plane.  The magenta line represents those values of $m$ and $f$ that lead to the proper relic abundance of DM for $\mathcal{O}(1)$ initial misalignment if the axions are treated independently.  As we have shown here, energy density can be resonantly transferred to axions with smaller decay constants (illustrated by blue arrows).  We have studied the case of two axions with nearby masses (``2-axion Friendship''), both when the pair comprise the totality of DM (``DM'') and when they are only a subcomponent (``$\Omega < 1$'') but there are other possibilities in a realistic axiverse.  For example multiple axions with nearby masses could transfer energy in a sequence (``Friendly Cascade'') or collections of axions could dynamically synchronize and lock onto a rational resonance, where no two frequencies match identically but they are related rationally.  These latter possibilities are likely to be less common than the two-axion case discussed in this work because they require more coincidences, but with $\mathcal{O}(100\text{s})$ of axions they may still be possible and further work is necessary to understand them.
    }
    \label{fig:mfcartoon}
\end{figure}

The example of autoresonance we have studied in this paper is not the only type of nonlinear resonance possible in a coupled oscillator system, and future work is needed to understand whether other types of resonance can show up in the axiverse.  For example, even a pendulum can resonate in a qualitatively different way than we have studied so far: at energies above the potential barrier, it can make complete circuits about its pivot, which opens up a large window of higher frequencies to autoresonance due to the nonlinearities of the oscillator.  This is illustrated in Fig.~\ref{fig:swing_resonance}. These circular resonances may be accessed if one axion obtains an approximately constant velocity, which may occur because of the complicated geometry of multi-axion potentials or because of an explicit breaking of PQ symmetry in the early universe \cite{hall2020axion,chang2020new}.\footnote{Mechanisms like \cite{hall2020axion,chang2020new} would also result in large density perturbations because the axion kinetic energy delays the onset of harmonic oscillations (see Sec.~\ref{sec:perturbations}).}
We illustrate some of the other possibilities for axiverse axions in Fig.~\ref{fig:mfcartoon}, but further work is needed to understand which of these can be realized in realistic models. 

String theory remains the most successful attempt at a unified theory of quantum gravity, but unfortunately we lack many experimental probes of this possibility.  Nearly all new effects (particles, forces, nonlocality, etc.) within the theory are suppressed by the string scale, which in principle can lie quite close to the Planck scale, making it virtually impossible to test with current technology.  String theory axions are a notable exception, and observing several distinct axions in the particle spectrum would hint at string theory as a UV completion for the SM.
In many scenarios, string axions can be accurately approximated as weakly-interacting massive fields.
However, this approximate picture of axions as a collection of perturbatively-coupled oscillators misses 
something important: 
in such a system, \textit{exact} resonances are necessary for appreciable energy transfer between normal modes~\cite{ford1961equipartition}.
Since there is no reason to expect that axion masses obey simple integer relations (assuming the masses are temperature-independent and remain fixed as the universe evolves), such exact resonances are impossible, and one would conclude that no significant transfer of energy can happen between axions in an axiverse.  As we have shown here, quite the opposite is true: in a realistic system, exact resonance can be obtained \textit{dynamically}, because the frequency of a nonlinear oscillator is a function of its amplitude.  In other words, perturbative treatments can miss important features if they do not account for the full nonlinearity of the axiverse potential.
The two-axion case studied here should be considered a minimal example of the effects of nonlinear couplings in the axiverse, and it already provides exciting signatures in the reach of near-future experiments.

\begin{acknowledgments}
We would like to thank Mustafa Amin, Asimina Arvanitaki, Savas Dimopoulos, Robert Lasenby, Viraf Mehta, Glenn Starkman, Harikrishnan Ramani for many useful discussions. 
We thank Asimina Arvanitaki, Sebastian Baum, Alexander Gu{\ss}mann, Robert Lasenby, and Viraf Mehta for helpful comments on the draft, and in particular we thank Masha Baryakhtar for her detailed and thoughtful comments.
D.C.\ is grateful to Dmitriy Zhigunov for the initial conversations that led to this line of research, and for helpful conversations about pseudospectral methods. Some of the computing for this project was performed on the Sherlock cluster. We would like to thank Stanford University and the Stanford Research Computing Center for providing computational resources and support that contributed to these research results. D.C.\ is grateful for the support of the Stanford Institute for Theoretical Physics (SITP) and the Moore Foundation. T.G.-T.\ is grateful for the support of the Department of Energy under Grant No. DE-SC0020266. O.S.\ is supported by a DARE fellowship from the Office of the Vice Provost for Graduate Education at Stanford University. J.O.T.\ is supported by a William R.\ Hewlett Stanford Graduate Fellowship.
\end{acknowledgments}
\appendix

\section{Mass vs.\ interaction basis} \label{app:massVsInteractionBasis}
The $\phs$ and $\phl$ fields in Eq.~\ref{eq:twoAxionPotentialPhi} are not mass eigenstates (i.e.\ they are not stable under propagation in the non-interacting limit).  They are, however, the natural basis in which to consider most of the early-time dynamics and the late-time signatures (since any couplings to the SM likely descend from the UV theory).  In this appendix we clarify this point and include the transformation from the interaction basis ($\phs$, $\phl$) to the mass basis.  For $\mathcal{F} \gg 1/(1 - \mu^2 )$ the two bases are quite similar and so this discussion has very little effect on the interpretation of the dynamics studied in this paper, although we have included it in our results where relevant.

We wish to find the propagation eigenstates of $\phl$ and $\phs$.  Expanding the potential of Eq.~\ref{eq:twoAxionPotentialPhi} to quadratic order yields a mass mixing matrix:
\begin{align}
    \nonumber V(\phl, \phs) &\approx \frac{1}{2} m^2 \left( \phs^2 + \frac{1}{\mathcal{F}^2} \phl^2 + \frac{2}{\mathcal{F}} \phs \phl + \mu^2 \phl^2 \right) \\
    &= \frac{1}{2} m^2 \mat{cc}{\phs & \phl} \mat{cc}{1 & \mathcal{F}^{-1} \\ \mathcal{F}^{-1} & \mu^2 + \mathcal{F}^{-2}} \mat{c}{\phs \\ \phl}
\end{align}
which has off-diagonal elements suppressed by the ratio of decay constants.  This matrix is easy to diagonalize, yielding the following basis of heavy and light fields $\nu_h$ and $\nu_l$:
\begin{align}
    \nu_h &\equiv \phs \cos \eta + \phl \sin \eta \\
    \nu_l &\equiv - \phs \sin \eta + \phl \cos \eta \\
    m_h^2 &= m_0^2 + \Delta m^2 \\
    m_l^2 &= m_0^2 - \Delta m^2 \\
    m_0^2 &\equiv \frac{1}{2} m^2 (1 + \mu^2 + \mathcal{F}^{-2} ) \\
    \Delta m^2 &=\f12 m^2(1 - \mu^2 - \mathcal{F}^{-2})\sec2\eta \\
    \sin \eta &\equiv \frac{1}{\sqrt{2}} \sqrt{1 - \frac{1 - \mu^2 - \mathcal{F}^{-2}}{\sqrt{ 4 \mathcal{F}^{-2} + (1 - \mu^2 - \mathcal{F}^{-2} )^2 } } }
\end{align}
For the late-time Newtonian evolution of the axion energy density, it is most useful to describe the system in this basis, since the energy densities in the fields $\nu_l$ and $\nu_h$ are constant in the small-amplitude limit.  As mentioned above, for $\mathcal{F} \gg (1 - \mu^2)^{-1}$ we have $\sin \eta \approx 0$ and so $\nu_l \approx \phl$ and $\nu_h \approx \phs$.  The effects of this basis rotation are thus very slight for most of the parameter space discussed in this paper, but we nevertheless use this basis when performing the analysis of Sec.~\ref{sec:newtonianPert}.

In rotating to the mass basis, we have also modified the quartic interactions
\begin{align}
\notag
    V_\te{int}(\nu_h,\nu_l) &= \f{1}{4!}\lambda_{hh}\nu_{h}^4 +\f{1}{3!}\lambda_{hhhl}\nu_{h}^3\nu_l + \f{1}{4}\lambda_{hl}\nu_h^2 \nu_l^2 
    \\&+\f{1}{3!}\lambda_{hlll}\nu_h\nu_l^3 + \f{1}{4!}\lambda_{ll}\nu_l^4\,,
\end{align}
where the even-interactions are
\begin{align}
    \f{f^2}{m^2}\lambda_{hh}&=-b^4\mu^2\mathcal{F}^{-2} - (a + b\mathcal{F}^{-1})^4\,,\\
    \f{f^2}{m^2}\lambda_{hl}&=-a^2 b^2\mu^2 (\mathcal{F}^{-2} + \mu^2)\,,\\
    \f{f^2}{m^2}\lambda_{ll}&=-a^4\mu^2\mathcal{F}^{-2} - (b - a\mathcal{F}^{-1})^4\,,
\end{align}
with
\begin{align}
    a^2= \cos^2\eta\,,\hspace{1cm} b^2= \sin^2\eta\,.
\end{align}
It turns out that the leading order self-interactions in the non-relativistic theory only come from those terms that conserve parity under $\nu_h\to - \nu_h$ and $\nu_l\to - \nu_l$ independently.  Odd-parity interactions enter at the next-to-leading order in the non-relativistic approximation and so we do not take them into account in this work, but for completeness, we list them here:
\begin{align}
    \f{f^2}{m^2}\lambda_{hhhl}&=-(a + b\mathcal{F}^{-1})^3 (a\mathcal{F}^{-1} - b) - a b^3\mu^2\mathcal{F}^{-2}\,,\\
    \f{f^2}{m^2}\lambda_{hlll}&=-(a\mathcal{F}^{-1} - b)^3(a + b\mathcal{F}^{-1}) - a^3 b\mu^2\mathcal{F}^{-2}\,.
\end{align}

\section{A more detailed study of autoresonance} \label{app:autoresInDetail}

The dynamics of autoresonating axions are rich, and in this appendix we focus on building analytic intuition for their behavior.
Even though the oscillators are quite nonlinear, it turns out that we can get good approximations for several interesting quantities by searching for stable autoresonant solutions and perturbing around them.

\subsection{Adiabatic evolution of resonance curves} \label{app:NLrescurve}

\begin{figure}
    \centering
    \includegraphics[width = \columnwidth]{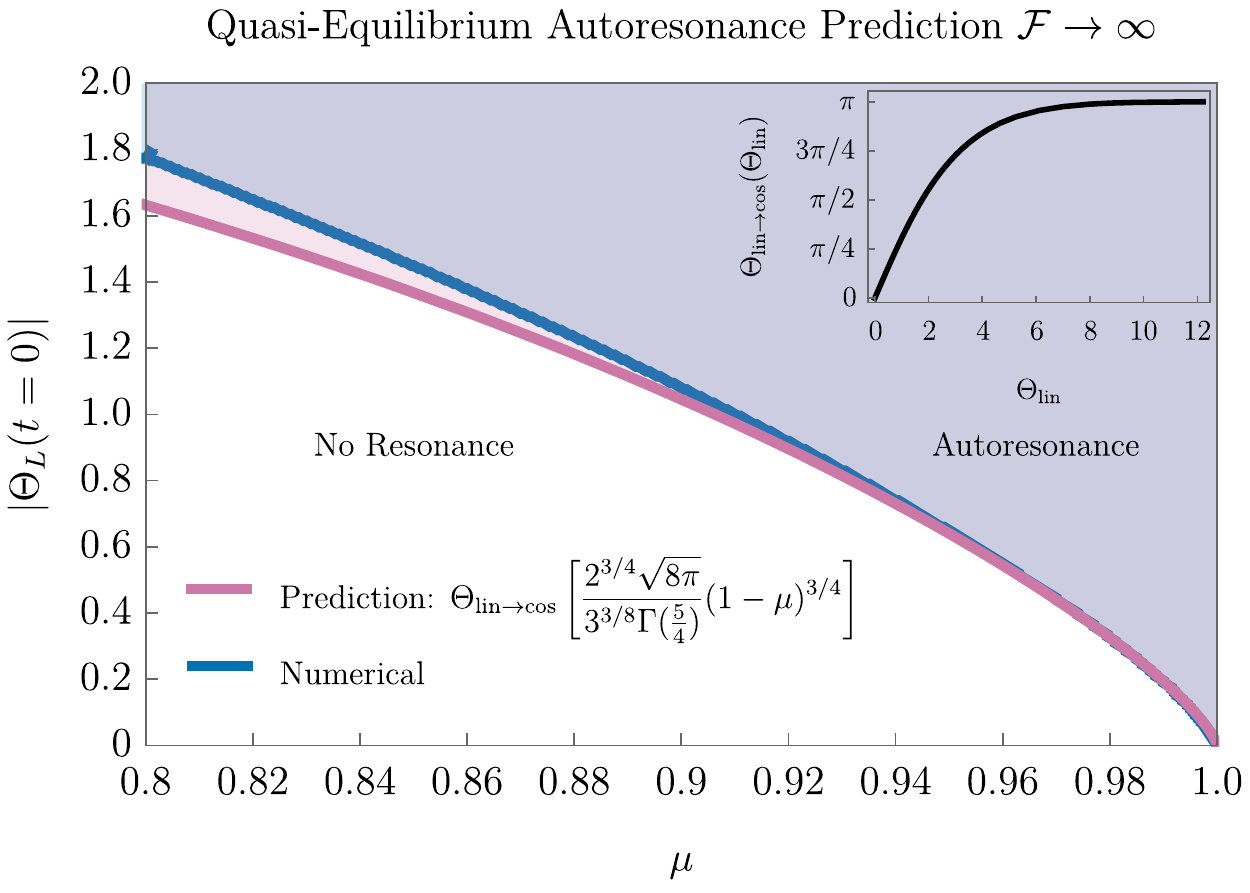}
    \caption{The set of parameters for which $\Ths$ ends up autoresonating for $\Ths(0) = 0$ in the $\calF\to\infty$ limit. We compare a numerical evaluation (Blue) to the analytic adiabatic prediction of the critical driver amplitude. The numeric autoresonance region corresponds to those parameters for which $\Ths$ has finite amplitude as $t\to\infty$. The analytic contour is obtained as the minimum driver amplitude for which a quasi-equilibrium configuration connects the zero amplitude linear resonance at $t = 0$ with the finite amplitude nonlinear resonance at $t = \infty$ as in Fig.~\ref{fig:equilibriumTrajectory}. Note that the analytic estimate improves as $\mu \to 1$, where the evolution of the resonance curve is slowest, and thus is most accurately described by an adiabatic approximation. \textbf{\underline{Inset:}} A plot of the function $\Theta_{\te{lin}\to\te{cos}}$, which takes as input the initial misalignment of a harmonic oscillator, and outputs the misalignment of a cosine oscillator that yields the same late-time relic abundance. Note that this function is the identity at small $\Theta_\te{lin}$.}
    \label{fig:quasiEquilibriumFullPrediction}
\end{figure}

\begin{figure*}[t]
    \centering
    \subfigure{\includegraphics[width=0.49\textwidth]{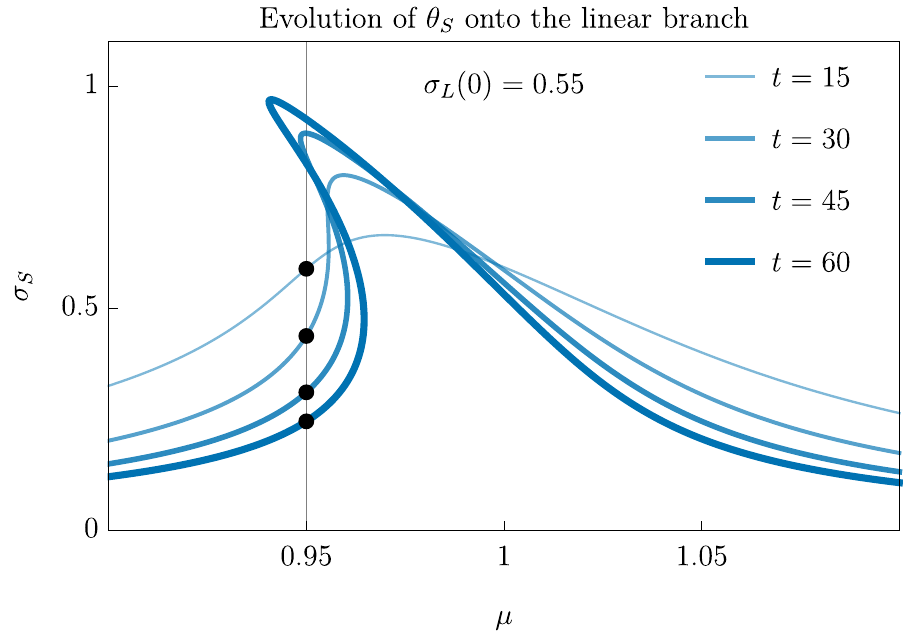}
    }
    \hfill
    \subfigure{\includegraphics[width=0.49\textwidth]{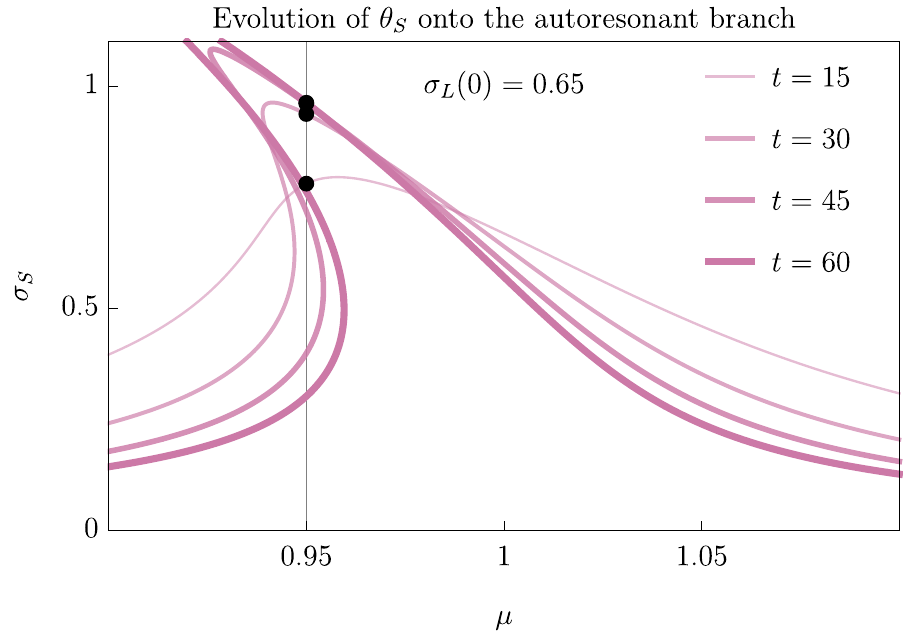}
    }
    \caption{Short axion resonance curves over a sequence of times for two different values of initial long amplitude $\sil(0)$.  Black dots represent the adiabatic evolution of an axion system with $\mu = 0.95$.  The short axion always begins on the linear branch at early times, but its final amplitude is determined by the evolution of the resonance curve.
    \textbf{\underline{Left:}} For $\sil(0) = 0.55$, the resonance curve ``tongue'' grows over the instantaneous equilibrium, leaving $\sis$ on the linear branch. 
    \textbf{\underline{Right:}} For $\sil(0) = 0.65$, the resonance curve narrows under the instantaneous equilibrium, leaving $\sis$ elevated on the nonlinear branch (autoresonance). Note: these resonance curves are made using Eq.~\ref{eq:resonanceCurve} to enable direct comparison with Fig.~\ref{fig:equilibriumTrajectory}; utilizing Eq.~\ref{eq:trueResonanceCurve} does not change the qualitative features of these two classes of evolution history.}
    \label{fig:resCurveEvolution}
\end{figure*}

Here we review the details of the calculations behind the resonance curve results in Sec.~\ref{sec:zeroModes}. For this purpose, we are interested in the $\calF \to \infty$ limit, for which $\Thl$ decouples from $\Ths$ and obeys a simple pendulum equation of motion:
\begin{eq}
    \label{eq:ThetaLEOM}
    \ddot \Theta_L + \frac{3}{2t}\dot \Theta_L +\mu^2 \sin \Theta_L = 0\,.
\end{eq}
For small initial conditions $\sin \Thl \approx \Thl$, the solutions can be obtained analytically:
\begin{eq}
    \Thl(t) = \Thl(0)\,\Gamma(5/4)2^{1/4}\f{J_{1/4}(\mu t)}{(\mu t)^{1/4}}\,,
\end{eq}
which, at late times, is approximately:
\begin{eq}
\Thl(t) = \f{\Thl(0)\Gamma(\f54)2^{3/4}}{\sqrt{\pi}}\f{\sin\p{\mu t + \f{\pi}{8}}}{(\mu t)^{3/4}}\,.
\end{eq}
At large initial conditions comparable to $\pi$, this approximation fails and we must correct the initial amplitude $\Thl(0)$ to account for the delay in oscillations caused by the flatness of the cosine potential. To this end, we define the function $\Theta_{\te{lin}\to\te{cos}}$, which takes as input the initial amplitude of a linear oscillator, and returns the corresponding initial amplitude of a cosine oscillator which results in the same energy density at late times. This function is shown in the inset of Fig.~\ref{fig:quasiEquilibriumFullPrediction}.  At small amplitudes it is approximately the identity, while at large amplitudes it asymptotes to $\pi$. The late-time amplitude of the full non-linear solution for $\Thl$ in the cosine potential can be written as:
\begin{eq}
    \label{eq:adiabaticSigmaLAmplitude}
    \sil(t) = \Theta^{-1}_{\te{lin}\to\te{cos}}(\Thl(0))\f{\Gamma(\f54)2^{3/4}}{\sqrt{\pi}(\mu t)^{3/4}}\,.
\end{eq}
This instantaneous amplitude $\sil$ will drive the short axion $\Ths$ at the long frequency $\mu$. Specifically, the equation of motion for $\Ths$, obtained in the small-$\Thl$ regime from Eq.~\ref{eq:eomLS}, becomes:
\begin{eq}
    \ddot \Theta_S + \f{3}{2t}\dot\Theta_S + \sin\Ths = - \cos\Ths \, \sil \cos(\mu t - \Phi)\,,
\end{eq}
where $\Phi$ is the relative phase between $\Ths$ and $\Thl$ (which is constant at leading order in the adiabatic approximation).
Note that compared to a standard driven pendulum (Eq.~\ref{eq:LandauOscillator}), the driver is suppressed by a $\cos \Ths$ correction. To leading order, the resonance curve can be obtained by only treating the terms oscillating at the driver frequency $\mu$, which is equivalently thought of as the small-amplitude limit. We thus take $\Ths \approx \sis \cos(\mu t)$ and expand $\sin \Ths$ and $\cos \Ths$ using the Jacobi-Anger formulae:
\begin{subequations}
    \label{eq:JacobiAnger}
    \begin{equation}
        \cos ( \sis \cos \mu t ) = J_0 (\sis) + 2 \sum\limits_{n=1}^{\infty} (-1)^n J_{2 n} (\sis) \cos ( 2 n \mu t ) \,,
    \end{equation}
    \begin{equation}
        \sin ( \sis \cos \mu t ) = 2 \sum\limits_{n=0}^{\infty} (-1)^{n} J_{2n+1}(\sis) \cos ( (2 n + 1) \mu t )\,.
    \end{equation}
\end{subequations}
Keeping only terms of frequency $\mu$, we collect the terms proportional to $\sin\mu t$ and $\cos\mu t$, leading to the equations
\begin{subequations}
    \label{eq:nocomplexnumbers}
    \begin{equation}
        - \mu^2 \sis + 2 J_1(\sis) + J_0(\sis)\sil\cos\Phi = 0\,,
    \end{equation}
    \begin{equation}
        \f{3\mu}{2t}\sis - J_0(\sis)\sil\sin\Phi = 0\,,
    \end{equation}
\end{subequations}
which, upon eliminating the phase shift $\Phi$, lead to the condition defining the resonance curve:
\begin{eq}
    \label{eq:trueResonanceCurve}
    \p{2J_1(\sis) - \mu^2 \sis}^2 + \p{\f{3\mu}{2t}}^2\sis^2 = J_0(\sis)^2 \sil^2\,.
\end{eq}
Expanding the Bessel function to leading non-linear order we arrive at the following approximate small-amplitude resonance curve:
\begin{eq}
    \label{eq:practicalResonanceCurve}
    \sis = \f{\sil(t)\p{1 - \f{\sis^2}{4}}}{\sqrt{\p{1 - \f{\sis^2}{8} - \mu^2}^2 + \f{9\mu^2}{4t^2}}}\,.
\end{eq}
Given a fixed frequency $\mu$, we are interested in tracking the equilibrium solution of $\sis$ given by this resonance condition when the driver amplitude $\sigma_L$ also varies slowly with time, as given by Eq.~\ref{eq:adiabaticSigmaLAmplitude}.

At small $t$ friction dominates, and there is only one real solution to Eq.~\ref{eq:practicalResonanceCurve}. At large $t$, the curve narrows around the free frequency curve as shown in Fig.~\ref{fig:landau curves} and, over a range of frequencies $\mu < 1$, can support two stable solutions on either the nonlinear branch (which asymptotes to a finite amplitude $\sis \approx 4\sqrt{(1 - \mu)}$), or on the linear branch (which tends to zero). In this adiabatic view of the evolving resonance curve, {\em autoresonance} is considered to occur when the early-time solution, which starts on the linear branch, is continuously connected to a late-time solution on the upper, non-linear branch. Autoresonance does not happen if the system remains on the linear branch. To find the critical point between these two regimes, it suffices to look for the largest $\Thl(0)$ for which the linear branch at $\mu$ never becomes complex. Solving for this condition in Eq.~\ref{eq:practicalResonanceCurve} leads to the $\Thl(0)$ amplitude cutoff
\begin{align}
    \label{eq:analyticalAutoresonanceCutoff}
    \Thl(0)\geq\Theta_{\te{lin}\to\te{cos}}\ps{\f{2^{3/4}\sqrt{8\pi}}{3^{3/8}\Gamma(\f{5}{4})}(1 - \mu)^{3/4}}\,.
\end{align}
We plot the resulting critical initial misalignment $\Thl(0)$ versus $\mu$ in Fig.~\ref{fig:quasiEquilibriumFullPrediction}, where we find excellent agreement between the analytical threshold (Magenta) and the numerical threshold (Blue).  In Fig.~\ref{fig:resCurveEvolution} we show an explicit comparison between an initial $\Thl(0)$ amplitude that results in autoresonance and one that does not.  The values are chosen to match those shown in Fig.~\ref{fig:equilibriumTrajectory}.

\subsection{Expected relic density ratio}
In this section, we derive the maximum relic abundance ratio at the end of autoresonance. In essence, this calculation assumes that autoresonance carries on until $\thl$ is small enough that flavor oscillations dominate its energy density $\rho_L$. Thus our goal in this section is to calculate the minimal size of flavor oscillations.

To begin, consider the mass eigenstates $\nu_l$ and $\nu_h$ with masses $m_l$ and $m_h$ respectively (see App.~\ref{app:massVsInteractionBasis}). We will assume that the total energy density is fixed at a constant $\rho_0$:
\begin{align}
    \rho_0 &= \f12\gen{\dot\nu_l^2} + \f12 m_l^2\gen{\nu_l^2} + \f12\gen{\dot\nu_h^2} + \f12 m_h^2\gen{\nu_h^2} \\
    &=m_l^2\gen{\nu_l^2} + m_h^2\gen{\nu_h}^2\,.
\end{align}
We will maximize the ratio $\rho_S/\rho_L$ subject to fixed $\rho_0$. Using our definition of $\rho_S$ and $\rho_L$ in Eq.~\ref{eq:arbitraryPartition}, expanding in the small $\Theta_S,\Theta_L$ limit, and using the fact that $m_l\neq m_h$ so that $\gen{\nu_l\nu_h} = 0$, we have
\begin{align}
\notag
    \rho_S&=\f14\gen{\nu_l^2}\p{2(1 - \sin(2\eta)) + m_l^2(1 - \cos(2\eta))}\\& + \f14\gen{\nu_h^2}\p{2(1 + \sin(2\eta)) + m_h^2(1 +\cos(2\eta))}\,,\\
    \notag
    \rho_L&=\f12\gen{\nu_l^2}(m_l^2 + \mu^2)\cos^2\eta \\&+ \f12\gen{\nu_h^2}(m_h^2 + \mu^2)\sin^2\eta\,.
\end{align}
One can check that the ratio $\rho_S/\rho_L$ is maximized for $\nu_l = 0$ provided $\mu < 1$,
\begin{align}
    \max\f{\rho_S}{\rho_L}&=\f{(1 + m_h^2)\csc^2\eta + 2\cot\eta - m_h^2}{\mu^2 + m_h^2}\,.
\end{align}
Expanding for $\calF\gg(1 - \mu^2)^{-1}$ and $1-\mu\to 0$, we find
\begin{align}
    \max\f{\rho_S}{\rho_L}\approx 4 \calF^2(1 - \mu)^2\,.
\end{align}
This estimate is essentially the envelope of the hourglass shape in Fig.~\ref{fig:autoresParamSpace}.  Our numeric results nearly saturate this bound, indicating that autoresonance transfers virtually all energy density out of the long field. An additional factor of $1/2$ appears to do a good job matching the measured final ratio:
\begin{equation}
    \left. \frac{\rhos}{\rhol} \right|_{\text{observed}} \approx 2 \calF^2 (1 - \mu)^2\,.
\end{equation}
\subsection{Stability of autoresonance}
We now derive a set of equations for the amplitude and phase of both axions during autoresonance and use them to compute the evolution of excitations on top of the autoresonance solution. We begin with the coupled equations of motion of Eq.~\ref{eq:eomLS}, where we are measuring time in units of $m^{-1}$.  Because we are expecting approximately periodic solutions, we make the following ansatz for $\Thl(t)$ and $\Ths(t)$:
\begin{subequations}
\begin{equation}
    \Ths = \sis \Re [e^{i \vps}]\,,
\end{equation}
\begin{equation}
    \Thl = \sil \Re [e^{i \vpl}]\,.
\end{equation}
\end{subequations}
We will assume that $\sis, \sil, \vpsd, \vpld$ (where dots denote time derivatives) all vary slowly compared to the oscillatory timescale $1/m$.  We can then insert our ansatz into the equations of motion and expand, keeping only the lowest order in $\sil$ and only the lowest harmonics of $\vps$ to obtain:
\begin{widetext}
\begin{subequations} \label{eq:eomLSansatz}
\begin{equation}
    \sildd + 2 i \vpld \sild + i \vpldd \sil - \vpld^2 \sil + \frac{3}{2 t} ( \sild + i \vpld \sil ) + (\mu^2 + \mathcal{F}^{-2} ) \sil = - 2 \frac{1}{\mathcal{F}^2} J_1 ( \sis ) e^{i \Phi}\,,
\end{equation}
\begin{equation} \label{eq:sisEqAnsatz}
    \sisdd + 2 i \vpsd \sisd + i \vpsdd \sis - \vpsd^2 \sis + \frac{3}{2 t} ( \sisd + i \vpsd \sis ) + 2 J_1 (\sis) = - \sil J_0 (\sis) e^{- i \Phi }\,,
\end{equation}
\end{subequations}
\end{widetext}
where $\Phi \equiv \vps - \vpl$ is the relative phase of the two oscillators, $J_n$ are Bessel functions, and we have used the Jacobi-Anger identities Eq.~\ref{eq:JacobiAnger}. Note that from Eq.~\ref{eq:sisEqAnsatz} we can see that if $\sis$ becomes so large that $J_0(\sis) = 0$, then $\sis$ is no longer driven. This critical $\sis$ determines a critical driving frequency $\mu\approx 0.64$ below which autoresonance is no longer possible, given by the first zero of $J_0(4\sqrt{1-\mu})$.

Now we may take the real and imaginary parts of Eq.~\ref{eq:eomLSansatz} to obtain a set of four coupled equations:
\begin{subequations} \label{eq:ansatzEqns}
\begin{equation} \label{eq:B18a}
    \sildd - \vpld^2 \sil + \frac{3}{2 t} \sild + (\mu^2 + \mathcal{F}^{-2}) \sil = - \frac{2}{\mathcal{F}^2} J_1 (\sis) \cos \Phi \,,
\end{equation}
\begin{equation} \label{eq:B18b}
    \vpldd + \left( \frac{3}{2 t} + 2 \frac{\sild}{\sil} \right) \vpld = - \frac{2}{\mathcal{F}^2} \frac{J_1 (\sis)}{\sil} \sin \Phi \,,
\end{equation}
\begin{equation} \label{eq:B18c}
    \sisdd - \vpsd^2 \sis + \frac{3}{2 t} \sisd + 2 J_1 (\sis) = - \sil J_0 (\sis) \cos \Phi\,,
\end{equation}
\begin{equation} \label{eq:B18d}
    \vpsdd + \left( \frac{3}{2 t} + 2 \frac{\sisd}{\sis} \right) \vpsd = \sil \frac{J_0 (\sis)}{\sis} \sin \Phi\,.
\end{equation}
\end{subequations}
These equations are interesting in their own right, and can be numerically integrated more efficiently than a rapidly-oscillating system such as Eq.~\ref{eq:eomLS}, but for now we will focus on further simplification.  We wish to find the state of the system when it is stably autoresonating, by which we mean we are looking for a solution for which $\sis$ is roughly constant and $\vpsd \approx \vpld$ so the oscillators are synchronized with each other.  We thus approximate $\sisd \approx \sisdd \approx \vpsdd \approx \vpldd \approx 0$ and obtain:
\begin{subequations}
\begin{equation} \label{eq:B5a}
    (\vpld^2 - \mu^2 - \mathcal{F}^{-2} ) = \frac{3}{2 t} \frac{\sild}{\sil} + \frac{2}{\mathcal{F}^2} \frac{J_1 (\sis)}{\sil} \cos \Phi\,,
\end{equation}
\begin{equation} \label{eq:B5b}
    \left( \frac{3}{2 t} + 2 \frac{\sild}{\sil} \right) \vpld = - \frac{2}{\mathcal{F}^2} \frac{J_1 (\sis)}{\sil} \sin \Phi \,,
\end{equation}
\begin{equation} \label{eq:B5c}
    2 \frac{J_1 ( \sis) }{\sis} - \vpsd^2 = - \sil J_0 (\sis) \cos \Phi\,,
\end{equation}
\begin{equation} \label{eq:B5d}
    \frac{3}{2 t} \vpsd = \sil \frac{J_0 (\sis)}{\sis} \sin \Phi\,.
\end{equation}
\end{subequations}
From these equations we may read off a few things. First, provided $t \gg \sil / \sild$ and $\mathcal{F}^2 \gg \sil^{-1}$, the long oscillator undergoes roughly free motion at its fundamental frequency: $\vpld^2 = \mu^2 + \mathcal{F}^{-2} \approx \mu^2$.  If we demand that $\vpsd \approx \vpld$ to ensure we are in autoresonance, that then implies that $\vpsd \approx \mu$, and from Eq.~\ref{eq:B5d} we can read off an expression for the relative phase of the two oscillators:
\begin{equation} \label{eq:Phib}
    \sin \Phib \approx \frac{3 \mu}{2 t} \frac{\sis}{\sil} \frac{1}{J_0 (\sis)}\,,
\end{equation}
where we have used the bar to denote the fact that this is the relative phase in steady-state autoresonance.

We now turn to the question of how excitations on top of this steady-state solution behave.  This will provide an analytic justification for the numeric observation that autoresonance is a stable condition.  We work in the limit $\calF \to \infty$, which implies $\vpld = \mu$ and $\sild / \sil = -3/(4 t)$.  We may then combine Eqs.~\ref{eq:B18b}, \ref{eq:B18c}, and \ref{eq:B18d} to obtain:
\begin{subequations}
\begin{equation}
    \sisdd + \frac{3}{2t} \sisd - \vpsd^2 \sis + 2 J_1 (\sis) + \sil \cos \Phi = 0\,,
\end{equation}
\begin{equation}
    \Phidd + \frac{3}{2 t} \Phid - \frac{\sil}{\sis} \sin \Phi + \frac{3 \mu}{2 t} + 2 \vpsd \frac{\sisd}{\sis} = 0\,.
\end{equation}
\end{subequations}

We now perturb around the equilibrium autoresonance solution, defining:
\begin{subequations}
\begin{equation}
    \Phi \equiv \Phib + \dPhi\,,
\end{equation}
\begin{equation}
    \sis \equiv \sisb + \dsis\,,
\end{equation}
\end{subequations}
with $\Phib$ defined in Eq.~\ref{eq:Phib} and $\sisb$ defined by the autoresonance condition:
\begin{equation}
    \frac{ 2 J_1(\sisb)}{\sisb} = \mu^2\,.
\end{equation}
Expanding and linearizing yields the pair of equations:
\begin{subequations}
\begin{equation}
    \dsisdd + \frac{3}{2 t} \dsisd - 2 \mu \sisb \dPhid - 2 J_2(\sisb) \dsis - \frac{3 \mu}{2 t} \sisb \dPhi = - \sil \cos \Phib\,,
\end{equation}
\begin{equation}
    \dPhidd + \frac{3}{2 t} \dPhid - \frac{\sil}{\sisb} \cos \Phib \dPhi + 2 \mu \frac{\dsisd}{\sisb} + \frac{3 \mu}{2 t} \frac{\dsis}{\sisb} = 0\,,
\end{equation}
\end{subequations}
where we have substituted in for $\sin \Phib$ with Eq.~\ref{eq:Phib} and approximated $J_0 (\sisb) \approx \sisb$.  For $t \gg 1$, we may neglect several terms, simplifying to:
\begin{subequations}
\begin{equation}
    \dsisdd - 2 J_2 ( \sisb ) - 2 \mu \sisb \dPhid - \frac{3 \mu}{2 t} \sisb \dPhi = 0\,,
\end{equation}
\begin{equation}
    \dPhidd - \frac{\sil}{\sisb} \dPhi + 2 \mu \frac{\dsisd}{\sisb} + \frac{3 \mu}{2 t}\,. \frac{\dsis}{\sisb} = 0
\end{equation}
\end{subequations}
To analyze the stability of autoresonance we can search for first-order perturbative solutions of the form:
\begin{equation}
    \dsis = \adsis e^{i \omp t} \qquad \dPhi = \adPhi e^{i \omp t}\,,
\end{equation}
where we will assume (and then confirm) that $| \omp | \ll 1$.  Plugging this into the above relations and solving yields:
\begin{equation}
    \omp \stackrel{t \gg 1}{\approx} \left( \frac{2 \sil}{\sisb} \frac{J_2 (\sisb)}{4 \mu^2 - 2 J_2 (\sisb)} \right)^{1/2} + i \frac{3}{4 t}\,,
\end{equation}
where the imaginary part in particular demonstrates that fluctuations about the autoresonant solution should damp away as $t^{-3/4}$ at large times.  As predicted, $| \omp | \ll 1$ so our approximations above were safe.

\subsection{Chaotic parameter space} \label{app:chaos}
As we've discussed in the previous sections, there is a wide range of parameter space where the zero-mode is well-described by a slowly varying amplitude and phase. This description neglects the initial phase of transient oscillations, and as we have seen in App.~\ref{app:NLrescurve}, transients often do not play a significant role in the evolution of $\Ths$. This is no longer true if the long axion delivers enough energy to the short axion that it can roll over many vacua, exploring the saddle points of the potential. If $\Ths$ happens to slow down near one of the saddle points, the direction it rolls off will depend sensitively on the details of its trajectory, and consequently, on its initial conditions. In other words, if $\Thl$ starts with enough energy, then the short axion exhibits classical chaos, leading to the intricate striations in Fig.~\ref{fig:energyPaisley} near $\Thl = \pi$. The possibility of chaotic evolution in this type of potential was first pointed out in Ref.~\cite{daido2015domain,daido2016level}.

During chaotic evolution, the short axion receives substantial energy from the long axion, leading to many of the same signatures we have described in Sec.~\ref{sec:signatures}. In particular, the chaotic rolling of $\Ths$ necessarily delays the onset of near-harmonic oscillations, generating large $\Ths$ perturbations, as described in Sec.~\ref{sec:perturbations}. Further, although it is no longer guaranteed, an $\mathcal{O}(1)$ fraction of chaotic initial conditions lead to autoresonance, and consequently enhanced direct detection prospects.

A new behavior is also possible for initial conditions sufficiently close to the boundary between striations in Fig.~\ref{fig:energyPaisley}. For these initial conditions, $\Ths$ spends a long time very close to the apex of the saddle point, causing rapid perturbation growth. If $\Ths$ gets close enough to the hilltop for long enough, the axion field in different parts of space can roll off to either side, creating a network of vacuum bubbles. The cosmological implications of this scenario require further investigation.

\section{Perturbations in detail} \label{app:perturbations_in_detail}

In this section, we provide the details of the perturbation growth rate calculations referenced in Sec.~\ref{sec:perturbations}. We first review the general formalism to numerically compute the full spectrum of axion perturbations. We then go on to describe the analytic approximations made in Sec.~\ref{sec:perturbations_one} and Sec.~\ref{sec:perturbations_two}.

\subsection{General formalism}
Consider a theory of $N$ interacting axions $\phi_1,\dots,\phi_N$ with scalar potential $V$
\begin{align}
    {\cal L}(\phi_1,\dots,\phi_N)&=\f12\p{\sum_{i = 1}^N\partial_\mu\phi_i\partial^\mu\phi_i} - V(\phi_1,\dots,\phi_N)\,.
\end{align}
To study the strongly self-interacting regime of this theory, it is helpful to change variables from the canonically normalized fields $\phi_i$ to the fields $\theta_i \equiv \phi_i/f_i$, where $f_i$ is chosen so that $\theta_i \approx 1$ is the scale of self-interaction. In the two-axion model, there is no ambiguity regarding the choice of $f_i$. However, one generally must take more care in choosing the scales $f_i$ if there are more instantons than axions \cite{mehta2021superradiance}. 

The axion field evolves in the background of the perturbed FLRW metric Eq.~\ref{eq:perturbedFLRW}
where $\Phi(t,{\bf x}) = \sum_{{\bf k}}\Phi_k(t,{\bf k}) e^{\I {\bf k} \cdot{\bf x}}$ is the adiabatic scalar perturbation with spectral components given by Eq.~\ref{eq:Phik0}.
Breaking $\theta_i$ down into homogeneous modes $\Theta_i$ and perturbations $\delta\theta_i$
\begin{align}
    \theta_i(t,{\bf x})&=\Theta_i(t) + \sum_{{\bf k}}\delta\theta_i(t,{\bf k})\,,
\end{align}
we arrive at the following set of equations of motion
\begin{subequations}
\label{eq:full_perturbation_equations}
\begin{equation}
    \ddot\Theta_i + 3 H\dot\Theta_i + \f{1}{f_i^2}\f{\partial V}{\partial\Theta_i}{}= 0\,,
\end{equation}
\begin{equation}
    \delta\ddot{\theta}_{ i} + 3 H\delta\dot{\theta}_{i}  + \f{\tilde k^2}{t}\delta\theta_{ i} + \f{1}{f_i^2}\f{\partial^2V}{\partial\Theta_i\partial\Theta_j}{}\delta\theta_j = {\cal S}_i\,,
\end{equation}
\begin{equation}
    {\cal S}_i \equiv 2\p{\f{t_k}{t}\td{\Phi_k}{t_k}{}\dot\Theta_{ i} -\Phi_k\f{1}{f_i^2}\f{\partial V}{\partial\Theta_i}{}}\,,
\end{equation}
\end{subequations}
where we've specialized to the case of radiation domination, and $\ktw$ and $t_k$ are defined in Eq.~\ref{eq:ktwDef} and Eq.~\ref{eq:Phik0} respectively.
For definiteness, we assume that inflation lasts long enough that the $\delta\theta_i$ initial conditions are well approximated as $\delta\theta_i = \delta\dot\theta_i = 0$.

\subsection{Analytical approximations}
Having reviewed the full set of perturbation equations, we now specialize to the case of the one-axion potential Eq.~\ref{eq:oneAxionPotential}.
As we have described in Sec.~\ref{sec:perturbations_one}, perturbations do not grow at an appreciable rate until after the homogeneous oscillations have settled down near the vacuum. At this point, the potential is well approximated by the leading nonlinear terms
\begin{align}
    \f{1}{f^2}V(f\theta) = \f12\theta^2 + \f{1}{3!}A\theta^3 + \f{1}{4!}B\theta^4\,,
\end{align}
where $A$ and $B$ are constants that may be determined from the full potential by Taylor expanding around the vacuum. The homogeneous mode $\Theta$ then satisfies the following equation of motion
\begin{align}
    0&=\ddot\Theta + \f{3}{2t}\dot\Theta + \p{1 + \f{1}{2}A\Theta + \f{1}{3!}B\Theta^2}\Theta\,.
\end{align}
For $\Theta$ oscillating with small amplitude $\sigma$, its waveform and frequency at leading nontrivial order in $\sigma$ are
\begin{align}
\label{eq:one_axion_solution_theta}
    \Theta(t) &= -\f14 A\sigma^2 + \sigma\cos\omega t + \f{1}{12}A\sigma^2\cos2\omega t\,,\\
\label{eq:one_axion_solution_omega}
    \delta\omega &= \f{3B-5 A^2}{48}\sigma^2\,.
\end{align}
From these formulas, we see that the cubic and quartic interactions have qualitatively distinct effects on the $\Theta$ oscillations. The sign of $B$ controls whether the quartic interaction is attractive or repulsive, leading to slower or faster oscillations respectively. On the other hand, $A$ always works to decrease the fundamental frequency of the oscillations. For positive $A$, the cubic interaction is repulsive for positive $\Theta$ and attractive for negative $\Theta$, ultimately causing $\Theta$ to spend more time at negative values where it is oscillating slower.  For negative $A$ the sides are switched, but in either case the net effect is to decrease the oscillator's frequency.

In the background of the homogeneous $\Theta$ oscillations, the equation of motion for the perturbation $\delta\theta$ is
\begin{subequations}
\label{eq:single_particle_perturbation}
\begin{equation}
    \delta\ddot\theta(t,\tilde k) + \f{3}{2t}\delta\dot\theta(t,\tilde k) + \p{1 + \f{\tilde k^2}{t} + A\Theta + \f12 B\Theta^2}\delta\theta(t,\tilde k) = {\cal S}\,,
\end{equation}
\begin{equation}
    {\cal S} \equiv 2\ps{\f{t_k}{t}\td{\Phi_{k}}{t_k}{}\dot\Theta + \Phi_{k}\p{\Theta + \f12A\Theta^2 + \f{1}{3!}B\Theta^3}}\,.
\end{equation}
\end{subequations}
The source ${\cal S}$ provides the axion with the initial fluctuations that will grow because of parametric resonance. Soon after the exponential growth starts, ${\cal S}$ becomes irrelevant and the perturbation growth rate may be computed from the homogeneous equation
\begin{align} \label{eq:pertSourceless}
    \delta\ddot\theta + \f{3}{2t}\delta\dot\theta + \p{1 + \f{\tilde k^2}{t} + A\Theta + \f12 B\Theta^2}\delta\theta \approx 0\,.
\end{align}
Modes will only grow once Hubble friction is small $H\ll 1$, i.e.\ once $t\gg1$. This allows us to treat the time variation of the Hubble friction, the zero-mode amplitude $\sigma\propto t^{-3/4}$, and the changing frequency $\delta\omega\propto \sigma^2\propto t^{-3/2}$ adiabatically. Thus, we may change variables
\begin{align}
    \delta\theta = e^{- \f{3}{4 t}}\psi\,,
\end{align}
so that $\psi$ obeys the frictionless version of Eq.~\ref{eq:pertSourceless} up to order $t^{-2}$. Inserting the known zero-mode evolution Eq.~\ref{eq:one_axion_solution_theta}, we arrive at the following equation for $\psi$
\begin{align}
\label{eq:mathieu_type}
    \ddot\psi + (1 + \alpha + 2\beta\cos t + 2\gamma\cos2t)\psi = 0\,,
\end{align}
where 
\begin{subequations}
\begin{eq}
    \alpha=\f{\tilde k^2}{t}-\f{1}{24}\p{A^2 - 3B}\sigma^2\,,
\end{eq}
\begin{eq}
    \beta=\f12A \sigma\,,
\end{eq}
\begin{eq}
    \gamma=\f{1}{24}\left( A^2 + 3 B\right)\sigma^2\,.
\end{eq}
\end{subequations}
This Mathieu-type equation can be solved directly by applying a Fourier transformation $t\to\omega_t$:
\begin{align}
\label{eq:floquet_fourier_transform}
\notag
    0&=-\omega_t^2\psi(\omega_t)+ (1+\alpha)\psi(\omega_t) + \beta(\psi(\omega_t+1) + \psi(\omega_t-1))\\&\hspace{0.5cm} +\gamma(\psi(\omega_t+2) + \psi(\omega_t-2))\,.
\end{align}
In this equation, only frequencies related by integer multiples of $m$ couple to one another, and thus this problem can be rephrased in terms of an infinite matrix. To see this, we define $\Gamma_\psi \in [0,1) +  \I \R$, so that its real part represents the non-integer real part of $\omega_t$. We can then label harmonics as:
\begin{align}
    \psi_n(\Gamma_\psi) &\equiv \psi (\Gamma_\psi + n) = \psi (\omega_t)\,.
\end{align}
The Fourier transformed Eq.~\ref{eq:floquet_fourier_transform} is thus equivalent to the matrix equation
\begin{align}
\notag
    0&=\p{-(\Gamma_\psi + n )^2 + 1 + \alpha}\psi_n + \beta\p{\psi_{n + 1} + \psi_{n - 1}}\\&\hspace{0.5cm} + \gamma \p{\psi_{n + 2} + \psi_{n - 2}}\,.
\end{align}
The eigenvalues $\Gamma_\psi$ characterize the growth-rate $\Im \Gamma_\psi$ and frequency $\Re \Gamma_\psi$ of the $\psi$ oscillations. 

To solve for $\Gamma_\psi$, we look for solutions with $|\Gamma_\psi| \to 0$, which corresponds to the principal instability branch of the Mathieu-type equation Eq.~\ref{eq:mathieu_type}. Thus, we approximate $(\Gamma_\psi + n)^2\approx 2n\Gamma_\psi + n^2$, leading to the following eigenvalue problem
\begin{widetext}
\begin{align} \label{eq:eigenProblem}
    0&=\det\mat{ccccccc}{\ddots\\
    &-4 + 4\Gamma_\psi + 1 + \alpha&\beta&\gamma\\
    &\beta&-1 + 2\Gamma_\psi + 1 + \alpha & \beta&\gamma\\
    &\gamma&\beta& 1 + \alpha & \beta&\gamma\\
    &&\gamma&\beta&-1 - 2\Gamma_\psi + 1 + \alpha & \beta\\
    &&&\gamma&\beta&-4 - 4\Gamma_\psi + 1 + \alpha\\
    &&&&&&\ddots}\,.
\end{align}
\end{widetext}
By truncating Eq.~\ref{eq:eigenProblem} at the leading $5\times 5$ elements, we arrive at the following expression for the $\psi$ growth rate
\begin{align}
\label{eq:one_particle_floquet_exponent}
\notag
    \Gamma_\psi &=- \I \sqrt{ \left(\frac{\gamma ^2}{4}-\frac{\alpha ^2}{4}\right)+ \left(\frac{\alpha }{3}-\frac{\gamma }{2}\right)\beta ^2+\frac{5}{36}\beta ^4 + {\cal O}(\sigma^5)}\,,\\
    &=- \I \abs{\delta\omega}\sqrt{1 - \p{1 + \f{\tilde k^2}{2 t_m\delta\omega}}^2}\,,
\end{align}
to order $\sigma^4$ in the root (using the fact that $\tilde k^2/t\sim \sigma^2$), and where $\delta\omega$ is as in Eq.~\ref{eq:one_axion_solution_omega}. Re-introducing the $-3/4t$ term we had absorbed into $\psi$, we arrive at Eq.~\ref{eq:growthRate} for the growth rate of $\delta \theta$: $\Gamma =\Re(-3/4t +\I \Gamma_\psi)$.

The perturbations begin growing when $\Gamma\geq 0$, which we define as the time $t_\te{init}$. Prior to $t = t_\te{init}$, the source term ${\cal S}$ holds $\delta\theta$ at an approximately constant initial amplitude given by Eq.~\ref{eq:dthetaInit},
and thus we arrive at the expression Eq.~\ref{eq:oneParticleGrowthRate} for the amplitude of $\delta\theta$, which we reproduce here for ease of reference:
\begin{subequations}
\begin{eq}
\label{eq:GrowthRateApp}
    \gen{\delta\theta(t,\tilde k)^2}\approx\gen{\delta\theta(t_\te{init},\tilde k)^2}\exp\ps{2\int_{t_\te{osc}}^t\diff t'\Gamma(t',\tilde k)}\,,
\end{eq}
\begin{eq}
    \gen{\delta\theta(t,\tilde k)^2}&\approx \f{\Phi_{k,0}^2}{\p{1 + \f{mt\tilde k^2}{\pi^2}}^2}\,.
\end{eq}
\end{subequations}
The integral in Eq.~\ref{eq:GrowthRateApp} can be evaluated exactly, but the resulting expression is hardly useful. To make clean analytic progress, it is helpful to first compute the integral by ignoring Hubble friction, and then to re-introduce Hubble friction at the end by adding $-(3/4)\log(t_\te{end}/t_\te{init})$. Because $\Theta$ is oscillating at the bottom of the potential with decaying amplitude proportional to $t^{-3/4}$, the frequency shift is $\delta\omega = \delta\omega_\te{osc} (t/t_\te{osc})^{-3/2}$, where $t_\te{osc}$ represents the time at which the zero-mode amplitude starts decaying as $\sigma\propto t^{-3/4}$, and thus $t_\te{osc} = t_\te{init}$ for the single axion model.  In the case of autoresonant axions, $t_\te{osc}$ is the time at which autoresonance ends, which is in general much larger than the time $t_\te{init}$ when perturbations start growing. Substituting our expression for $\delta\omega$ into Eq.~\ref{eq:one_particle_floquet_exponent} and plugging into Eq.~\ref{eq:GrowthRateApp} we arrive at the integrated growth rate (neglecting Hubble friction):
\begin{widetext}
\begin{align} \label{eq:autoresonant_growth}
    \int_{t_\te{osc}}^{\infty}\diff t'\Re\p{\Gamma(t',\tilde k) + \f{3}{4 t_m}}&= 2\tilde k^2 \p{ \p{\f{4(1 - \mu)t_{\text{osc}}}{\tilde k^2} -1}^{1/2}-\arccot\ps{\p{\f{4(1 - \mu)t_{\text{osc}}}{\tilde k^2} -1}^{-1/2}}}\,.
\end{align}
\end{widetext}
The parametric resonance ends at the time
\begin{align}
    t_\te{end}&=\f{t_\te{osc}^3\delta\omega_\te{osc}}{(\tilde k/2)^4}\,,
\end{align}
so we add $-(3/4)\log(t_\te{osc}^2\delta\omega_\te{osc}/(\tilde k/2)^4)$ to account for Hubble damping.

We have thus accounted for perturbation growth in the single-particle model Eq.~\ref{eq:oneAxionPotential}. As we discussed in Sec.~\ref{sec:perturbations_two}, this calculation carries through unchanged in the two-axion model (Eq.~\ref{eq:twoAxionPotential}) for the perturbations of $\ths$ that accrue after the end of autoresonance, where $\delta\omega = (\mu - 1) (t/t_\te{osc})^{-3/2}$. Further, the same physics applies to perturbations that grow during autoresonance, except that $\delta\omega(t)$ is simply constant, fixed by the frequency shift between the long axion and short axion $\delta\omega(t) = \mu - 1$. In our approximate treatment of Hubble friction, the integrated growth rate during autoresonance turns out to be exactly one half the integrated growth rate after autoresonance, although this growth occurs over only roughly $2\%$ of the time.

\section{Simulations of non-perturbative structure growth during radiation domination} \label{app:nonperturbative_in_detail}
In Sec.~\ref{sec:NonpertAutoresonance}, we outlined the results of $3+1d$ numerical simulations in which the collapse of non-perturbative fluctuations lead to the breakdown of autoresonance. In this appendix, we provide the details of these simulations and outline improvements that can be made in future work.

\subsection{Metric perturbations and the equations of motion}
In this first section, we review the equations of motion for a set of scalar particles $\phi_1,\dots,\phi_n$ in a potential $V(\phi_1,\dots,\phi_n)$ in the background of an FLRW spacetime in the presence of adiabatic scalar perturbations $\Phi(t,{\bf x})$ Eq.~\ref{eq:perturbedFLRW}. 
As in previous sections, we work in terms of the variables $\theta_i(t,{\bf x}) = \phi_i(t,{\bf x})/f_i$, where $f_i$ is the scale of self-interaction for $\phi_i$.
Treating the metric perturbations at first order, the $\theta_i$ equations of motion are
\begin{widetext}
\begin{equation}
    \label{eq:full}
    \ps{\p{1 - 2\Phi}\partial_t^2 + \p{3 H\p{1 - 2\Phi} - 4\dot\Phi}\partial_t - \p{1 + 2\Phi}\f{1}{a^2}\nabla^2} \theta_i + \f{1}{f_i^2}\pd{V}{\theta_i}{} = 0\,.
\end{equation}
\end{widetext}
The axion fields $\theta_i$ are endowed with order 1 initial misalignment and homogeneous initial conditions by a sufficiently long period of inflation. 

Unlike in the linearized equations, where each wavelength of $\theta_i$ evolves independently, large $\theta_i$ fluctuations couple different modes, and therefore the relative size of perturbations on different scales becomes important. In other words, we may no longer be agnostic to the phase and amplitude of the metric perturbations $\Phi_k$: a particular realization of the metric perturbation $\Phi$ must be generated from its dimensionless power spectrum inside our integration volume.

Our simulation takes place inside a symmetric box of size $2L$ and resolution $dL$, corresponding to a momentum resolution of $k_\te{max} = \pi/dL$ and $dk = \pi/L$. The dimensionless metric power spectrum is defined in terms of $\Phi_{k}$ as
\begin{subequations}
\begin{eq}
    {\cal P}_\Phi(k) = \p{\f{k}{k_0}}^{n_s-1} \Phi_k^2\,,
\end{eq}
\begin{eq}
    {\cal P}_{\dot \Phi}(k) = \p{\f{k}{k_0}}^{n_s-1} \p{\f{t_k}{2 t}}^2 \p{\td{\Phi_k}{t_k}{}}^2\,,
\end{eq}
\end{subequations}
(see discussion around Eq.~\ref{eq:Phik0} for definitions of $\Phi_k$, $n_s$, and $t_k$).
The dimensionful power spectrum $P$ is defined in terms of the dimensionless power spectrum ${\cal P}$ as
\begin{align}
    P = \f{2\pi^2}{k^3}{\cal P}\,.
\end{align}

A particular realization of the field is then generated from the dimensionful power spectrum with the procedure of Ref.~\cite{pen1997generating}. First, for each point ${\bf k}$ in the momentum grid, generate two random numbers $R_1({\bf k})$ and $R_2({\bf k})$ uniformly distributed on the interval $[0,1]$. Then define
\begin{align}
    \rho_{{\bf k}} = -2\log R_1({\bf k})\,,\hspace{0.5cm}\varphi_{{\bf k}} = 2\pi R_2({\bf k})\,.
\end{align}
A particular realization of the $\Phi$ and $\dot \Phi$ Fourier coefficients is then computed as
\begin{subequations}
\begin{equation}
    \Phi_{{\bf k}}(t) = s \sqrt{V\rho_{{\bf k}} P_\Phi(t,{\bf k})} e^{\I\varphi_{{\bf k}}}\,,\\
\end{equation}
\begin{equation}
    \dot\Phi_{{\bf k}}(t) = s' \sqrt{V\rho_{{\bf k}} P_{\dot \Phi}(t,{\bf k})} e^{\I\varphi_{{\bf k}}}\,,
\end{equation}
\end{subequations}
where $s$ and $s'$ denote the signs of $\Phi_k(t_k)$ and $d\Phi_k(t_k)/dt_k$ respectively and $V = (2 L)^3$ is the comoving integration volume. The zero-momentum terms represent a constant shift of $\Phi_{{\bf k}}$ and $\dot\Phi_{{\bf k}}$, which we remove by setting $\Phi_{\bf 0} = \dot\Phi_{\bf 0} = 0$. The real-space fields $\Phi(t,{\bf x})$ and $\dot\Phi(t,{\bf x})$ are then
\begin{subequations}
\begin{eq}
    \Phi(t,{\bf x}) = \Re \p{\f{dk}{2\pi}}^3\sum_{{\bf k}} \Phi_{{\bf k}}(t) e^{-\I {\bf k}\cdot{\bf x}}\,,
\end{eq}
\begin{eq}
    \dot\Phi(t,{\bf x}) = \Re \p{\f{dk}{2\pi}}^3\sum_{{\bf k}} \dot\Phi_{{\bf k}}(t) e^{-\I {\bf k}\cdot{\bf x}}\,.
\end{eq}
\end{subequations}
These expressions can be written in terms of the fast Fourier transform (FFT), or \texttt{fftn(fftshift($\Phi_k$))} in \texttt{Matlab}.

Finally, we discuss the process of measuring the power spectrum of a real field $F({\bf x})$ at an instant in time. Measuring the density power spectrum is especially important when verifying the $3+1d$ code, since the density power spectrum can be directly compared to the output of the linearized theory of Sec.~\ref{app:perturbations_in_detail}.

In order to measure the power spectrum of a real field $F({\bf x})$, we first compute its Fourier transform
\begin{align}
    F_{{\bf k}} = dL^3 \sum_{{\bf k}} F({\bf x}) e^{-\I {\bf k}\cdot{\bf x}}\,.
\end{align}
The power spectrum of $F({\bf x})$ is the average of $F_{{\bf k}}^2$ over concentric spherical momentum shells. The fact that small $|{\bf k}|$ shells contain fewer momentum grid points means that the one should not trust the the low-frequency power spectrum to reflect the statistical properties of the field.
Define the magnitude of the momentum vector $k_r = \sqrt{{\bf k}^2}$. Let $k_r(n) = k_r$ when $k_r$ is in the interval $[(n-1),n]dk$, and zero otherwise. Let $N_n$ be the number of non-zero elements in $k_r(n)$. The power spectrum is then
\begin{align}
    P_F(n\, dk) = \sum_{k\in k_r(n)} \f{\abs{F_k}^2}{N_n V}\,.
\end{align}

\subsection{Numerical methods}
To evolve the equations of motion Eq.~\ref{eq:full}, we use Runge-Kutta 4th-order (RK4) time-integration, with pseudospectral derivative operators. Here, we provide a brief review of pseudospectral methods.

The Laplacian operator in Eq.~\ref{eq:full} poses a computational challenge: in position space, it represents matrix multiplication, which can be an inefficient process. The pseudospectral method recognizes that the potential is best computed in position-space, where it acts as a pointwise operator, and derivatives are best computed in momentum space, where they acts as pointwise operators. The pseudospectral algorithm to compute derivatives is as follows:
\begin{enumerate}
    \item compute the FFT of $\theta_i$,
    \item apply the derivative operator in momentum space (pointwise multiplication),
    \item compute the inverse FFT (IFFT).
\end{enumerate}
We note that the pseudospectral method is well suited to GPU acceleration, since it makes use of pointwise matrix multiplication and the FFT, both of which have efficient GPU implementations.
Because the FFT is an extremely optimized algorithm, converting between position space and momentum space is an extremely efficient process, in essence making the pseudospectral method an efficient implementation of multiplication that would otherwise need to take place to compute the action of a differential operator. The numerical Laplacian is computed as:
\begin{align}
    \nabla^2\theta = \f{1}{V}\sum_{{\bf k}} (-{\bf k}^2)e^{-\I {\bf k}\cdot{\bf x}}\sum_{{\bf k}}\theta e^{\I {\bf k}\cdot{\bf x}}\,.
\end{align}

We note that to ensure convergence of the pseudospectral method, it is often helpful to suppress the numerical instability of high-frequency modes by truncating momentum space somewhat below the maximum possible resolution of the spatial grid ($k_\te{max} = \pi/dL$). In our calculations, we take this cutoff to be $k_\te{max}/2$.

\subsection{Future directions}
While our preliminary simulations shed some light on the possible consequences of nonlinear fluctuations during autoresonance, we recognize an opportunity to develop higher resolution simulations in order to reach a definitive conclusion. In particular, our simulations are limited in the range of comoving momenta they can resolve, $|\tilde {\bf k}|\in[m,30 m]$, which is particularly restrictive at the time of oscillon formation. Because our simulations take place on a comoving grid, oscillons, whose physical size does not redshift, appear to get smaller, requiring higher and higher momenta to fully resolve. This, combined with the fact that oscillons already have very broad momentum spectra, means that our simulations are substantially less reliable after oscillons have formed, and our observation that autoresonance is terminated by oscillon formation may not hold up to higher resolution simulations.

We end on a tangentially related note that there are additional questions which will only be resolved by $3+1d$ simulations. In particular, it need not be the case that the axion rolls to the true vacuum. For example, the potential Eq.~\ref{eq:oneAxionPotential} will in general have many false vacua, and it may be more likely that the axion rolls there than the true vacuum. In this case, there are two possibilities. First, the axion can quantum tunnel out of the false vacuum into the true vacuum, nucleating vacuum bubbles that quickly expand to fill the universe. Second, the axion can \emph{classically} tunnel out of the false vacuum, also nucleating vacuum bubbles that expand to fill the universe, but potentially on a very different timescale. Classical tunnelling occurs when the axion perturbations become large enough that the field must explore adjacent vacua, and the rapid perturbation growth experienced in potentials such as Eq.~\ref{eq:oneAxionPotential} may make this the dominant tunneling mechanism. Both classical and quantum tunnelling require detailed simulations to resolve signatures such as gravitational wave production and the matter power spectrum.

\bibliography{bibliography}
\end{document}